\documentclass[12pt]{article}

\usepackage{amsfonts}
\usepackage{amssymb}
\usepackage{amsthm}
\usepackage[fleqn]{amsmath}
\usepackage[mathcal]{euscript}
\usepackage{mathrsfs}

\newtheorem{corollary}{Corollary}

\newtheorem{lemma}{Lemma}
\newtheorem{proposition}{Proposition}

\newtheorem{theorem}{Theorem}

\renewcommand{\iint}{{\int\!\!\!\!\int}}
\renewcommand{\iiiint}{{\int\!\!\!\!\int\!\!\!\!\int\!\!\!\!\int}}
\newcommand{\Ref}[1]{(\ref{#1})}

\newcommand{\supp}{\mathrm{supp}\,}

\newcommand{\Prob}{\mathrm{Prob}}

\newcommand{\Vol}{\mathrm{Vol}\,}

\newcommand{\kB}{k_{\mathrm{B}}} % Boltzmann's constant

\newcommand{\eps}{\epsilon}
\newcommand{\veps}{\varepsilon}
\newcommand{\vepskin}{\varepsilon_{\mathrm{kin}}}
\newcommand{\invveps}{{\textstyle{\frac{1}{\veps}}}}

\newcommand{\vect}[1] {\boldsymbol{{ #1}} }

\newcommand{\pV}{{\vect{p}}}           % 3-momentum
\newcommand{\qV}{{\vect{q}}}           % 3-position

\DeclareMathAlphabet{\mathpzc}{OT1}{pzc}{m}{it}

\newcommand\pzcE{{\mathpzc{E}}}
\newcommand\pzcF{{\mathpzc{F}}}

\newcommand\pzcH{{\mathpzc{H}}}
\newcommand\pzcHB{\pzcH_{\mathrm{B}}}

\newcommand\pzcQ{{\mathpzc{Q}}}
\newcommand\pzcR{{\mathpzc{R}}}
\newcommand\pzcS{{\mathpzc{S}}}

\newcommand{\abs}[1]{\left| #1 \right|}
\newcommand{\norm}[1]{\left\| #1 \right\| }

\newcommand{\oli}[1]{\overline #1 }
\newcommand{\uli}[1]{\underline #1 }

\newcommand{\dKR}{d_{\mathrm{KR}}}
\newcommand{\dd}{\mathrm{d}}

\newcommand{\withNto}{\stackrel{\textrm{\tiny N}\to\infty}{\longrightarrow}}

\newcommand{\mun}{{{}^{n}\!{\mu}}}
\newcommand{\dotmun}{{{}^{n}\!{\dot\mu}}}
\newcommand{\varrhok}{{{}^{k}\!{\varrho}}}
\newcommand{\varrhom}{{{}^{m}\!{\varrho}}}
\newcommand{\varrhon}{{{}^{n}\!{\varrho}}}
\newcommand{\dotvarrhon}{{{}^{n}\!{\dot\varrho}}}

\newcommand{\Nset}{\mathbb{N}}
\newcommand{\Rset}{\mathbb{R}}
\newcommand{\Sset}{\mathbb{S}}

\newcommand{\Asp}{\mathfrak{A}}

\newcommand{\Csp}{\mathfrak{C}}

\newcommand{\Lsp}{\mathfrak{L}}

\newcommand{\Psp}{\mathfrak{P}}

\newcommand{\cE}{{\cal E}}

\newcommand{\cS}{{\cal S}}

\newcommand{\cU}{{\cal U}}

\newcommand{\bP}{{\mathbf P}}
\newcommand{\bQ}{{\mathbf Q}}

\newcommand{\bX}{{\mathbf X}}

\newcommand{\bTheta}{{\mathbf \Theta}}

\newcommand{\frN}{{\textstyle{\frac{1}{N}}}}
\newcommand{\frnN}{{\textstyle{\frac{n}{N}}}}
\newcommand{\frNT}{{\textstyle{\frac{1}{N\vartheta}}}}
\newcommand{\tfrhalf}{{\textstyle{\frac{1}{2}}}}
\newcommand{\tfrN}{{\textstyle{\frac{1}{N}}}}

\begin{document}

\title{The\! Vlasov\! continuum\! limit\! for\! the\\ classical\!
	 microcanonical\! ensemble}

\vspace{-0.3cm}
\author{\normalsize \sc{Michael K.-H. Kiessling}\\[-0.1cm]
	\normalsize Department of Mathematics, Rutgers University\\[-0.1cm]
	\normalsize Piscataway NJ 08854, USA}
\vspace{-0.3cm}
\date{$\phantom{nix}$}
\maketitle
\vspace{-1.6cm}

\begin{abstract}
\noindent
	For classical Hamiltonian $N$-body systems with mildly regular pair interaction potential 
(in particular, $\Lsp^2_{loc}$ integrability is required) it is shown that when $N\to\infty$ in a
fixed bounded domain $\Lambda\subset\Rset^3$, with energy $\cE$ scaling as $\cE\propto N^2$, then 
Boltzmann's ergodic ensemble entropy $S_\Lambda(N,\cE)$ has the asymptotic expansion 
$S_\Lambda(N,N^2\veps) = - N\ln N + s_{_\Lambda}(\veps) N + o(N)$;
here, the $N\ln N$ term is combinatorial in origin and independent of the rescaled Hamiltonian, 
while $s_{_\Lambda}(\veps)$ is the system-specific Boltzmann entropy per particle, i.e.
$-s_{_\Lambda}(\veps)$ is the minimum of Boltzmann's $H$ function for a perfect gas of energy $\veps$ subjected
to a combination of externally and self-generated fields.
	It is also shown that any limit point of the $n$-point marginal ensemble measures is a 
linear convex superposition of $n$-fold products of the $H$-function-minimizing one-point functions.
	The proofs are direct, in the sense that (a)  the map
$\cE\mapsto S(\cE)$ is studied rather than its inverse $\cS\mapsto E(\cS)$;
(b) no regularization of the microcanonical measure $\delta(\cE-H)$ is invoked, and (c) 
no detour via the canonical ensemble.
	The proofs hold irrespective of whether microcanonical and canonical ensembles are equivalent or not.
\end{abstract}

%%%%%%%%%%%%%%%%%%%%%%%%%%%%%%%%%%%%%%%%%%%%%%%%%%%%%%%%%%%%%%%%%
\vfill
\hrule
\smallskip\noindent
{\small 
Typeset in \LaTeX\ by the author. Version of August 11, 2009.% orig. subm.: February 12, 2009. 

\smallskip\noindent
Appeared in: {\sl Reviews in Mathematical Physics} {\bf 21}, 1145--1195 (2009).

\smallskip\noindent
\copyright 2009 The author.
This preprint may be reproduced for non-commercial purposes.}
%%%%%%%%%%%%%%%%%%%%%%%%%%%%%%%%%%%%%%%%%%%%%%%%%%%%%%%%%%%%%%%%%%%%%%%%%%%%%%%%%%%%%%%%%
%%%%%%%%%%%%%%%%%%%%%%%%%%%%%%%%%%%%%%%%%%%%%%%%%%%%%%%%%%%%%%%%%%%%%%%%%%%%%%%%%%%%%%%%%
%%%%%%%%%%%%%%%%%%%%%%%%%%%%%%%%%%%%%%%%%%%%%%%%%%%%%%%%%%%%%%%%%%%%%%%%%%%%%%%%%%%%%%%%%
\section{Introduction}
%%%%%%%%%%%%%%%%%%%%%%%%%%%%%%%%%%%%%%%%%%%%%%%%%%%%%%%%%%%%%%%%%%%%%%%%%%%%%%%%%%%%%%%%%
%%%%%%%%%%%%%%%%%%%%%%%%%%%%%%%%%%%%%%%%%%%%%%%%%%%%%%%%%%%%%%%%%%%%%%%%%%%%%%%%%%%%%%%%%
%%%%%%%%%%%%%%%%%%%%%%%%%%%%%%%%%%%%%%%%%%%%%%%%%%%%%%%%%%%%%%%%%%%%%%%%%%%%%%%%%%%%%%%%%
\vskip-.2truecm
\noindent
	The rigorous foundations of equilibrium statistical mechanics have largely been laid long ago
\cite{RuelleBOOK,PenroseBOOK,BattelleSeattleRecontres,MartinLoef}, but the most basic problem 
in classical statistical mechanics, namely the rigorous asymptotic evaluation of Gibbs' 
microcanonical ensemble \cite{Gibbs} in the limit of a large number $N$ of particles, has 
only been treated in an approximate way.
	The standard way of dealing with the microcanonical ensemble (a.k.a. Boltzmann's ergodic ensemble 
\cite{Boltzmann}) in a rigorous manner \cite{RuelleBOOK,Lanford,MartinLoef}
has been to replace its singular ensemble measure by a regularized measure (usually also 
referred to as microcanonical, although quasi-microcanonical would seem a better name).
	In these approaches one \emph{cannot} take the limit of vanishing regularization; yet,
since one can approximate the singular measure as closely as one pleases, 
``this is not completely unsatisfactory from a conceptual point of view'' (\cite{Lanford}, p.4).
	All the same, Lanford's wording makes it plain that it is desirable to find a way to remove
the regularization or to  avoid it altogether.

	Recently \cite{KieRUELLEpaper} the author noticed that after only minor modifications, Ruelle's
method \cite{RuelleBOOK} to establish the thermodynamic limit for Boltzmann's ergodic ensemble entropy, 
taken per volume (or per particle), works without the need for any regularization of the ensemble measure;
a follow-up work on the thermodynamic limit of the correlation functions is planned.
	Taking ``\emph{the} thermodynamic limit'' \cite{RuelleBOOK} means that the domain $\Lambda$ 
grows ``evenly'' with $N$ and such that $N/\Vol(\Lambda)\to \rho$ with $\rho$ a fixed number density, and 
the energy $\cE$ scales such that $\cE/\Vol(\Lambda)\to\veps$ (or $\cE/N\to\veps$, abusing notation), 
with $\veps\in\Rset$ a fixed energy density (or energy per particle) --- this limit covers 
systems of interest in condensed matter physics or chemical physics, such as those with hard core or 
Lennard-Jones interactions.

	In the present paper we will be concerned with another limit $N\to\infty$, where $\Lambda$ is  
fixed and $\cE$ scales such that $\cE/ N^2\to\veps$. 
	This limit covers systems of interest in 
plasma and astrophysics, such as those with Coulomb or (mollified) Newton interactions.
	It is variably known\footnote{The first two names refer to a Weiss-type ``mean-field approximation'' 
		becoming exact in the limit, but we will not invoke any such approximation and speak of the \emph{Vlasov limit}.}
as a ``thermodynamic mean-field limit,'' a ``self-averaging limit,'' or ``Vlasov limit.'' 
	We will study the Boltzmann ensemble entropy {and} the correlation functions.

	The remainder of this paper is structured as follows.
	In section 2 we collect the defining formulas of the ergodic / microcanonical ensemble for finite $N$ and 
explain which probabilistic quantities are of physical interest.
	In section 3 we give a heuristic motivation for the Vlasov  limit. 
	In section 4 we state our main theorems, ordered by increasing depth.
	Their proofs are given in sections 5.1 to 5.3.
	Section 6 lists some spin-offs of our results, and section 7 closes our paper with an outlook on some 
open problems.
%
%%%%%%%%%%%%%%%%%%%%%%%%%%%%%%%%%%%%%%%%%%%%%%%%%%%%%%%%%%%%%%%%%%%%%%%%%%%%%%%%%%%%%%%%%
%%%%%%%%%%%%%%%%%%%%%%%%%%%%%%%%%%%%%%%%%%%%%%%%%%%%%%%%%%%%%%%%%%%%%%%%%%%%%%%%%%%%%%%%%
%%%%%%%%%%%%%%%%%%%%%%%%%%%%%%%%%%%%%%%%%%%%%%%%%%%%%%%%%%%%%%%%%%%%%%%%%%%%%%%%%%%%%%%%%
\section{A brief review of the ergodic ensemble}
%%%%%%%%%%%%%%%%%%%%%%%%%%%%%%%%%%%%%%%%%%%%%%%%%%%%%%%%%%%%%%%%%%%%%%%%%%%%%%%%%%%%%%%%%
%%%%%%%%%%%%%%%%%%%%%%%%%%%%%%%%%%%%%%%%%%%%%%%%%%%%%%%%%%%%%%%%%%%%%%%%%%%%%%%%%%%%%%%%%
%%%%%%%%%%%%%%%%%%%%%%%%%%%%%%%%%%%%%%%%%%%%%%%%%%%%%%%%%%%%%%%%%%%%%%%%%%%%%%%%%%%%%%%%%
\vskip-.2truecm
\noindent
	For a Newtonian $N$-body system\footnote{All particles belong to a single specie.
		We use units of $mc^2$ for energy,  $mc$ for momentum, and
		$h/mc$ for length, where $m$ is particle mass, $c$ the speed of light, $h$ Planck's constant.}
in a domain $\Lambda\subset\Rset^3$ with Hamiltonian\footnote{It is understood that 
		$W_\Lambda$ is symmetric, i.e. $W_\Lambda(\qV,\tilde{\qV})=W_\Lambda(\tilde{\qV},\qV)$,
		and not reducible to a sum of one-body terms; other details of $W$ and $V$ will be 
		specified in the next section.}
\begin{equation} 
H^{(N)}_\Lambda (\pV_1,\dots,\qV_N)
=
 \sum_{1\leq i\leq N}
{\tfrhalf}\abs{\pV_i}^2
+
\sum\sum_{\hskip-.7truecm 1\leq i < j\leq N} W_\Lambda(\qV_i,\qV_j)
+
 \sum_{1\leq i\leq N} V_\Lambda^{(N)}(\qV_i)\,,
\label{HAM}
\end{equation}	
\vskip-.2truecm
\noindent
the ergodic / microcanonical ensemble is a family $\{\bX^{(N)}_k|k\in\Nset\}$ of i.i.d. copies of a random vector
$\bX^{(N)} = (\bP_1,\bQ_1;\dots;\bP_N,\bQ_N) \in (\Rset^{3}\!\times\!\Lambda)^{N}$  distributed according to
the stationary\footnote{Stationarity is defined w.r.t. the flow generated by the 
			Hamiltonian $H^{(N)}_\Lambda(\pV_1,\dots,\qV_N)$.} 
single-system a-priori probability measure 
\begin{equation}
\mu^{(N)}_\cE(\dd^{6N}\!X)
= 
\big(N!{\Omega^\prime_{H^{(N)}_\Lambda}(\cE)}\big)^{-1}\,\delta\big(\cE-H^{(N)}_\Lambda(X^{(N)})\big)\dd^{6N}\!X\,,
\label{BOLTZMANNsERGODE}
\end{equation}
\vskip-.2truecm
\noindent
where $X^{(N)}: =(\pV_1,\qV_1;...;\pV_N,\qV_N)\in(\Rset^3\!\times\!\Lambda)^N$ and
$\dd^{6N}\!X$ is $6N$-dimensional Lebesgue measure, and where
\begin{equation}
\Omega^\prime_{H^{(N)}_\Lambda}(\cE) 
= 
{\textstyle{\frac{1}{N!}}}
\int \delta\!\left(\cE-H^{(N)}_\Lambda(X^{(N)})\right)\dd^{6N}\!X
\label{STRUKTURfunktion}
\end{equation}
\vskip-.1truecm
\noindent
is known as the \textit{structure function};\footnote{The $N!$ term was supplied by Gibbs to resolve Gibbs' paradox.
			It cancels out in \Ref{BOLTZMANNsERGODE}.}
here, the $^\prime$ means derivative w.r.t. $\cE$ of
\begin{equation}
\Omega_{H^{(N)}_\Lambda}(\cE) 
= 
{\textstyle{\frac{1}{N!}}}\int \chi_{\left\{H^{(N)}_\Lambda<\cE\right\}}\dd^{6N}\!X\,,
\label{STRUKTURfunktionPRIMITIVE}
\end{equation}
\vskip-.1truecm
\noindent
where $\chi_{\{H^{(N)}_\Lambda<\cE\}}$ is the characteristic function of the  set 
$\{H^{(N)}_\Lambda(X^{(N)})<\cE\}\in (\Rset^3\!\times\!\Lambda)^N$, over which the integrals extend.
	Thus, if $\mathscr{B}$ denotes the Borel sets of $(\Rset^3\!\times\!\Lambda)^N\subset\Rset^{6N}$, then
$\bigl((\Rset^3\!\times\!\Lambda)^N,\mathscr{B},\mu^{(N)}_\cE\bigr)$ is the single-system probability space;
so if $B\in\mathscr{B}$ is a Borel set, then the probability of $\bX^{(N)}$ being in $B$ 
is\footnote{It is tacitly understood that whenever one encounters a
		physically interesting subset $L$ of a Borel null-set which is not itself
		Borel measurable, then we use the Lebesgue $\sigma$-algebra.}
\begin{equation}
\Prob\left(\bX^{(N)}\in B \right) = \mu^{(N)}_\cE(B)\,.
\end{equation}
\vskip-.1truecm
\noindent
	Clearly, $\Prob\left(\bX^{(N)}\in B \right) = 0$ unless $B\cap \{H^{(N)}_\Lambda =\cE\}\neq \emptyset$;
put differently, $\Prob\bigl(H^{(N)}_\Lambda (\bX^{(N)}_k)=\cE\bigr) = 1\ \forall\ k\in \Nset$.
	Moreover, $\Prob\left(\bX^{(N)}\in \dd^{6N}\!X\right) = \mu^{(N)}_\cE(\dd^{6N}\!X)$ 
is the a-priori probability for $\bX^{(N)}$ to be in $\dd^{6N}\!X$ about $X^{(N)}$.

	The ergodic ensemble is probabilistically meaningful for all $N\in\Nset$, yet its thermodynamic significance 
emerges only in the large $N$ regime (Avogadro's $N\approx 10^{23}$) when it makes sense to speak of a solid, 
a liquid, a plasma (etc.) on macroscopic scales of space and time.
	Since the typical physical characteristics of solids and liquids (etc.) are 
not revealed by ``picturing'' such systems as individual points in $\Rset^{6N}$, one associates each microstate 
$X^{(N)}$ with a unique family of empirical $n$-point ``densities'' on~$(\Rset^3\!\times\!\Lambda)^n$, 
$n=1,2,...,N$.
	The \emph{normalized one-point ``density'' with $N$ atoms} (empirical measure) is given by
\begin{equation}
\uli\Delta^{(1)}_{X^{(N)}}(\pV,\qV)
 = 
{\textstyle{\frac{1}{N}}} \sum_{1\leq i\leq N} \delta(\pV-\pV_i)\delta(\qV-\qV_i)
\label{normalEMPmeasONE}
\end{equation}
\vskip-.15truecm
\noindent
and the \emph{normalized two-point density with $N$ atoms} ($U$-statistic of order 2) by
\begin{equation}
\uli\Delta^{(2)}_{X^{(N)}}(\pV,\qV;\pV',\qV')
 = 
{\textstyle{\frac{1}{N(N-1)}}}
\sum\sum_{\hskip-.7truecm 1\leq i \neq j\leq N}
\delta(\pV-\pV_i)\delta(\qV-\qV_i)\delta(\pV'-\pV_j)\delta(\qV'-\qV_j);
\label{normalEMPmeasTWO}
\end{equation} 			
\vskip-.2truecm
\noindent
similarly the empirical $n$-point densities with $n =3,...,N$ are defined.
	The map $X^{(N)}\to\{\uli\Delta^{(n)}_{X^{(N)}}\}_{n=1}^N$ is bijective if we insist that the particular
labeling given to us algebraically with r.h.s.\Ref{normalEMPmeasONE} or r.h.s.\Ref{normalEMPmeasTWO} etc.
has an intrinsic meaning; however, considered purely measure theoretically as ``density'' on $\Rset^{6n}$ each 
$\uli\Delta^{(n)}_{X^{(N)}}$ is invariant under the permutation group applied to the particular labeling, and 
since there are $N!$ distinct $X^{(N)}$s obtained by permuting the particle labels, the map 
$X^{(N)}\to\{\uli\Delta^{(n)}_{X^{(N)}}\}_{n=1}^N$ is many-to-one in this sense.
	Understood in this measure theoretic way the empirical $n$-point densities do not depend on 
the unphysical (though mathematically convenient) labeling of the particles,\footnote{So, physically 
		we can identify these $N!$ distinct $X^{(N)}$s 
		with a single \emph{$N$-point configuration in $\Rset^3\!\times\!\Lambda$}, which is a point 
		$\widetilde{X}^{(N)}\in\Rset^{3N}\!\times\!\Lambda^N_{\neq}/S_N$. 
		The subscript $_{\neq}$ means
		that coincidence points are removed, and $S_N$ is the symmetric group of order $N$.
		We should also write $\uli\Delta^{(n)}_{\widetilde{X}^{(N)}}$, 

\vskip-.1truecm
\noindent
		with the understanding that as measure $\uli\Delta^{(n)}_{\widetilde{X}^{(N)}}$ is given by
		$\uli\Delta^{(n)}_{X^{(N)}}$ for \emph{any} of the $N!$ points

\vskip-.1truecm
\noindent
		$X^{(N)}$ in the pre-image in $\Rset^{6N}$ of $\widetilde{X}^{(N)}$.
		The  map $\widetilde{X}^{(N)}\to \{\uli\Delta^{(n)}_{\widetilde{X}^{(N)}}\}_{n=1}^N$
		is bijective.}
and so are physically more natural than points in $\Rset^{6N}$; when $n$ is small, say $n=1$ or $2$, then 
$\uli\Delta^{(n)}_{X^{(N)}}$ is also ``physically more manifest'' than a point in $\Rset^{6N}$.

	Hence, the probabilities of interest to physicists will be of the form
\begin{equation}
\Prob\left(\uli\Delta^{(n)}_{\bX^{(N)}}\in \widetilde{B}\right)
\label{physPROBnormalEMPmeas}
\end{equation}
\vskip-.2truecm
\noindent
for physically significant measurable sets $\widetilde{B}$ in $\Psp^s((\Rset^3\!\times\!\Lambda)^n)$,
the permutation-symmetric probability measures on $(\Rset^3\!\times\!\Lambda)^n$.
	Among the physically significant sets are balls (w.r.t. a suitable topology, still to be chosen) 
centered at a representative $n$-point density function for a solid, liquid, ... , or complements of such balls.
	As for the topology, the fine ($TV$) topology for $\Psp^s((\Rset^3\!\times\!\Lambda)^n)$ is \emph{not}
suitable as it is equivalent to discriminating between different $\uli\Delta^{(n)}_{X^{(N)}}$ w.r.t. 
the Borel sigma algebra of $\Rset_{\neq}^{6N}/S_N$ (see footnote 6).
	Practically accessible\footnote{Even if the balls in $TV$ topology \emph{were} practically accessible, 
		for $N\gg 1$ the amount of information would be sheer overwhelming and not very illuminating.}
are only some considerably less finely resolved events, such as the empirical $n$-point densities 
$\uli\Delta^{(n)}_{X^{(N)}}$ distinguished w.r.t. the weak topology, quantified by a convenient 
Kantorovich-Rubinstein metric $\dKR$ on $\Psp^s((\Rset^3\!\times\!\Lambda)^n)$. 
	Two very different points in $\Rset_{\neq}^{6N}/S_N$, say $\widetilde{X}^{(N)}$ and $\widetilde{Y}^{(N)}$, 
can map into two densities $\uli\Delta^{(n)}_{X^{(N)}}$ and 
$\uli\Delta^{(n)}_{Y^{(N)}}$ which in weak topology 
on $\Psp^s((\Rset^3\!\times\!\Lambda)^n)$ are virtually indistinguishable; here $X^{(N)}$ and $Y^{(N)}$ are any
two representative points out of the $N!$ points each, which constitute the pre-image in $\Rset^{6N}$ of 
$\widetilde{X}^{(N)}$, respectively $\widetilde{Y}^{(N)}$.
	So the probabilities of interest to physicists are  typically of the form
\begin{equation}
\Prob\left(\dKR\left(\uli\Delta^{(n)}_{\bX^{(N)}},f^{(n)}_{\mathrm{eq}}\right) > \delta\right) ,
\label{physPROBnormalEMPmeasDISTfeq}
\end{equation}
where 
$f^{(n)}_{\mathrm{eq}}(\pV^{(1)},\qV^{(1)};...;\pV^{(n)},\qV^{(n)}) 
		\in (\Psp\cap \Csp^0_b)((\Rset^3\!\times\!\Lambda)^n)$ 
is an \emph{equilibrium} density function, defined --- in the simplest of all cases ---  
implicitly as the \emph{unique} function for which (after rescaling of variables and parameters, if necessary)
\begin{equation}
\Prob\left(\dKR\left(\uli\Delta^{(n)}_{\bX^{(N)}},f^{(n)}_{\mathrm{eq}}\right) > \delta\right) \withNto 0
\quad \forall \delta>0.
\label{physPROBnormalEMPmeasDISTfeqTOzero}
\end{equation}
	In these simplest of all cases, \Ref{physPROBnormalEMPmeasDISTfeqTOzero} also explains what is 
meant by a ``representative $n$-point density function;'' and whether $f^{(n)}_{\mathrm{eq}}$ represents a solid, 
liquid, gas, etc., depends on the specific configurational correlations exhibited by $f^{(n)}_{\mathrm{eq}}$.
	In more complicated (and more interesting) situations, several ``competing'' equilibrium functions 
$f^{(n)}_{\mathrm{eq}}$ may exist, and \Ref{physPROBnormalEMPmeasDISTfeqTOzero} has to be 
modified accordingly.

	The ``simplest case'' scenario just described was discovered by Boltzmann (p.442 of \cite{Boltzmann}),
based on his explicit evaluation of \Ref{BOLTZMANNsERGODE} for the perfect gas.
	He realized that when $H^{(N)}_\Lambda$ is the perfect gas Hamiltonian and $N\gg 1$, then basically 
every point of $\{H^{(N)}_\Lambda=\cE\}$ (identified with an $n$-pt. density through the map 
$X^{(N)}\to\uli\Delta^{(n)}_{X^{(N)}}$) lies in the vicinity (w.r.t. weak topology) of one and the same 
equilibrium density function $f^{(n)}_{\mathrm{eq}}$ at that energy $\cE$, and given $n$.
	When  $H^{(N)}_\Lambda$ sports non-trivial pair interactions, Boltzmann's description needs to be modified 
slightly to account for the phenomenon of phase transitions.

	While there can hardly be a doubt that Boltzmann's insight into \Ref{BOLTZMANNsERGODE} is correct, the 
rigorous results which support his assessment have been obtained not for \Ref{BOLTZMANNsERGODE} but for 
some regularized approximation of this singular measure \cite{RuelleBOOK,BattelleSeattleRecontres,MartinLoef}.
	In this paper we will finally vindicate Boltzmann's ideas in the Vlasov regime of the relevant class of
Hamiltonians \Ref{HAM}.
%
%%%%%%%%%%%%%%%%%%%%%%%%%%%%%%%%%%%%%%%%%%%%%%%%%%%%%%%%%%%%%%%%%%%%%%%%%%%%%%%%%%%%%%%%%
%%%%%%%%%%%%%%%%%%%%%%%%%%%%%%%%%%%%%%%%%%%%%%%%%%%%%%%%%%%%%%%%%%%%%%%%%%%%%%%%%%%%%%%%%
%%%%%%%%%%%%%%%%%%%%%%%%%%%%%%%%%%%%%%%%%%%%%%%%%%%%%%%%%%%%%%%%%%%%%%%%%%%%%%%%%%%%%%%%%
\section{Heuristic considerations on the Vlasov limit}
%%%%%%%%%%%%%%%%%%%%%%%%%%%%%%%%%%%%%%%%%%%%%%%%%%%%%%%%%%%%%%%%%%%%%%%%%%%%%%%%%%%%%%%%%
%%%%%%%%%%%%%%%%%%%%%%%%%%%%%%%%%%%%%%%%%%%%%%%%%%%%%%%%%%%%%%%%%%%%%%%%%%%%%%%%%%%%%%%%%
%%%%%%%%%%%%%%%%%%%%%%%%%%%%%%%%%%%%%%%%%%%%%%%%%%%%%%%%%%%%%%%%%%%%%%%%%%%%%%%%%%%%%%%%%
\vskip-.2truecm
\noindent
        For the ergodic ensemble to exhibit a Vlasov regime the Hamiltonian \Ref{HAM} needs to satisfy 
additional conditions.
	In particular, a necessary condition on the symmetric and irreducible pair potential $W_\Lambda$ is local 
integrability, i.e. $W_\Lambda(\qV,\,\cdot\,)\in\Lsp^1(B_r(\qV)\cap\Lambda)\; \forall\qV\in\Lambda$.
	We remark that for the existence of a \emph{dynamical} Vlasov regime the local integrability of the 
\emph{forces} derived from $W_\Lambda$ is mandatory, viz.
$\nabla_\qV W_\Lambda(\qV,\,\cdot\,)\in\Lsp^1(B_r(\qV)\cap\Lambda)\; \forall\qV\in\Lambda$.
        Coulomb's electrical and Newton's gravitational interactions belong in either class.
	Physically meaningful external potentials $V_\Lambda^{(N)}$ are continuous for $\qV\in\Lambda$; it
has minor technical advantages to assume that $V_\Lambda^{(N)}$ is actually continuous also at
the boundary, i.e. $\lim_{\qV'\to\qV}V_\Lambda^{(N)}(\qV') =V_\Lambda^{(N)}(\qV)$ for all $\qV\in\partial\Lambda$
and $\qV'\in\Lambda$.
	For convenience we assume that $\inf H^{(N)}_\Lambda (\pV_1,\dots,\qV_N) = 
\min H^{(N)}_\Lambda (\pV_1,\dots,\qV_N) = \cE_g(N)>-\infty$, and call $\cE_g(N)$ the $N$-body ground state 
energy;\footnote{Presumably boundedness below is not technically necessary.
		We expect that pair interactions which diverge logarithmically to $-\infty$ can be accommodated 
		but require additional weak compactness estimates, e.g. in some $\Lsp^p$ space; cf. 
		\cite{KieLebLMP}.}
Newton's gravitational interactions need to be regularized to achieve $\cE_g(N)>-\infty$.

	In the introduction we have already mentioned that the Vlasov limit scaling for such interactions
is $\cE \asymp N^2\veps$ for $N\gg 1$.
	We now explain why.
	Integrating \Ref{normalEMPmeasONE} over $\pV$-space $\Rset^3$ gives a 
normalized one-point ``density'' (empirical measure) on $\Lambda$ with $N$ atoms, which by abuse 
of notation we denote as follows,
\begin{equation}
	\uli\Delta^{(1)}_{X^{(N)}}(\qV)
\equiv
	\int_{\Rset^3}\uli\Delta^{(1)}_{X^{(N)}}(\pV,\qV)\dd^3p
= 
{\textstyle{\frac{1}{N}}}\sum_{1\leq i\leq N} \delta(\qV-\qV_i).
\label{normalEMPmeasOFq}
\end{equation}
	Whenever Boltzmann's simplest scenario holds, then there is an equilibrium
density $\rho_{\cE,N}\in (\Psp\cap \Csp^0_b)(\Lambda)$, depending on $N(\gg 1)$ and $\cE$, such that
$\uli\Delta^{(1)}_{X^{(N)}}(\qV) \approx \rho_{\cE,N}(\qV)$ for overwhelmingly most $X^{(N)}$ distributed by 
\Ref{BOLTZMANNsERGODE}, where ``$\approx$'' means the two ``densities'' do not differ by much in a 
conventional Kantorovich-Rubinstein metric $\dKR$.
	This suggests that when $\Lambda\subset\Rset^3$ is fixed and
$N\to\infty$ together with $\cE\to\infty$ such that $\cE/N^\alpha\to\veps$ for a yet-to-be determined $\alpha$, 
then $\rho_{\cE,N}\stackrel{N\to\infty}{\longrightarrow}\rho_\veps\in (\Psp\cap \Csp^0_b)(\Lambda)$ 
and $\uli\Delta^{(1)}_{X^{(N)}}(\qV)\dd^3p\stackrel{N\to\infty}{\longrightarrow} \rho_\veps(\qV)$, weakly.
	Implementing this law-of-large-numbers type scenario inevitably leads to $\alpha=2$, as 
is most easily seen if we assume for a moment that $W_\Lambda\in\Csp^0_b(\Lambda\!\times\!\Lambda)$.
	Then $\qV\mapsto W_\Lambda({\qV,\qV})$ is a bounded continuous function in $\Lambda$ and we can write
\begin{eqnarray}
H^{(N)}(X^{(N)}) \!\!\! &=&\!\!\!  
N \iint {\tfrhalf}|\pV|^2 \uli\Delta^{(1)}_{X^{(N)}}(\pV,\qV)\dd^3p\dd^3q
\nonumber\\
&&\!\! +
N\iint\!\! 
\Bigl(V_\Lambda^{(N)}(\qV) -{\tfrhalf}W_\Lambda({\qV,\qV}) \Bigr)\uli\Delta^{(1)}_{X^{(N)}}(\pV,\qV)\dd^3p\dd^3q 
\label{HasAfctOFempMEAS}\\
&&\!\! +  N^2
\iiiint
{\tfrhalf} 
W_\Lambda({\qV,\tilde{\qV}})\uli\Delta^{(1)}_{X^{(N)}}(\pV,\qV)\dd^3p\dd^3q\uli\Delta^{(1)}_{X^{(N)}}(\tilde{\pV},\tilde{\qV})
\dd^3\tilde{p}\dd^3\tilde{q},
\nonumber
\end{eqnarray}
and when
$\int_{\Rset^3}\uli\Delta^{(1)}_{X^{(N)}}(\pV,\qV)\dd^3p \approx \rho_\veps
(\qV)$, we find
\begin{eqnarray}
H^{(N)}(X^{(N)})
 \!\!\!&\approx &\!\!\!  N \iint 
{\tfrhalf}|\pV|^2 \uli\Delta^{(1)}_{X^{(N)}}(\pV,\qV)\dd^3p\dd^3q  
\nonumber\\
&&\!\! + N\int\!\! 
\Bigl(V_\Lambda^{(N)}(\qV) 
-{\tfrhalf}W_\Lambda({\qV,\qV}) \Bigr)\rho_\veps(\qV)\dd^3q 
\label{HasAfctOFempMEASuRHO}\\
&&\!\! + N^2\int\!\!\!\int{\tfrhalf} 
W_\Lambda({\qV,\tilde{\qV}})\rho_\veps(\qV)\rho_\veps(\tilde{\qV})\dd^3q \dd^3\tilde{q}.
\nonumber
\end{eqnarray}
	The last term clearly scales $\propto N^2$ because $W_\Lambda$ and $\rho_\veps$ 
are independent of $N$.
	In a sense this already establishes the $\cE\propto N^2$ scaling.
	However, we have yet to consider the terms on the first two lines on the r.h.s. of 
\Ref{HasAfctOFempMEASuRHO}.
	It would seem that these scale $\propto N$ and so, for large $N$, would become insignificant as 
compared to the 
one in the last line, but only the $N\int \frac{1}{2}W_\Lambda({\qV,\qV})\rho_\veps(\qV)\dd^3q$ contribution 
will surely become insignificant\footnote{Incidentally, this indicates that the Vlasov limit does not require
		the continuity of $W_\Lambda$, the only purpose of which was to furnish identity 
		\Ref{HasAfctOFempMEAS} which involves $W_\Lambda(\qV,\qV)$.}
for large $N$, for the same reasons for why the last one scales $\propto N^2$
($W_\Lambda$ and $\rho_\veps$ do not depend on $N$).
	As for the external potential $V_\Lambda^{(N)}(\qV)$, the superscript $^{(N)}$ indicates that we
may want to adjust it to the number of particles in the system on which it acts in order to retain 
a noticeable effect when $N$ becomes large. 
	So in particular we can set $V_\Lambda^{(N)}(\qV) = N V_\Lambda(\qV)$ [or $=(N-1)V_\Lambda(\qV)$], 
with $V_\Lambda(\qV)$ independent of $N$,  and find
$N\int V_\Lambda^{(N)}(\qV) \rho_\veps(\qV)\dd^3q = N^2\int V_\Lambda(\qV) \rho_\veps(\qV)\dd^3q$
[$+O(N)$], scaling $\propto N^2$ [in leading order], hence remaining significant in \Ref{HasAfctOFempMEASuRHO} 
as $N$ becomes large.
	And as to the kinetic energy term, it is important to realize that 
$\int_{\Rset^3}\uli\Delta^{(1)}_{X^{(N)}}(\pV,\qV)\dd^3p \approx \rho_\veps(\qV)\in (\Psp\cap \Csp^0_b)(\Rset^3)$
does \textit{not} imply that 
$\uli\Delta^{(1)}_{X^{(N)}}(\pV,\qV)\approx f_\veps(\pV,\qV) \in (\Psp\cap \Csp^0_b)(\Rset^3\!\times\!\Lambda)$.
	For instance, we can have that
$N^{3/2}\uli\Delta^{(1)}_{X^{(N)}}(N^{1/2}\pV,\qV) \approx f_\veps(\pV,\qV)\in (\Psp\cap \Csp^0_b)(\Rset^3\!\times\!\Lambda)$
so that a significant fraction of the energy will be distributed over the kinetic 
degrees of freedom,\footnote{Unless $\cE$ is the ground state energy for which all particle 
		momenta vanish, indeed.}
and then, up to terms of $O(N)$, we find
\begin{eqnarray}
H^{(N)}(X^{(N)}) \!\! &\approx &\!\!  
N^{2}\left(\iint \left({\tfrhalf}|\pV|^2 + V_\Lambda(\qV)\right) f_\veps(\pV,\qV)\dd^3p\dd^3q\right.
\label{HasAfctOFf}\\
&& \ + 
\left.\iiiint {\tfrhalf} W_\Lambda({\qV,\tilde{\qV}})
f_\veps(\pV,\qV)f_\veps(\tilde{\pV},\tilde{\qV})\dd^3p\dd^3q \dd^3\tilde{p}\dd^3\tilde{q}\right).
\nonumber
\end{eqnarray}
	This scaling scenario can be verified explicitly for the perfect gas ($W_\Lambda\equiv 0$) 
by inspecting Boltzmann's calculations, and it is reasonable to expect that it will continue to hold 
for a physically interesting class of $W_\Lambda\not\equiv 0$.

	To summarize, the Vlasov limit for the Hamiltonian \Ref{HAM} with $V^{(N)}=NV$
means that $N^{3/2}\uli\Delta^{(1)}_{\bX^{(N)}}(N^{1/2}\pV,\qV)\stackrel{N\to\infty}{\longrightarrow} 
f_\veps(\pV,\qV)$ weakly in $\Psp(\Rset^3\!\times\!\Lambda)$, with 
$f_\veps(\pV,\qV)\in(\Psp\cap \Csp^0_b)(\Rset^3\!\times\!\Lambda)$,
and $N^{-2} H^{(N)}(X^{(N)})\stackrel{N\to\infty}{\longrightarrow} \pzcE(f_\veps) =\veps > \veps_g$,
where
\begin{eqnarray}
\pzcE(f)\!\! &=&\!\! \iint \left( {\tfrhalf}|\pV|^2 +V_\Lambda(\qV)\right) f(\pV,\qV)\dd^3p\dd^3q
\nonumber\\
&&\!\! + \iiiint{\tfrhalf}
W_\Lambda(\qV,\tilde{\qV})f(\pV,\qV)f(\tilde{\pV},\tilde{\qV})\dd^3p\dd^3q \dd^3\tilde{p}\dd^3\tilde{q}
\label{eOFf}
\end{eqnarray}
is the ``energy of $f$,'' and where $\veps_g = \inf_{f\in\Psp(\Rset^3\!\times\!\Lambda)}\pzcE(f)$ is given by
\begin{equation}
\veps_g
= \inf_{\rho\in\Psp(\Lambda)}
\Big(\int V_\Lambda(\qV)\rho(\qV)\dd^3q
+ \iint{\tfrhalf}
W_\Lambda(\qV,\tilde{\qV})\rho(\qV)\rho(\tilde{\qV})\dd^3q\dd^3\tilde{q}\Big).
\label{Enull}
\end{equation}
%
%%%%%%%%%%%%%%%%%%%%%%%%%%%%%%%%%%%%%%%%%%%%%%%%%%%%%%%%%%%%%%%%%%%%%%%%%%%%%%%%%%%%%%%%%
%%%%%%%%%%%%%%%%%%%%%%%%%%%%%%%%%%%%%%%%%%%%%%%%%%%%%%%%%%%%%%%%%%%%%%%%%%%%%%%%%%%%%%%%%
%%%%%%%%%%%%%%%%%%%%%%%%%%%%%%%%%%%%%%%%%%%%%%%%%%%%%%%%%%%%%%%%%%%%%%%%%%%%%%%%%%%%%%%%%
\section{The Vlasov limit for Boltzmann's Ergode} 
%%%%%%%%%%%%%%%%%%%%%%%%%%%%%%%%%%%%%%%%%%%%%%%%%%%%%%%%%%%%%%%%%%%%%%%%%%%%%%%%%%%%%%%%%
%%%%%%%%%%%%%%%%%%%%%%%%%%%%%%%%%%%%%%%%%%%%%%%%%%%%%%%%%%%%%%%%%%%%%%%%%%%%%%%%%%%%%%%%%
%%%%%%%%%%%%%%%%%%%%%%%%%%%%%%%%%%%%%%%%%%%%%%%%%%%%%%%%%%%%%%%%%%%%%%%%%%%%%%%%%%%%%%%%%
	We now state our main results about the Vlasov scaling limit for Boltzmann's ergodic 
{ensemble} of $N$-body systems in a format which will be recognized as the familiar folklore
by anyone with a joint expertise in Vlasov theory and statistical mechanics.
	We will also utilize some less familiar notions.

	In the following, $\Lambda\subset\Rset^3$ is  a bounded, connected domain (open) which
does not depend on $N$.
	The upshot of the previous section is that if we want the external potential to remain significant
when $N$ gets large, then our $N$-body dynamics in $\Lambda$ will be governed
by Hamiltonians \Ref{HAM} of the special type
\begin{equation} 
H^{(N)}_\Lambda (\pV_1,\dots,\qV_N)
=
 \sum_{1\leq i\leq N}\left(
{\tfrhalf}\abs{\pV_i}^2 
+ (N-1)V_\Lambda(\qV_i)\right)
+ \sum\sum_{\hskip-.7truecm 1\leq i < j\leq N} W_\Lambda(\qV_i,\qV_j),
\label{HAMwithNV}
\end{equation}	
with the single particle potential $V_\Lambda$ and the pair interaction $W_\Lambda$  independent of $N$.
	We choose $N-1$ rather than $N$ as scaling for $V_\Lambda^{(N)}$ because then we can absorb 
$V_\Lambda$ and $W_\Lambda$ together in a new $N$-independent \emph{effective pair interaction}
$U_\Lambda(\qV,\tilde{\qV}):=W_\Lambda(\qV,\tilde{\qV})+V_\Lambda(\qV)+ V_\Lambda(\tilde{\qV})$.
	This doesn't affect any of our results (as we will prove), but the Hamiltonian \Ref{HAMwithNV} 
can be recast shorter as 
\begin{equation} 
{H}^{(N)}_\Lambda (\pV_1,\dots,\qV_N)
=
 \sum_{1\leq i\leq N} {\tfrhalf}\abs{\pV_i}^2 
+ \sum\sum_{\hskip-.7truecm 1\leq i < j\leq N} {U}_\Lambda(\qV_i,\qV_j).
\label{HAMwithU}
\end{equation}	
	Since heuristically we expect for a Hamiltonian system with Hamiltonian \Ref{HAMwithU}
under Vlasov scaling  that
$N^{3/2}\uli\Delta^{(1)}_{\bX^{(N)}}(N^{1/2}\pV,\qV)\stackrel{N\to\infty}{\longrightarrow} f_\veps(\pV,\qV)$
weakly, with $f_\veps(\pV,\qV)\in(\Psp\cap \Csp^0_b)(\Rset^3\!\times\!\Lambda)$, we also expect that the 
rescaled particle momentum random vectors $N^{-1/2}\bP_i$ converge in distribution, implying that 
$\sum_{i} \frac{1}{2}\abs{\bP_i}^2 \approx N^{2}\vepskin$ for $\mu^{(N)}_\cE$-most $\bX^{(N)}$, where 
$\vepskin$ is the kinetic energy contribution to $\veps$.
	We find it more convenient to work with random variables which themselves converge in distribution
and so re-scale the momentum variables as $\pV_k = N^{1/2}\tilde\pV_k$ in \Ref{HAMwithU};
or in more economical notation: we replace $\pV_k \to N^{1/2}\pV_k$ in \Ref{HAMwithU}. 
	With this minor additional abuse of notation our  Hamiltonian finally reads
\begin{equation} 
{H}^{(N)}_\Lambda (\pV_1,\dots,\qV_N)
=
N \sum_{1\leq i\leq N} {\tfrhalf}\abs{\pV_i}^2 
+ \sum\sum_{\hskip-.7truecm 1\leq i < j\leq N} {U}_\Lambda(\qV_i,\qV_j).
\label{HAMrescaled}
\end{equation}	

	Our main results will be proved under the following hypotheses on $U_\Lambda(\qV,\tilde{\qV})$:
\begin{eqnarray}
&&(H1)\quad   {\mbox{\textit{Symmetry}:}}
		\ U_\Lambda(\check{\qV},\hat{\qV})=U_\Lambda(\hat{\qV},\check{\qV}) 
\nonumber\\
&&(H2)\quad   {\mbox{\textit{Lower\ Semi-Continuity}:}}
	\ U_\Lambda(\check{\qV},\hat{\qV})\ {\rm is\ l.s.c.\ on\ }  \oli{\Lambda}\!\times\!\oli{\Lambda} 
\nonumber\\
&&
(H3)\quad  {\mbox{\textit{Sublevel\ Set\ Regularity}:}}\,
	\iint\chi_{\big\{
U_\Lambda(\check{\qV},\hat{\qV})- \min U_\Lambda < \eps
			\big\}}\dd^3\check{q}\dd^3\hat{q}
>0 
\nonumber\\
&&
(H4)\quad  {\mbox{\textit{Local\ Square\ Integrability}:}}
	\ U_\Lambda(\qV,\,\cdot\,)\in \Lsp^2\left(B_r(\qV)\cap\Lambda\right)\ \forall\ \qV\in{\Lambda}
\nonumber\\
&&
(H5)\quad {\mbox{\textit{Confinement}:}}\
	U_\Lambda(\check{\qV},\hat{\qV})=
	+\infty\ \mathrm{whenever}\ \check{\qV}\not\in\oli{\Lambda}\ \mathrm{or}\ \hat{\qV}\not\in\oli{\Lambda}
\nonumber
\end{eqnarray}
	Hypothesis $(H1)$ is a consequence for $W_\Lambda$ of Newton's ``actio equals re-actio,'' plus the 
symmetrized added contribution of $V_\Lambda$, both of which need no further commentary.
	Hypothesis $(H2)$ is satisfied by many important pair interactions invoked in physics, though not by all.
	For instance, the Coulomb pair potential 
	$U_\Lambda^{Coul}(\check{\qV},\hat{\qV}) = 1/|\check{\qV}-\hat{\qV}|$ for 
$\check{\qV}\neq\hat{\qV}$ satisfies $(H2)$ after also setting $U_\Lambda^{Coul}(\qV,\qV) \equiv u$ for any particular $u\in \Rset$.
	On the other hand, the Newton pair potential 
$U_\Lambda^{Newt}(\check{\qV},\hat{\qV}) = -U_\Lambda^{Coul}(\check{\qV},\hat{\qV})$
does not satisfy $(H2)$ for any choice of $u$; however, the regularized 
Newton pair potential 
$U_{\Lambda, reg}^{Newt}(\check{\qV},\hat{\qV}) =
 -(\chi_{_{B_r}}*U_\Lambda^{Coul}*\chi_{_{B_r}})(\check{\qV},\hat{\qV})$
(where $f*g$ denotes the conventional convolution product of $f$ and $g$)
does satisfy $(H2)$.
	By $(H2)$, there exists an $N$-dependent ground state energy $\cE_g(N)$, i.e. 
$H^{(N)}_\Lambda \geq \cE_g(N)>-\infty$, but the ground state configuration can have
some unwanted features.\footnote{For instance, in our example of the amended Coulomb
		pair potential one can choose $u=0$, but then Thomson's problem on $\Sset^2\subset\Rset^3$ 
		\cite{Thomson} yields as ground state configuration always the 
		spurious one (up to $SO(3)$ action) for which all particle positions coincide.
		To avoid these spurious ground state configurations it is advisable to choose $u>0$ huge.}
	Hypothesis $(H3)$ eliminates the possibility of energetically isolated ground states, thus guaranteeing
the existence of a fat set of minimizing sequences of configurations. 
	Hypothesis $(H4)$ is a little stronger than necessary, but it allows us to make convenient use of 
Chebychev's inequality to prove a law of large numbers for the pair-specific interaction energy; the important
Coulomb potential satisfies $(H4)$.
	Note that $(H4)$ implies local $\Lsp^1$ integrability of $U_\Lambda$, which is needed in various
integrals featuring in the Vlasov limit.  
	Note also that by $(H2)\&(H4)$ there exists an $N$-independent $\veps_g\in\Rset$ defined by \Ref{Enull}.
	In Appendix A we show that
$(H1)\& (H2)$ guarantee that the pair-specific ground state energy $\cE_g(N)/[N(N-1)] \equiv\veps_g(N)$ is monotonic 
increasing with $N$, and using also $(H3)$ and $(H4)$ we show that $\veps_g(N)\nearrow\veps_g$ as $N\to\infty$.
	In Appendix A we also show that if $U_\Lambda\geq 0$, then also 
$\cE_g(N)/N^2 \equiv\tilde\veps_g(N)\nearrow\veps_g$ as $N\to\infty$.
	Hypothesis $(H5)$ is really inherited from the dynamical theory of $N$ particles in 
$\Lambda\subset\Rset^3$, where one sets $V_\Lambda^{(N)}=+\infty$ for $\qV\not\in\oli{\Lambda}$
to dynamically model confinement in a container; $(H5)$ has a minor notational advantage
by allowing us to treat physical space integrals like momentum space integrals as over all $\Rset^3$, 
the spatial cutoff to $\Lambda$ automatically being provided by the potential $V_\Lambda$ through $U_\Lambda$. 
	Usually, $(H5)$ is not listed explicitly as a hypothesis on the interactions even when spatial 
integrations are explicitly restricted to $\Lambda$. 
	This concludes our commentary on the list of hypotheses $(H1)-(H5)$.

	All our results (except Proposition \ref{prop:UperNsqMinusN} in Appendix A) will be formulated 
and proved under the convenient assumption that  $U_\Lambda\geq 0$, so that $\veps_g \geq 0$.
	Since $U_\Lambda$ has a minimum in $\oli\Lambda^2$, by $(H2)$, and since the physics
of our dynamical system does not change if we simply add a constant to $U_\Lambda$, 
we may assume that  $U_\Lambda\geq 0$ without loss of generality.
	We emphasize that this choice is merely for convenience, given $(H2)$, and so is not
listed as another hypothesis.

	The simplest objects of interest are the \emph{thermodynamic functions}.
	In the 1960s and hence, techniques based on monotonicity, convexity and super-additivity estimates
have been developed to prove their existence and regularity 
in the limit $N\to\infty$ which avoids having to control the more sophisticated objects of interest, 
which are the correlation functions.
	For the traditional thermodynamic limit scaling, see Ruelle's book \cite{RuelleBOOK} and 
\cite{KieRUELLEpaper} for a recent extension of Ruelle's arguments to Boltzmann's Ergode proper.
	For the Vlasov scaling of the canonical ensemble, see \cite{KieCPAM}. 
	To extend these arguments to Boltzmann's Ergode proper with Vlasov scaling, 
our first goal is to show that the logarithm of the structure function 
\Ref{STRUKTURfunktion} for the Hamiltonian \Ref{HAMrescaled}, which 
yields Boltzmann's ergodic ensemble entropy\footnote{Entropy 
		is measured in units of $\kB$, where $\kB$ is Boltzmann's constant.}
(cf. eq.(305) in \cite{Gibbs}),
\begin{equation}
S_{H^{(N)}_\Lambda}(\cE) 
= \ln \Omega^\prime_{H^{(N)}_\Lambda}(\cE),
\label{BOLTZMANNentropy}
\end{equation}
{admits} the correct type of asymptotic expansion for $N\to\infty$ with 
$\cE=N^2\veps$, and has the correct qualitative $\veps$ dependence.
	The usual strategy can be put to work if we assume just a little more than $(H1)$--$(H5)$.
	In this vein we state:
\begin{theorem}
\label{thm:ENTROPYperPARTICLElimit}
	Let $H^{(N)}_\Lambda$ be given in \Ref{HAMrescaled}, with $U_\Lambda$ 
satisfying conditions $(H1)$ and $(H5)$, but with $(H2)$, $(H3)$, $(H4)$ replaced by the 
single stronger condition:
\begin{equation}
(H6)\qquad {\mbox{\textit{Continuity}:}}
	\ U_\Lambda(\check{\qV},\hat{\qV})\ {\rm is\ continuous\ on\ }  \oli{\Lambda}\!\times\!\oli{\Lambda} .
\end{equation}
	Let  $\veps >\veps_g$, with $\veps_g\geq 0$ defined as before. 
	Then the ergodic ensemble entropy \Ref{BOLTZMANNentropy} has the following asymptotic expansion
for $N\gg 1$,
\begin{equation}
S_{H^{(N)}_\Lambda}(N^2\veps) 
= - N\ln N + N s_{_\Lambda}(\veps) + o(N)\,,
\label{MCentropyASYMPexpansionI}
\end{equation}
where $s_{_\Lambda}(\veps)$ is the system-specific Boltzmann entropy per particle.
	The function $\veps\mapsto s_{_\Lambda}(\veps)$ is continuous and strictly increasing for $\veps>\veps_g$.
\end{theorem}

	We remark that the leading term of r.h.s.\Ref{MCentropyASYMPexpansionI} is purely combinatorial 
in origin and independent of the Hamiltonian $H^{(N)}_\Lambda$ --- it is solely due to the $N!$ in 
\Ref{STRUKTURfunktion}.
	System-specific information begins to show in the next to leading term, which is $O(N)$.
	The $o(N)$ term in \Ref{MCentropyASYMPexpansionI} is presumably $O(\ln N)$.

	We will also prove two upgrades of Theorem $1$ (Theorems $1^{+}$ and $1^{++}$) which involve the 
decomposition of the system-specific Boltzmann entropy per particle $s_{_\Lambda}(\veps)$ into a ``kinetic'' 
and an ``interaction'' contribution.
	The discussion of this more technical material is postponed until section 5.1.

	While they do yield valuable qualitative information about the thermodynamic functions for the systems under 
study, in this case $s_{_\Lambda}(\veps)$, existence theorems such as Theorem \ref{thm:ENTROPYperPARTICLElimit} 
and their ``proofs by sub-additivity'' have the disadvantage that they do not characterize the limit objects in 
a way which would allow their systematic evaluation for physically interesting irreducible pair potentials 
$W_\Lambda$ and external one-body potentials $V_\Lambda$.
	It is this type of characterization that we are after, and in section 5.2 we prove that 
$s_{_\Lambda}(\veps)$ satisfies the familiar maximum entropy variational principle for the entropy per 
particle of a perfect gas in a combination of self- and externally generated fields. 
	More precisely, we prove the following strengthening of Theorem \ref{thm:ENTROPYperPARTICLElimit}.
\begin{theorem}
\label{thm:maxENTROPYvp}
	Let $H^{(N)}_\Lambda$ be given in \Ref{HAMrescaled}, with $U_\Lambda\geq 0$
satisfying $(H1)$--$(H5)$.
	Let  $\veps >\veps_g$. 
	Then the Boltzmann entropy \Ref{BOLTZMANNentropy} has the asymptotic expansion
\begin{equation}
S_{H^{(N)}_\Lambda}(N^2\veps) = - N\ln N + N s_{_\Lambda}(\veps) +o(N)
\label{MCentropyASYMPTOTICS}
\end{equation}
for $N\gg 1$, and the system-specific Boltzmann entropy per particle is given by
\begin{equation}
s_{_\Lambda}(\veps) = -\pzcHB(f_\veps),
\label{specENTROPYisHfuncMINUS}
\end{equation}
where $\pzcHB(f)$ is ``Boltzmann's $H$ function'' of $f$, which reads\footnote{We remark that Euler's number
			$e$ in \Ref{HfuncOFf} is inherited from the $N!$ term in \Ref{BOLTZMANNentropy}.}
\begin{equation}
\pzcHB(f)
 = \iint f(\pV,\qV) \ln (f(\pV,\qV) /e)\dd^3p\dd^3q,
\label{HfuncOFf}
\end{equation}
and where $f_\veps$ is any minimizer of this $H$ functional over the set of trial densities
$\Asp_\veps = \{f \in (\Psp\cap\Lsp^1\cap\Lsp^1\ln\Lsp^1)(\Rset^3\!\times\!\Lambda): \pzcE(f) = \veps \}$,
where $\pzcE(f)$ now reads
\begin{equation}
\hskip-.4truecm
\pzcE(f)\! = \!\!
 \iint\! {\tfrhalf}|\pV|^2  f(\pV,\qV)\dd^3p\dd^3q
+\!\! 
\iiiint\!{\tfrhalf}
U_\Lambda(\qV,\tilde{\qV})f(\pV,\qV)f(\tilde{\pV},\tilde{\qV})\dd^3p\dd^3q \dd^3\tilde{p}\dd^3\tilde{q}.\!
\label{eOFfU}
\end{equation}
	Any minimizer $f_\veps$ of $\pzcHB(f)$ over the set $\Asp_\veps$ is of the form
\begin{equation}
{f_\veps}(\pV,\qV) 
= 
\sigma_\veps(\pV)\rho_\veps(\qV),
\label{fMaxwellBoltzmannMC}
\end{equation}
where $\rho_\veps(\qV)$ solves the following fixed point equation on $\qV$ space,
\begin{equation}
\rho_\veps(\qV)
=
\frac{
		\exp\left(
-\vartheta_\veps(\rho_\veps)^{-1}\int_\Lambda U_\Lambda(\qV,\tilde{\qV})\rho_\veps(\tilde{\qV})\dd^3\tilde{q} 
			\right)}
   {\int_\Lambda\exp\left(
-\vartheta_\veps(\rho_\veps)^{-1} \int_\Lambda U_\Lambda(\hat{\qV},\tilde{\qV})\rho_\veps(\tilde{\qV})\dd^3\tilde{q}
			\right)
\dd{\hat{q}}}
\label{fixPOINTEQrhoU}
\end{equation}
with $\vartheta_\veps(\rho)$ given by
\begin{equation}
\!\!
{\textstyle{\frac{3}{2}}}\vartheta_\veps(\rho)
=
\veps -\!\! \iint{\tfrhalf}U_\Lambda(\qV,\tilde{\qV})\rho(\qV)\rho(\tilde{\qV})\dd^3q \dd^3\tilde{q},
\label{ThetaEasFNCTLrhoU}
\end{equation}	
and where 
$\sigma_\veps(\pV) =\sigma(\rho_\veps)(\pV)$, with $\sigma(\rho)(\pV)$ defined whenever $\vartheta_\veps(\rho)>0$, by
\begin{equation}
\sigma(\rho)(\pV)=
\left({\textstyle{2\pi\vartheta_\veps(\rho)}}\right)^{-\frac{3}{2}}
\exp\bigl(-\textstyle{\frac{1}{2}} \left|\pV\right|^2/\vartheta_\veps(\rho)\bigr).
\label{sigmaOFrho}
\end{equation}	
\end{theorem}
	Evidently, every minimizer of $\pzcHB(f)$ over $\Asp_\veps$ factors into a product of a \emph{Maxwellian} 
on $\pV$ space and a purely space-dependent ``self-consistent \emph{Boltzmann factor}.''\footnote{The
		expression conventionally known as ``Boltzmann factor'' results when $W_\Lambda\equiv 0$ 
		so that $U_\Lambda(\qV,\tilde{\qV})= V_\Lambda(\qV)+V_\Lambda(\tilde{\qV})$, i.e. 
		for the perfect gas acted on by an external potential $V_\Lambda$.}
	However, the Maxwellian in \Ref{fMaxwellBoltzmannMC} is not autonomous from the 
Boltzmann factor in \Ref{fMaxwellBoltzmannMC}, as is manifest by the functional dependence of the (rescaled)
temperature $\vartheta=\vartheta_\veps(\rho_\veps)$ on $\rho_\veps$, see \Ref{sigmaOFrho}.
	For a subset of $\veps$ values the minimizer of $\pzcHB(f)$ over $\Asp_\veps$ may not be unique,
but all minimizers produce the same asymptotic formula \Ref{MCentropyASYMPTOTICS}.
	In such a case of non-uniqueness of minimizers, they always seem to constitute
either a finite set (typically a first order phase transition)
or a continuous group orbit of a compact group (e.g., when $\Lambda$ is invariant
under $SO(2)$ or $SO(3)$ and a minimizer breaks that symmetry), to the best of our knowledge;
this seems to cover all physically relevant possibilities.

	In addition to the minimizers of $\pzcHB(f)$ there may be non-minimizing critical points of $\pzcHB(f)$ 
satisfying \Ref{fMaxwellBoltzmannMC}--\Ref{sigmaOFrho}, but these are irrelevant for \Ref{MCentropyASYMPTOTICS}.

	Our Theorem \ref{thm:deFinettiERGODE}, proved in section 5.3 with input from section 5.2,
characterizes the Vlasov limit $N\to\infty$ of the marginal measures
\begin{equation}
\mun^{(N)}_\cE\big(\dd^{6n}\!X\big) 
= \mu^{(N)}_\cE\big(\dd^{6n}\!X\times (\Rset^3\!\times\!\Lambda)^{N-n}\big),\qquad n=1,2,...\ (n\ \mathrm{fixed})
\label{marginalERGODEn}
\end{equation}
in terms of the $f_\veps$.
	We note that the object of interest in (mathematical) physics is not 
\Ref{BOLTZMANNsERGODE} itself but only the collection of its first few marginal measures \Ref{marginalERGODEn}.
	To state our theorem, we introduce $\Psp^s((\Rset^3\!\times\!\Lambda)^\Nset)$,
the permutation-symmetric probability measures on the set of infinite sequences in $\Rset^3\!\times\!\Lambda$. 
	A theorem of de Finetti \cite{deFinetti}, Dynkin \cite{Dynkin}, and Hewitt--Savage
\cite{HewittSavage} (see also \cite{EllisBOOK}, App.A.9.) states that $\Psp^s((\Rset^3\!\times\!\Lambda)^\Nset)$ 
is uniquely presentable as an average of infinite product measures; i.e., for each
$\mu \in \Psp^s((\Rset^3\!\times\!\Lambda)^\Nset)$ there exists a unique probability measure 
$\nu(d\tau|\mu)$ on $\Psp(\Rset^3\!\times\!\Lambda)$, such that 
\begin{equation}
\mun (\dd^{3n}p\dd^{3n}q) =
	\int_{\Psp(\Rset^3\!\times\!\Lambda)} 
	\tau^{\otimes n}(\dd^{3}p_1\dd^{3}q_1\cdots\dd^{3}p_n\dd^{3}q_n)\, \nu(\dd\tau|\mu) \qquad
\forall n\in \Nset,
\label{deFinettiDECOMP}
\end{equation}
where $\mun$ is the $n$-th marginal measure of $\mu$, and 
$\tau^{\otimes n}(\dd^{3}p_1\dd^{3}q_1\cdots\dd^{3}p_n\dd^{3}q_n)
		\equiv  \tau(\dd^{3}p_1\dd^{3}q_1)\otimes \cdots\otimes\tau(\dd^{3}p_n\dd^{3}q_n)$.
	Equation \Ref{deFinettiDECOMP} is also the extremal decomposition for 
the convex set $\Psp^s((\Rset^3\!\times\!\Lambda)^\Nset)$, see \cite{HewittSavage}.

\begin{theorem}
\label{thm:deFinettiERGODE}
	Under the same assumptions as in Theorem \ref{thm:maxENTROPYvp}, consider \Ref{BOLTZMANNsERGODE} 
with Hamiltonian \Ref{HAMrescaled} as extended to a probability on $(\Rset^3\!\times\!\Lambda)^\Nset$.
	Then the sequence $\{\mu^{(N)}_{N^2\veps}\}_{N\in\Nset}$ is tight, so one can 
extract a subsequence $\{\mu^{(\dot{N}[N])}_{N^2\veps}\}_{N\in\Nset}$ such that
\begin{equation}
\lim_{N\to\infty} \mun^{(\dot{N}[N])}_{{\dot{N}}^2\veps} (\dd^{3n}p\dd^{3n}q)
= 
\dotmun_\veps(\dd^{3n}p\dd^{3n}q) 
\in\Psp^s((\Rset^3\!\times\!\Lambda)^n)\quad \forall n\in \Nset\,.
\label{VLASOVlimptERGODE}
\end{equation}
	The decomposition measure  $\nu(d\tau|\dot\mu_\veps)$ of each such limit point $\dot\mu_\veps$ 
is supported by the subset of $\Psp(\Rset^3\!\times\!\Lambda)$ which consists of the probability measures 
$\tau_\veps(\dd^{3}p\dd^{3}q) ={f_\veps}(\pV,\qV)\dd^{3}p\dd^{3}q$
which minimize the $H$ functional $\pzcHB(f)$ over $\Asp_\veps$.
\end{theorem}
%
%%%%%%%%%%%%%%%%%%%%%%%%%%%%%%%%%%%%%%%%%%%%%%%%%%%%%%%%%%%%%%%%%%%%%%%%%%%
%%%%%%%%%%%%%%%%%%%%%%%%%%%%%%%%%%%%%%%%%%%%%%%%%%%%%%%%%%%%%%%%%%%%%%%%%%%
%%%%%%%%%%%%%%%%%%%%%%%%%%%%%%%%%%%%%%%%%%%%%%%%%%%%%%%%%%%%%%%%%%%%%%%%%%%
\section{Proofs}\label{sec:proofs}
%%%%%%%%%%%%%%%%%%%%%%%%%%%%%%%%%%%%%%%%%%%%%%%%%%%%%%%%%%%%%%%%%%%%%%%%%%%
%%%%%%%%%%%%%%%%%%%%%%%%%%%%%%%%%%%%%%%%%%%%%%%%%%%%%%%%%%%%%%%%%%%%%%%%%%%
%%%%%%%%%%%%%%%%%%%%%%%%%%%%%%%%%%%%%%%%%%%%%%%%%%%%%%%%%%%%%%%%%%%%%%%%%%%
	We have stated our Theorems 1,2,3 entirely in terms of the familiar quantities of kinetic theory.
	These are the one-body density function $f_\veps(\pV,\qV)$ which minimizes Boltzmann's $H$-function $\pzcH(f)$
under the familiar energy functional constraint $\pzcE(f)=\veps$, and the system-specific Boltzmann entropy 
per particle $s_{_\Lambda}(\veps)$ which is given as the negative of Boltzmann's $H$-function evaluated with 
$f_\veps$.
	However, in this format our theorems give essentially symmetric weight to the $\pV$ and $\qV$ variables,
which ignores the fact that the $\pV$-space integrations involved in \Ref{marginalERGODEn} and 
\Ref{BOLTZMANNentropy} can be carried out explicitly in the same fashion as for the perfect gas.
	As a consequence the problem reduces to studying the large $N$ asymptotics of the expressions which result 
from these $\pV$-space integrations.\footnote{All Boltzmann needed for this was that $(1+x/n)^n \asymp e^x$; 
		cf. \cite{Boltzmann}, part II, ch. 3.
		Of course, things are not quite as straightforward with an irreducible $W_\Lambda\not\equiv 0$,
		or else Boltzmann would not have had to have $W_\Lambda\not\equiv 0$ excluded from his analysis.}
	In fact, all the hard analytical work goes into controlling the $\qV$-space integrations.
	This is certainly the case as far as the entropy per particle goes, yet also each minimizer $f_\veps$ of 
$\pzcHB(f)$ over the set $\Asp_\veps$ is uniquely determined by $\rho_\veps$, which signals that all of our 
Theorems 1 to 3 will be essentially straightforward corollaries of theorems about certain $\qV$-space expressions.
	Those theorems take a less familiar form, presumably, which is why their statements have been
relegated into this section where we prove Theorems 1 to 3. 
%
%%%%%%%%%%%%%%%%%%%%%%%%%%%%%%%%%%%%%%%%%%%%%%%%%%%%%%%%%%%%%%%%
%%%%%%%%%%%%%%%%%%%%%%%%%%%%%%%%%%%%%%%%%%%%%%%%%%%%%%%%%%%%%%%%
\subsection{Proof of Theorem \ref{thm:ENTROPYperPARTICLElimit} and its two upgrades} 
%%%%%%%%%%%%%%%%%%%%%%%%%%%%%%%%%%%%%%%%%%%%%%%%%%%%%%%%%%%%%%%%
%%%%%%%%%%%%%%%%%%%%%%%%%%%%%%%%%%%%%%%%%%%%%%%%%%%%%%%%%%%%%%%%
%
%%%%%%%%%%%%%%%%%%%%%%%%%%%%%%%%%%%%%%%%%%%%%%%%%%%%%%%%%%%%%%%%
% \subsubsection{Proof of Theorem \ref{thm:ENTROPYperPARTICLElimit}}
%%%%%%%%%%%%%%%%%%%%%%%%%%%%%%%%%%%%%%%%%%%%%%%%%%%%%%%%%%%%%%%%
	To prove Theorem \ref{thm:ENTROPYperPARTICLElimit} we first formulate and then prove an upgraded version 
(Theorem \ref{thm:ENTROPYperPARTICLElimit}$^+$), whose proof also proves Theorem \ref{thm:ENTROPYperPARTICLElimit}.
%
%%%%%%%%%%%%%%%%%%%%%%%%%%%%%%%%%%%%%%%%%%%%%%%%%%%%%%%%%%%%%%%%
\subsubsection{Theorem \ref{thm:ENTROPYperPARTICLElimit}$^+$ and its proof}
%%%%%%%%%%%%%%%%%%%%%%%%%%%%%%%%%%%%%%%%%%%%%%%%%%%%%%%%%%%%%%%%
	Carrying out the $\pV$ integrations\footnote{It is understood that $\dd^{6N}X$ etc.
		now involves the $\pV$ variables used in \Ref{HAMrescaled}.}
in $\Omega^\prime_{H^{(N)}_\Lambda}(\cE)$ given by \Ref{STRUKTURfunktion}, with ${H}^{(N)}_\Lambda$ 
given in
\vskip-.1truecm
\noindent
\Ref{HAMrescaled}, Boltzmann's ergodic ensemble entropy \Ref{BOLTZMANNentropy} becomes
\begin{equation}
\hskip-.2truecm
S_{H^{(N)}_\Lambda}(\cE) 
= 
\ln\!\left({\textstyle{\frac{(2/N)^{3N/2}}{3N}}}\left|\Sset^{3N-1}\right|\!\Psi^\prime_{I^{(N)}_\Lambda}(\cE) \right)
\label{MCentropyINTEGRATEDinP}
\end{equation}
with $\left|\Sset^{3N-1}\right|$ the standard measure of the unit $3N-1$ sphere $\Sset^{3N-1}$, and with
\begin{equation}
\Psi^\prime_{I^{(N)}_\Lambda}(\cE) 
=
{\textstyle{\frac{(3/2)}{(N-1)!}}}
\int\!\!\!
	\left(\cE-I_\Lambda^{(N)}(\qV_1,...,\qV_N)\right)^{\frac{3N}{2}-1}\!\chi_{\left\{I^{(N)}_\Lambda<\cE\right\}}
 \dd^{3N}\!q,
\label{MCconfigINTEGRAL}
\end{equation}
where we introduced the \textit{interaction Hamiltonian} 
\begin{equation} 
{I}^{(N)}_\Lambda (\qV_1,\dots,\qV_N)
=
\sum\sum_{\hskip-.7truecm 1\leq i < j\leq N} {U}_\Lambda(\qV_i,\qV_j).
\label{IntHAM}
\end{equation}	
	Implementing the Vlasov limit scaling, i.e. setting $\cE=N^2\veps$ with  $\veps>\veps_g\geq 0$, 
recalling that $\left|\Sset^{3N-1}\right| = \pi^{3N/2}/\Gamma(3N/2)$, 
and using Stirling's formula for Euler's $\Gamma$ function, we obtain the following asymptotic 
expansion for \Ref{MCentropyINTEGRATEDinP},
\begin{eqnarray}
\!\!S_{{H}^{(N)}_\Lambda}(N^2\veps) 
=\!\!\!&-&\!\!\!\!\! N\ln N +
N \ln\left(|\Lambda|\left({\textstyle{\frac{4\pi e}{3}\veps}}\right)^{3/2} \right) +O(\ln N)
\nonumber\\
\!\!&+&\!\!\! 
\ln \!\int\!\!\!
\left(1 - {\textstyle{\frac{1}{\veps N^2}}}{I}_\Lambda^{(N)}(\qV_1,...,\qV_N) \right)_+^{\frac{3N}{2}-1}\!
\lambda(\dd^{3N}\!q).
\label{MCentropyINTEGRATEDinPasympREV}
\end{eqnarray}
where $(\cdots)_+$ means the positive part of $(\cdots)$; moreover, 
$\lambda(\dd^{3N}\!q)$ is the $N$-fold product of the normalized Lebesgue measure 
$\lambda(\dd^{3}q) = |\Lambda|^{-1}\dd^{3}q$ on $\Lambda$.
	For brevity we wrote $|\Lambda|$ for the volume $\Vol(\Lambda)$ of $\Lambda$.

	When $I^{(N)}_\Lambda\equiv 0$ in $\Lambda^{N}$, then $H^{(N)}_\Lambda$ becomes the Hamiltonian 
of the perfect gas without external fields,\footnote{It is tacitly understood that the cutoff provided by 
		$I^{(N)}_\Lambda$ remains effective, so that the configurational integrations in 
		\Ref{MCentropyINTEGRATEDinPasympREV}
		are still over $\Lambda^{N}$.}
abbreviated as ${K}^{(N)}_\Lambda$ (for \emph{kinetic} Hamiltonian).
	In this case the second line  in \Ref{MCentropyINTEGRATEDinPasympREV} vanishes, 
and \Ref{MCentropyINTEGRATEDinPasympREV} becomes the asymptotic expansion of the entropy of the spatially 
uniformly distributed perfect gas, viz.
\begin{equation}
\!\!S_{{K}^{(N)}_\Lambda}(N^2\veps) 
= - N\ln N +
N \ln\left(|\Lambda|\left({\textstyle{\frac{4\pi e}{3}\veps}}\right)^{3/2} \right) +O(\ln N).
\label{MCentropyPERFECTgas}
\end{equation}
	The coefficient of the $O(N)$ term in \Ref{MCentropyPERFECTgas} gives the system-specific Boltzmann 
entropy per particle of the spatially uniform perfect gas, which we denote by
\begin{equation}
s_{_\Lambda,{}_K}(\veps) 
=
 \ln\left(|\Lambda|\left({\textstyle{\frac{4\pi e}{3}\veps}}\right)^{3/2} \right).
\label{sK}
\end{equation}
	Whenever interactions ${I}_\Lambda^{(N)}\not\equiv 0$ of the admitted type are present, 
Theorem \ref{thm:ENTROPYperPARTICLElimit} follows if we can show that the second line in 
\Ref{MCentropyINTEGRATEDinPasympREV} is $O(N)$ and so contributes additively to the system-specific 
Boltzmann entropy per particle, and provided it has the right monotonicity and regularity.
	This is expressed in 

\begin{proposition}
\label{prop:configENTROPYperPARTICLElimit}
	Under the assumptions stated in Theorem \ref{thm:ENTROPYperPARTICLElimit}, there holds
\begin{equation}
\lim_{N\to\infty} {\textstyle{\frac{1}{N}}}
\ln \!\int\!\!\!
\left(1 - {\textstyle{\frac{1}{\veps N^2}}}{I}_\Lambda^{(N)}(\qV_1,...,\qV_N) \right)_+^{\frac{3N}{2}-1}\!
\lambda(\dd^{3N}\!q)
= 
s_{_\Lambda,{}_I}(\veps).
\label{sI}
\end{equation}
	The function $\veps\mapsto s_{_\Lambda,{}_I}(\veps)$ is continuous and increasing for $\veps>\veps_g\geq 0$.
\end{proposition}
	This concludes the pretext for our first upgrade of Theorem \ref{thm:ENTROPYperPARTICLElimit}, stated
next.
	
\setcounter{theorem}{0}
\begin{theorem}
\label{thm:ENTROPYperPARTICLElimitUP}$^{\hskip-.2truecm +}$
	Theorem \ref{thm:ENTROPYperPARTICLElimit} holds, with
\begin{equation}
s_{_\Lambda}(\veps) 
=
s_{_\Lambda,{}_K}(\veps) 
+ 
s_{_\Lambda,{}_I}(\veps) ,
\label{sKsI}
\end{equation}
where $s_{_\Lambda,{}_K}(\veps)$ is given in \Ref{sK}, and $s_{_\Lambda,{}_I}(\veps)$ in \Ref{sI}.
\end{theorem}
\medskip\noindent
{\textit{Proof of Theorem \ref{thm:ENTROPYperPARTICLElimitUP}$^{+}$:}} 

	Clearly, Proposition \ref{prop:configENTROPYperPARTICLElimit} and formula 
\Ref{MCentropyINTEGRATEDinPasympREV} imply Theorem \ref{thm:ENTROPYperPARTICLElimit} and the splitting of
the system-specific Boltzmann entropy per particle $s_{_\Lambda}(\veps)$ in 
\Ref{MCentropyASYMPexpansionI} into a sum of a kinetic and an interaction component, \Ref{sKsI}.
	Proposition \ref{prop:configENTROPYperPARTICLElimit} also adds a piece of information about 	
$s_{_\Lambda,{}_I}(\veps)$ which does not just re-express what is stated in Theorem \ref{thm:ENTROPYperPARTICLElimit}.
	In fact, by the known strict increase of $\veps\mapsto\ln\veps$, the increase of 
$\veps\mapsto s_{_\Lambda,{}_I}(\veps)$ implies the strict increase of $\veps\mapsto s_{_\Lambda}(\veps)$, 
but the increase of $\veps\mapsto s_{_\Lambda,{}_I}(\veps)$ does not follow from the properties of 
$\veps\mapsto\ln\veps$ and the strict increase of $\veps\mapsto s_{_\Lambda}(\veps)$.
	So Theorem \ref{thm:ENTROPYperPARTICLElimitUP}$^+$ holds and extends 
Theorem \ref{thm:ENTROPYperPARTICLElimit}.
\qed

\medskip\noindent
{\textit{Proof of Proposition \ref{prop:configENTROPYperPARTICLElimit}:}} 

	By hypothesis $(H6)$, $U_\Lambda$ is bounded continuous on $\oli\Lambda\!\times\!\oli\Lambda$, 
so we can write
\begin{eqnarray}
N^{-2}{I}_\Lambda^{(N)}(\qV_1,...,\qV_N)
\!\!\! &=&\!\!\!  
  \iint{\tfrhalf}
U_\Lambda({\check{\qV},\hat{\qV}})\uli\Delta^{(1)}_{X^{(N)}}(\check{\qV})\uli\Delta^{(1)}_{X^{(N)}}(\hat{\qV})
	\dd^3\check{q}\dd^3\hat{q}
\nonumber\\
&&\!\! -
{\textstyle{\frac{1}{N}}}
\int\! {\tfrhalf} U_\Lambda({\qV,\qV}) \uli\Delta^{(1)}_{X^{(N)}}(\qV)\dd^3q ,
\label{IasBform}
\end{eqnarray}
and we may abbreviate the first term of r.h.s.\Ref{IasBform} in bilinear form notation, 
\begin{equation}
  \iint{\tfrhalf}
U_\Lambda(\check{\qV},\hat{\qV})\uli\Delta^{(1)}_{X^{(N)}}(\check{\qV})\uli\Delta^{(1)}_{X^{(N)}}(\hat{\qV})
	\dd^3\check{q}\dd^3\hat{q}
\equiv
\big\langle\uli{\Delta}^{(1)}_{X^{(N)}}\,,\,\uli{\Delta}^{(1)}_{X^{(N)}}\big\rangle.
\label{UbilinearFORM}
\end{equation}
	The above integrals extend over $\Rset^3$, and we set
${I}_\Lambda^{(N)}(\qV_1,...,\qV_N)=\infty$ as well as 
$\big\langle\uli{\Delta}^{(1)}_{X^{(N)}}\,,\,\uli{\Delta}^{(1)}_{X^{(N)}}\big\rangle=\infty$
if any $\qV_k\not\in\oli\Lambda$.
	Also by $(H6)$, the term in the second line of r.h.s.\Ref{IasBform} is $O(N^{-1})$ for all 
$X^{(N)}\in\oli\Lambda^N$. 
	Recalling our claim (which we promised to prove) that the limit $N\to\infty$ for the ensemble does not
change if the Hamiltonian is changed by an additive term of order $O(N^{-1})$ relative to the leading terms, 
we now introduce the configurational integral
\begin{equation}
\Upsilon_{\Lambda}^{(N)}(\veps)
\equiv
\ln \!\int\!\!\!
\left(1 - \invveps\big\langle\uli{\Delta}^{(1)}_{X^{(N)}}\,,\uli{\Delta}^{(1)}_{X^{(N)}}\big\rangle_{} \right)_+^{\frac{3N}{2}-1}\!
\lambda(\dd^{3N}\!q)
\label{logCONFIGintUPS}
\end{equation}
for all $N> N_{U}(\veps)$ (to be defined).
	Note that the integral \Ref{logCONFIGintUPS} is generally not well-defined for all $N\in\Nset$ because
$\big\langle\uli{\Delta}^{(1)}_{X^{(N)}}\,,\uli{\Delta}^{(1)}_{X^{(N)}}\big\rangle_{}$ is bigger than 
$N^{-2}{I}_\Lambda^{(N)}(\qV_1,...,\qV_N)$ by the absolute value of the second line of 
r.h.s.\Ref{IasBform}, which reads precisely $N^{-2}\sum_{k=1}^N {\tfrhalf} U_\Lambda({\qV_k,\qV_k})$.
	And while this term $=O(1/N)$, when $N$ is not large enough then it is possible that
$\big\langle\uli{\Delta}^{(1)}_{X^{(N)}}\,,\uli{\Delta}^{(1)}_{X^{(N)}}\big\rangle_{} >\veps$ {everywhere}
in $\oli\Lambda^N$, in which case the integral in \Ref{logCONFIGintUPS} vanishes, and its logarithm $=-\infty$, then.
	Yet, when  $N> N_{U}(\veps)$ the integral \Ref{logCONFIGintUPS} is well-defined, and we conclude that 
(modulo the proof of precisely the just re-uttered claim that $O(1/N)$ contributions
to the Hamiltonian drop out when $N\to\infty$) our proposition \ref{prop:configENTROPYperPARTICLElimit}
is proved if we can prove the following proposition.
\begin{proposition}
\label{prop:configENTROPYperPARTICLElimitSPECIAL}
	Under the hypotheses on $U_\Lambda$ in Thm.~\ref{thm:ENTROPYperPARTICLElimit}, when $N\!\gg\!N_U(\veps)$~then
\begin{equation}
\Upsilon_{\Lambda}^{(N)}(\veps)
= 
 N \gamma_{_\Lambda}(\veps) + o(N)\,.
\label{MCentropyASYMPexpansionIII}
\end{equation}
	The function $\veps\mapsto \gamma_{_\Lambda}(\veps)$ is continuous and increasing for $\veps>\veps_g\geq 0$.
\end{proposition}
\medskip\noindent
{\textit{Proof of Proposition \ref{prop:configENTROPYperPARTICLElimitSPECIAL}:}} 
	
	We will establish uniform bounds and super-additivity estimates.

	For $0<n<N$, we set $X^{(N)} \equiv (X^{(n)},Y^{(N-n)})$, which also defines $Y^{(N-n)}$. 
	We note the convex linear decomposition
\begin{equation}
\uli\Delta^{(1)}_{X^{(N)}}(\qV)  
= 
\frnN
\uli\Delta^{(1)}_{X^{(n)}}(\qV)  
+
(1-\frnN)
\uli\Delta^{(1)}_{Y^{(N-n)}}(\qV)  .
\label{normalEMPmeasOFqCONVEXdecomp}
\end{equation}
	Since  $U_\Lambda\geq 0$ is the kernel of a bilinear form which is positive definite when restricted to
the set of probability measures on $\Lambda$, Jensen's inequality gives us
\begin{equation}
\big\langle\uli{\Delta}^{(1)}_{X^{(N)}}\,,\,\uli{\Delta}^{(1)}_{X^{(N)}}\big\rangle_{}
\leq
\frnN
\big\langle\uli{\Delta}^{(1)}_{X^{(n)}}\,,\,\uli{\Delta}^{(1)}_{X^{(n)}}\big\rangle_{}
+
(1-\frnN)
\big\langle\uli{\Delta}^{(1)}_{Y^{(N-n)}}\,,\,\uli{\Delta}^{(1)}_{Y^{(N-n)}}\big\rangle_{}.
\label{IasBformJENSENdecomp}
\end{equation}
	We of course also have $1 = \frnN + (1-\frnN)$, and so we conclude that
\begin{eqnarray}
&&\hskip-1.2truecm
\left(1-\invveps\big\langle\uli{\Delta}^{(1)}_{X^{(N)}}\,,\,\uli{\Delta}^{(1)}_{X^{(N)}}\big\rangle_{}\right)_+
\geq 
  \nonumber\\
&&\hskip-1truecm
\left(
  \frnN\left[1-\invveps\big\langle\uli{\Delta}^{(1)}_{X^{(n)}}\,,\uli{\Delta}^{(1)}_{X^{(n)}}\big\rangle_{}\right]
 + 
  (1-\frnN)\!
   \left[1-\invveps\big\langle\uli{\Delta}^{(1)}_{Y^{(N-n)}}\,,\uli{\Delta}^{(1)}_{Y^{(N-n)}}\big\rangle_{}\right]
\right)_+.
\label{IasBformJENSENdecompU}
\end{eqnarray}

	Next we recall that, if $\varphi$ is some function on a domain $D$, and if $\Sigma(\varphi_+)$ denotes the 
support of its positive part, and $\chi_{_{\Sigma(\varphi_+)}}$ is the characteristic function of 
$\Sigma(\varphi_+)$, then the inclusion 
$\Sigma(\varphi_+)\cap\Sigma(\vartheta_+)\subset\Sigma((\varphi+\vartheta)_+)$ 
for any two such functions $\varphi$ and $\vartheta$ yields the estimate
\begin{equation}
\hskip-.1truecm
(\varphi+\vartheta)_+ 
=
(\varphi+\vartheta) \chi_{_{\Sigma((\varphi+\vartheta)_+)}}
\geq 
(\varphi+\vartheta) \chi_{_{\Sigma(\varphi_+)}}\chi_{_{\Sigma(\vartheta_+)}}
= 
(\varphi_++\vartheta_+) \chi_{_{\Sigma(\varphi_+)}}\chi_{_{\Sigma(\vartheta_+)}}.
\label{fgPLUSsupports}
\end{equation}
	Set
$\varphi = \frnN\left[1-\invveps \big\langle\uli{\Delta}^{(1)}_{X^{(n)}}\,,\uli{\Delta}^{(1)}_{X^{(n)}}\big\rangle_{}\right]$
and 
$\vartheta = 
(1-\frnN)\!\left[1 -\invveps \big\langle\uli{\Delta}^{(1)}_{Y^{(N-n)}}\,,\uli{\Delta}^{(1)}_{Y^{(N-n)}}\big\rangle_{}\!\right].$
	Then inequality \Ref{fgPLUSsupports} applies to r.h.s. \Ref{IasBformJENSENdecompU}.
	Applying next the classical inequality between the 
arithmetic and the geometric means of any two positive numbers $A$ and $B$, viz.
$\alpha A + (1-\alpha) B \geq A^{\alpha}B^{(1-\alpha)}$ for any $\alpha\in[0,1]$, 
we get
\begin{eqnarray}
&&\hskip-.8truecm
\left(1-\invveps\big\langle\uli{\Delta}^{(1)}_{X^{(N)}}\,,\,\uli{\Delta}^{(1)}_{X^{(N)}}\big\rangle_{}\right)_+
\geq 
  \nonumber\\
&&
\left[1-\invveps\big\langle\uli{\Delta}^{(1)}_{X^{(n)}}\,,\,\uli{\Delta}^{(1)}_{X^{(n)}}\big\rangle_{}\right]_+^\frnN
\left[1-\invveps\big\langle\uli{\Delta}^{(1)}_{Y^{(N-n)}}\,,\,\uli{\Delta}^{(1)}_{Y^{(N-n)}}\big\rangle_{}\right]_+^{1-\frnN}.
\label{IasBformJENSENarithgeomINEQU}
\end{eqnarray}

	We now use \Ref{IasBformJENSENarithgeomINEQU} to estimate r.h.s.\Ref{logCONFIGintUPS}.
	For this, let $N\gg N_U(\veps)$ and let $N_U(\veps)<n < N-N_U(\veps)$.
	Noting that the resulting integral over $\Lambda^{N}$ factors into two integrals, one over $\Lambda^{n}$
and another over $\Lambda^{N-n}$, and working out the powers, we find
\begin{eqnarray}
\ln \!\int\!\!\!
\left(1-\invveps\big\langle\uli{\Delta}^{(1)}_{X^{(N)}}\,,\uli{\Delta}^{(1)}_{X^{(N)}}\big\rangle_{} \right)_+^{\frac{3N}{2}-1}\!
\lambda(\dd^{3N}\!q)
\geq && 
\label{almostSUPERaddI}\\
&&\hskip-7truecm \ln \!\int\!\!\!
\left(
 1-\invveps\big\langle\uli{\Delta}^{(1)}_{X^{(n)}}\,,\uli{\Delta}^{(1)}_{X^{(n)}}\big\rangle_{}\right)_+^{\frac{3n}{2}-\frnN}\!
\lambda(\dd^{3n}\!q) +
\nonumber\\
&&\hskip-6.5truecm 
\ln \!\int\!\!\!
\left(1-\invveps\big\langle\uli{\Delta}^{(1)}_{X^{(N-n)}}\,,\uli{\Delta}^{(1)}_{X^{(N-n)}}\big\rangle_{} 
\right)_+^{\frac{3(N-n)}{2}-1+\frnN}\! \lambda(\dd^{3(N-n)}\!q),
\nonumber
\end{eqnarray}
where we also relabeled the integration variables under the second integral on r.h.s.\Ref{almostSUPERaddI} 
from $Y^{(N-n)}$ to $X^{(N-n)}$.
	Noting next that $0< \frnN < 1$,
we resort again to Jensen's inequality, this time w.r.t. the $\lambda$ measures in the two integrals 
on  r.h.s.\Ref{almostSUPERaddI}.
	Also using $\ln (\cdots)^a = a \ln (\cdots)$, we arrive at 
\begin{eqnarray}
\ln \!\int\!\!\!
\left(1-\invveps\big\langle\uli{\Delta}^{(1)}_{X^{(N)}}\,,\uli{\Delta}^{(1)}_{X^{(N)}}\big\rangle_{} \right)_+^{\frac{3N}{2}-1}\!
\lambda(\dd^{3N}\!q)
\geq && 
\label{almostSUPERaddII}\\
&&\hskip-8truecm 
\left(1+{\textstyle{\frac{2-2n/N}{3n-2}}}\right) 
\ln \!\int\!\!\!
\left(1 -\invveps \big\langle\uli{\Delta}^{(1)}_{X^{(n)}}\,,\uli{\Delta}^{(1)}_{X^{(n)}}\big\rangle_{} \right)_+^{\frac{3n}{2}-1}\!
\lambda(\dd^{3n}\!q) +
\nonumber\\
&&\hskip-8.5truecm 
\left(1+{\textstyle{\frac{2n/N}{3(N-n)-2}}}\right) 
\ln \!\int\!\!\!
\left(
	1-\invveps\big\langle\uli{\Delta}^{(1)}_{X^{(N-n)}}\,,\uli{\Delta}^{(1)}_{X^{(N-n)}}\big\rangle_{} 
\right)_+^{\frac{3(N-n)}{2}-1}\!
\lambda(\dd^{3(N-n)}\!q).
\nonumber
\end{eqnarray}
	Formula \Ref{almostSUPERaddII} writes shorter thusly,
\begin{equation}
\Upsilon_{\Lambda}^{(N)}(\veps) 
\geq
 \left(1+{\textstyle{\frac{2-2n/N}{3n-2}}}\right) \Upsilon_{\Lambda}^{(n)}(\veps) 
+ \left(1+{\textstyle{\frac{2n/N}{3(N-n)-2}}}\right) \Upsilon_{\Lambda}^{(N-n)}(\veps).
\label{UPSalmostSUPERadd}
\end{equation}
	So $N\mapsto \Upsilon_{\Lambda}^{(N)}(\veps)$ is almost super-additive.

	To be able to create a properly super-additive function we establish upper and lower bounds of
$\ell\mapsto \Upsilon_{\Lambda}^{(\ell)}(\veps)$ which are linear in $\ell$, whenever
$\ell>N_U(\veps)$; we will need those bounds with $\ell\in\{n,N-n\}$, with $\ell>1$.
	As a by-product, the upper bound with $\ell=N$ will also guarantee convergence of the constructed 
super-additive function.

	The upper bound is trivial.
	Recall that by hypothesis 
$\big\langle\uli{\Delta}^{(1)}_{X^{(\ell)}}\,,\uli{\Delta}^{(1)}_{X^{(\ell)}}\big\rangle_{} \geq 0$ for all $\ell\in\Nset$.
	So for $\ell>N_U(\veps)$ and $\veps>\veps_g\geq 0$ we find
\begin{equation}
{\textstyle{\frac{2}{3\ell-2}}} \ln \!\int\!\!\!
\left(
	1- \invveps\big\langle\uli{\Delta}^{(1)}_{X^{(\ell)}}\,,\uli{\Delta}^{(1)}_{X^{(\ell)}}\big\rangle_{} 
\right)_+^{\frac{3\ell}{2}-1}
\!\lambda(\dd^{3\ell}\!q)
\leq 0 .
\label{UPPERbound}
\end{equation}

	As for the lower bound, we distinguish two cases, 
(a): $\big\langle|\Lambda|^{-1},|\Lambda|^{-1} \big\rangle_{}<\veps$,
and
(b): $\big\langle|\Lambda|^{-1},|\Lambda|^{-1}\big\rangle_{}\geq \veps$.
	In case (a) we apply Jensen's inequality w.r.t. $\lambda$ to the convex map $x\mapsto (1-x)_+^\theta$ 
(for $\theta\geq 1$), and also use $\ell-1<\ell$, to get
\begin{eqnarray}
&& \ln \left[\!\int\!\!\!
\left(
	1 -\invveps \big\langle\uli{\Delta}^{(1)}_{X^{(\ell)}}\,,\uli{\Delta}^{(1)}_{X^{(\ell)}}\big\rangle_{} 
\right)_+^{\frac{3\ell}{2}-1}
\!\lambda(\dd^{3\ell}q)\right]^{{\frac{2}{3\ell-2}}}
\geq
\nonumber
\\
&&
\ln \left[1-
{\textstyle{\frac{1}{\ell}}}\invveps \int\! {\tfrhalf} U_\Lambda({\qV,\qV}) \lambda(\dd^3q )
-
\invveps \iint{\tfrhalf}U_\Lambda({\check{\qV},\hat{\qV}})\lambda(\dd^3\check{q})\lambda(\dd^3\hat{q})
\right]_+,
\label{lowerBOUNDa}
\end{eqnarray}
and r.h.s.\Ref{lowerBOUNDa}$\geq -C > -\infty$ when $\ell>\ell_{crit}(\veps)$ (given $U_\Lambda$), with 
$C>0$ independent of $\ell$. 
	Since the interaction entropy exists when $\ell> N_U(\veps)$, clearly $\ell_{crit}\geq N_U(\veps)$,
but after at most an adjustment of $C$, we can conclude that 
l.h.s.\Ref{lowerBOUNDa}$\geq -C > -\infty$ when $\ell>N_U(\veps)$, with $C>0$ independent of $\ell$. 
	In case (b), inequality \Ref{lowerBOUNDa} is still true but now trivial, for
r.h.s.\Ref{lowerBOUNDa}$=-\infty$ for all $\ell >1$, then.
	So instead we now proceed as follows.
	By hypothesis $(H6)$, the bilinear form
$\big\langle\uli{\Delta}^{(1)}_{X^{(\ell)}}\,,\uli{\Delta}^{(1)}_{X^{(\ell)}}\big\rangle_{}$
takes its minimum $\veps_g^*(\ell)\geq \veps_g$.
	Clearly, $\veps_g^*(\ell) = \tilde\veps_g(\ell) +O(\ell^{-1})$, where 
$\tilde\veps_g(\ell) := \min \ell^{-2}{I}_\Lambda^{(\ell)}(\qV_1,...,\qV_\ell)$, and since 
$\tilde\veps_g(\ell)\leq \veps_g$ (as proved in Appendix A), we have that
$\veps_g^*(\ell) \leq \veps_g +O(\ell^{-1})$;
of course, we also assume that $\ell>N_U(\veps)$ so that $\veps_g^*(\ell) <\veps$. 
	By permutation symmetry there are many equivalent minimizers, but possibly also several distinct
permutation group orbits of minimizers. 
	We pick any particular minimizer $Q_g^{(\ell)}$ and let $\qV_{g,k}^{(\ell)}\in\Lambda$ denote 
the $k$-th coordinate vector in $Q_g^{(\ell)}$.
	By $(H6)$ again, we can vary all the $\qV_k$ in the minimizing configuration a little bit, say, 
each $\qV_k$ in $B_\delta(\qV_{g,k}^{(\ell)})\cap\oli{\Lambda}$, where $B_\delta(\qV)$ is a ball centered at $\qV$, 
with radius $\delta>0$ independent of $k$ and $\ell$ but chosen small enough (given $\veps$) so that
$\big\langle\uli{\Delta}^{(1)}_{X^{(\ell)}}\,,\uli{\Delta}^{(1)}_{X^{(\ell)}}\big\rangle_{}$ does not change
by more than $(\veps-\veps_g+O(\ell^{-1}))/2$.
	For brevity we write $B_\delta[k]$ for $B_\delta(\qV_{g,k}^{(\ell)})$; let $\chi_{B_\delta[k]}$ 
be the characteristic function of $B_\delta[k]$.
	We use that 
$\lambda(\dd^3{q}_k)=\chi_{B_\delta[k]}\lambda(\dd^3{q}_k) +\chi_{B^c_\delta[k]}\lambda(\dd^3{q}_k)$
where $B^c_\delta[k]=\Lambda\backslash B_\delta[k]$ is the complement in $\Lambda$ of $B_\delta[k]$, 
then use that both terms in this decomposition are non-negative so that we get an upper estimate by dropping the contribution from $\chi_{B^c_\delta[k]}\lambda(\dd^3{q}_k)$ for each $k$. 
	After this step the restriction to the positive part of $(1-\invveps\langle\,.\,,\,.\,\rangle)$ is 
eventually tautological when $\ell$ is sufficiently large so that the $O(\ell^{-1})$ term has gotten sufficiently
small.
	We next apply Jensen's inequality w.r.t. the probability measure
$\prod_{1\leq k\leq\ell}(\int_{B_\delta[k]\cap\Lambda}\lambda(\dd^3{q}))^{-1}\chi_{B_\delta[k]}\lambda(\dd^3q_k)$ 
to the convex map $x_+\mapsto x_+^\theta$ (for $\theta\geq 1$), finally recall that 
$0\leq \veps_g<\veps$, and get
\begin{eqnarray}
\hskip-1truecm
&& \left[\!\int\!\!\!
\left(
	1 -\invveps \big\langle\uli{\Delta}^{(1)}_{X^{(\ell)}}\,,\uli{\Delta}^{(1)}_{X^{(\ell)}}\big\rangle_{} 
\right)_+^{\frac{3\ell}{2}-1}
\!\lambda(\dd^{3\ell}q)\right]^{{\frac{2}{3\ell-2}}}
\geq
\nonumber\\
&& \left[\!\int\!\!\!
\left(
	1 -\invveps \big\langle\uli{\Delta}^{(1)}_{X^{(\ell)}}\,,\uli{\Delta}^{(1)}_{X^{(\ell)}}\big\rangle_{} 
\right)^{\frac{3\ell}{2}-1}
\!\prod_{1\leq k\leq\ell}\chi_{B_\delta[k]}\lambda(\dd^3q_k)
\right]^{{\frac{2}{3\ell-2}}}
\geq
\nonumber\\
&&
|C_\delta|^{{\frac{2}{3-2/\ell}}}
\!\int\!\!\!
\left(1- \invveps\big\langle\uli{\Delta}^{(1)}_{X^{(\ell)}}\,,\uli{\Delta}^{(1)}_{X^{(\ell)}}\big\rangle_{} \right)
\!\prod_{1\leq k\leq\ell}
\textstyle{\frac{\chi_{B_\delta[k]}}{\int_{B_\delta[k]\cap\Lambda}\lambda(\dd^3{q})}}\lambda(\dd^3q_k)
\geq
\nonumber\\
&&
|C_\delta|^{{\frac{2}{3-2/\ell}}}
\left(1- \tfrhalf\left(1+\textstyle{\frac{\veps_g}{\veps}}\right)+O(\ell^{-1})\right)
\geq C > 0
\label{lowerBOUNDb}
\end{eqnarray}
\vskip-.3truecm
\noindent
for $\ell$ large enough; here
\begin{equation}
 C_\delta = \min_{\qV'}\int_{B_\delta(\qV')\cap\oli{\Lambda}}\lambda(\dd^3{q})>0.
\label{Cdelta}
\end{equation}

	In summary, our list of inequalities \Ref{UPPERbound}, \Ref{lowerBOUNDa} and  \Ref{lowerBOUNDb}, and 
the finiteness of the number of $\ell$ until ``$\ell$ is large enough,'' establishes 
that when $\ell>N_U(\veps)$, then for some $\ell$-independent constant $C_*>0$,
\begin{equation}
-\left({\textstyle{{\frac{3\ell}{2}-1}}}\right)C_*
\leq 
\Upsilon_{\Lambda}^{(\ell)}(\veps)
\leq
0;
\label{UPPERlowerBOUNDS}
\end{equation}
incidentally, our \Ref{lowerBOUNDa} and \Ref{lowerBOUNDb} produce an upper estimate for $N_U(\veps)$.

	Recall that in this proof we assume that $N\gg N_U(\veps)$, and that $N_U(\veps)<n < N-N_U(\veps)$.
	With the help of \Ref{UPPERlowerBOUNDS}, for $\ell\in\{n,N-n\}$, we conclude from \Ref{UPSalmostSUPERadd} 
that there exists a $C\in \Rset$ independent of $n$ and $N$ such that
\begin{equation}
\Upsilon_{\Lambda}^{(N)}(\veps)
\geq
\Upsilon_{\Lambda}^{(n)}(\veps)
+ \Upsilon_{\Lambda}^{(N-n)}(\veps)
+C.
\label{almostSUPERaddIII}
\end{equation}
	Adding that constant $C$ to both sides of the inequality \Ref{almostSUPERaddIII} shows that 
$N\mapsto \Upsilon_{\Lambda}^{(N)}(\veps)+C$ is a super-additive function for all $\veps>\veps_g\geq 0$.
	And using \Ref{UPPERbound} with $N$, we also see that 
$N^{-1}\bigl(\Upsilon_{\Lambda}^{(N)}(\veps)+C\bigr)$ is bounded above, and so, by standard facts about 
super-additive functions, $N^{-1}\bigl(\Upsilon_{\Lambda}^{(N)}(\veps)+C\bigr)$ converges as $N\to\infty$,
\begin{equation}
\lim_{N\to\infty} N^{-1}
\Bigl(\Upsilon_{\Lambda}^{(N)}(\veps)+C\Bigr) 
= 
\sup_{N\in\Nset} N^{-1}\Bigl(\Upsilon_{\Lambda}^{(N)}(\veps)+C\Bigr) ,
\label{SUPERaddPerNlim}
\end{equation}
and since $N^{-1}C\stackrel{N\to\infty}{\longrightarrow}0$, we conclude that $N^{-1}\Upsilon_{\Lambda}^{(N)}(\veps)$ 
converges as well, i.e.
\begin{equation}
\lim_{N\to\infty} {\textstyle{\frac{1}{N}}}\Upsilon_{\Lambda}^{(N)}(\veps)
= 
 \gamma_{_\Lambda}(\veps).
\label{UPSlim}
\end{equation}
	This proves \Ref{MCentropyASYMPexpansionIII}. 

	To prove continuity of $\gamma_{_\Lambda}(\veps)$, we establish upper and lower bounds on the
derivative of the functions $\veps\mapsto N^{-1}\Upsilon_{\Lambda}^{(N)}(\veps)$ which are uniform in $N>N_U(\veps)$.
	Differentiating the functions
$\veps\mapsto N^{-1}\Upsilon_{\Lambda}^{(N)}(\veps)+\left({\textstyle{{\frac{3}{2}-\frac{1}{N}}}}\right)\ln\veps$,
we obtain
\begin{equation}
{\textstyle{\frac{1}{N}}}{\Upsilon_{\Lambda}^{(N)}}^\prime(\veps)
= 
\left({\textstyle{{\frac{3}{2}-\frac{1}{N}}}}\right)
\frac{1}{\veps}
\left[
\frac{{\int}\!
\bigl(1-\invveps
\big\langle\uli{\Delta}^{(1)}_{X^{(N)}}\,,\uli{\Delta}^{(1)}_{X^{(N)}}\big\rangle_{}\bigr)_+^{\!\frac{3N}{2}-2}
\lambda(\dd^{3N}\!q)}
     {{\int}\!
\bigl(1-\invveps
\big\langle\uli{\Delta}^{(1)}_{X^{(N)}}\,,\uli{\Delta}^{(1)}_{X^{(N)}}\big\rangle_{} \bigr)_+^{\!\frac{3N}{2}-1}
\lambda(\dd^{3N}\!q)}-1
\right].
\label{UPSprime}
\end{equation}
	To get a lower bound, we split off a factor 
$\big(1-\invveps\big\langle\uli{\Delta}^{(1)}_{X^{(N)}}\,,\uli{\Delta}^{(1)}_{X^{(N)}}\big\rangle_{} \big)_+$ 
in the integrand of the denominator of r.h.s.\Ref{UPSprime}, and using that $\veps>\veps_g\geq 0$, 
the positivity of the bilinear form now gives
$\big(1- \invveps\big\langle\uli{\Delta}^{(1)}_{X^{(N)}}\,,\uli{\Delta}^{(1)}_{X^{(N)}}\big\rangle_{}\big)_+\leq 1$, 
and so 
\begin{equation}
{\textstyle{\frac{1}{N}}}{\Upsilon_{\Lambda}^{(N)}}^\prime(\veps)
\geq
{\textstyle{\left(\frac{3}{2}-\frac{1}{N}\right)\frac{1}{\veps}}}[1-1]
= 0;
\label{UPSprimeLOWERbound}
\end{equation}
incidentally, this shows once again monotonicity $\uparrow$ of $\veps\mapsto \gamma_{_\Lambda}(\veps)$.
	To get an $N$-independent upper bound to \Ref{UPSprime}, note 
that ${\frac{3N}{2}-2} = \big({\frac{3N}{2}-1}\big)\left(1- \frac{2}{3N-2}\right)$ and that
$0< \left(1- \frac{2}{3N-2}\right)<1$ for $N>1$, then apply
Jensen's inequality w.r.t. $\lambda$ to pull the power $\left(1- \frac{2}{3N-2}\right)$ out 
of the integral in the numerator, then note a cancellation versus the denominator.
	Since $0< \left(1- \frac{2}{3N-2}\right)<1$ for $N>1$,
\begin{equation}
\hskip-.5truecm
{\textstyle{\frac{1}{N}}}{\Upsilon_{\Lambda}^{(N)}}^\prime(\veps)
\leq
\frac{\textstyle{\frac{3}{2}-\frac{1}{N}}}{\veps}
\left[\!
\left[\int\!\!\!
\Big(1-\invveps\big\langle\uli{\Delta}^{(1)}_{X^{(N)}}\,,\uli{\Delta}^{(1)}_{X^{(N)}}\big\rangle_{} \Big)_+^{\!\frac{3N}{2}-1}
\!\!\lambda(\dd^{3N}\!q)\right]^{\!\!-\frac{2}{3N-2}}\!\!\!\!-1\right]\!\!
\label{UPSprimeUPPERbound}
\end{equation}
whenever $N>N_U(\veps)$ (so that the integral is non-zero).
	By the first inequality in \Ref{UPPERlowerBOUNDS} 
with $\ell=N$, the r.h.s.\Ref{UPSprimeUPPERbound} is bounded above independently of $N$.
	The continuity of $\veps\mapsto \gamma_{_\Lambda}(\veps)$ follows.

	Proposition \ref{prop:configENTROPYperPARTICLElimitSPECIAL} is proved. 
\qed

	To complete the proof of Proposition \ref{prop:configENTROPYperPARTICLElimit} we still need to show 
that the omission 
of ${\textstyle{\frac{1}{N}}}\int\!{\tfrhalf} U_\Lambda({\qV,\qV}) \uli\Delta^{(1)}_{X^{(N)}}(\qV)\dd^3q$ from
\Ref{IasBform} was justified. 
	This is now straightforward.
	By hypothesis $(H6)$, $U_\Lambda(\geq 0)$ is a bounded continuous function on 
$\oli\Lambda\!\times\!\oli\Lambda$. 
	So there exists an $N$-independent constant $B>0$ such that
\begin{equation}
0 \leq \int\!{\tfrhalf} U_\Lambda({\qV,\qV}) \uli\Delta^{(1)}_{X^{(N)}}(\qV)\dd^3q \leq B,
\label{abBOUNDonUqq}
\end{equation}
as long as $X^{(N)}\in\oli\Lambda^N$.
	Thus, and abbreviating the expression in the second line on r.h.s.\Ref{MCentropyINTEGRATEDinPasympREV} 
by $S_{{I}^{(N)}_\Lambda}(N^2\veps)$, we have the two-sided estimate
\begin{equation}
{\Upsilon_{\Lambda}^{(N)}}\left(\veps\right)
\leq
S_{{I}^{(N)}_\Lambda}(N^2\veps) 
\leq
{\Upsilon_{\Lambda}^{(N)}}\left(\veps+BN^{-1}\right).
\label{lowerUPPERboundONconfigENTROPY}
\end{equation}
	But
\begin{equation}
\abs{ \Upsilon_{\Lambda}^{(N)}\left(\veps+BN^{-1}\right) - \Upsilon_{\Lambda}^{(N)} 
(\veps)}
\leq 
\int_{\veps}^{\veps+BN^{-1}}\abs{{\Upsilon_{\Lambda}^{(N)}}^\prime(\varsigma)}\dd\varsigma
\leq 
BC,
\label{DIFFconfigENTROPY}
\end{equation}
the last inequality by \Ref{UPSprimeUPPERbound} and by the first inequality in \Ref{UPPERlowerBOUNDS}, 
with $\ell=N$, and by $\veps\leq\varsigma\leq 2\veps$.
	So we conclude that for any $B>0$ we have
\begin{equation}
\lim_{N\to\infty} {\textstyle{\frac{1}{N}}}\Upsilon_{\Lambda}^{(N)}\left(\veps + BN^{-1}\right)
=
\gamma_{_\Lambda}(\veps).
\label{gammaOFvepsPERTURBEDa}
\end{equation}
	Hence, and by \Ref{lowerUPPERboundONconfigENTROPY}, 
\begin{equation}
\lim_{N\to\infty} {\textstyle{\frac{1}{N}}}S_{{I}^{(N)}_\Lambda}(\veps)
=
\gamma_{_\Lambda}(\veps),
\label{gammaOFvepsPERTURBEDb}
\end{equation}
and Proposition \ref{prop:configENTROPYperPARTICLElimit} is proved, 
with $s_{_\Lambda,{}_I}(\veps)=\gamma_{_\Lambda}(\veps)$.
\qed

This also completes the proof of Theorem \ref{thm:ENTROPYperPARTICLElimit}.
\qed
%
%%%%%%%%%%%%%%%%%%%%%%%%%%%%%%%%%%%%%%%%%%%%%%%%%%%%%%%%%%%%%%%%
\subsubsection{Theorem \ref{thm:ENTROPYperPARTICLElimit}$^{++}$ and its proof}
%%%%%%%%%%%%%%%%%%%%%%%%%%%%%%%%%%%%%%%%%%%%%%%%%%%%%%%%%%%%%%%%
	Ruelle's proof \cite{RuelleBOOK} of the traditional thermodynamic limit for 
\Ref{BOLTZMANNentropy} per volume\footnote{Actually, Ruelle 
		discussed the entropy of a regularized microcanonical ensemble measure \cite{RuelleBOOK}.
		In \cite{KieRUELLEpaper} the author showed that a minor modification of Ruelle's approach establishes
		the thermodynamic limit for \Ref{BOLTZMANNentropy} per volume without regularization.}
proceeded along somewhat different lines, and when adapted to the Vlasov scaling it yields an interesting 
alternate proof of Theorem \ref{thm:ENTROPYperPARTICLElimit} which characterizes $s_{_\Lambda}(\veps)$ in
terms of a variational principle (VP) involving $s_{_\Lambda,{}_K}(\veps)$ and yet another (auxiliary) 
``interaction entropy,'' which we denote by $\oli{s}_{_\Lambda,{}_I}(\veps)$.
	For technical reasons we now need to assume that $\veps_g>0$ (rather than $\veps_g\geq 0$).

	So, following Ruelle \cite{RuelleBOOK} we introduce the configurational integral\footnote{Instead of 
		the normalized Lebesgue measure $\lambda(\dd^{3N}\!q)$, Ruelle \cite{RuelleBOOK}
		uses $N!^{-1}\dd^{3N}\!q$ which gives equivalent results in the thermodynamic limit; not
		so in the Vlasov limit.}
\begin{equation}
\Xi_{I^{(N)}_\Lambda}(\cE) 
= 
\int\!\!\chi_{\left\{I^{(N)}_\Lambda<\cE\right\}}
\lambda(\dd^{3N}\!q).
\label{RUELLEconfigINTEGRAL}
\end{equation}
	Up to a purely numerical factor, \Ref{RUELLEconfigINTEGRAL} is quasi the ``$3N/2$-th derivative'' w.r.t.
$\cE$ of $\Psi_{I^{(N)}_\Lambda}(\cE)$, the first derivative of which is given in \Ref{MCconfigINTEGRAL}.
	For convenience we rewrite \Ref{RUELLEconfigINTEGRAL}, with $\cE = N^2\veps$, as
\begin{equation}
\Xi_{I^{(N)}_\Lambda}(N^2\veps) 
= 
\int\!\!\Big(\veps - N^{-2}{I}_\Lambda^{(N)}(\qV_1,...,\qV_N) \Big)_+^{0}\!\lambda(\dd^{3N}\!q).
\label{RUELLEconfigREWRITE}
\end{equation}
\begin{proposition}
\label{prop:RUELLEconfigENTROPYperPARTICLElimitSPECIAL}
	Assume the hypotheses of Theorem \ref{thm:ENTROPYperPARTICLElimit}, but now let $\veps_g>0$.
	Then the following limit exists,
\begin{equation}
\lim_{N\to\infty} {\textstyle\frac{1}{N}}
\ln \Xi_{I^{(N)}_\Lambda}(N^2 \veps) 
= 
\oli{s}_{_\Lambda,{}_I}(\veps) ,
\end{equation}
and $\oli{s}_{_\Lambda,{}_I}(\veps) \leq 0$ is an increasing, right-continuous, function of 
$\veps>\veps_g$.
\end{proposition}
\medskip\noindent
{\textit{Proof of Proposition \ref{prop:RUELLEconfigENTROPYperPARTICLElimitSPECIAL}:}} 
	
	Simplest things first, we note that $\Xi_{I^{(N)}_\Lambda}(\cE) \leq 1$ (obviously), which proves that 
$\ln \Xi_{I^{(N)}_\Lambda}(N^2 \veps) \leq 0$ for all $N$, and so $\oli{s}_{_\Lambda,{}_I}(\veps)\leq 0$ whenever
this limit exists.
	The proof that this limit exists and is a monotonically increasing right-continuous function of 
$\veps>\veps_g>0$ consists of two main steps.

	First, as in our proof of Thm.\ref{thm:ENTROPYperPARTICLElimit}, 
we temporarily replace $N^{-2}I_\Lambda^{(N)}(\qV_1,...,\qV_N)$ by 
$\big\langle\uli{\Delta}^{(1)}_{X^{(N)}}\,,\uli{\Delta}^{(1)}_{X^{(N)}}\big\rangle_{}$
in \Ref{RUELLEconfigREWRITE} and study its logarithm.
	For this we need once again to assume that $N\gg N_U(\veps)$.
	Inspection of our proof of Proposition \ref{prop:configENTROPYperPARTICLElimitSPECIAL} reveals that we can
recycle inequality \Ref{IasBformJENSENarithgeomINEQU}, take its vanishing power, integrate and take logarithms, 
and for $N_U(\veps)<n < N-N_U(\veps)$, in place of \Ref{almostSUPERaddI} we now find
\begin{eqnarray}
\ln \!\int\!\!\!
\Big({\veps} - \big\langle\uli{\Delta}^{(1)}_{X^{(N)}}\,,\uli{\Delta}^{(1)}_{X^{(N)}}\big\rangle_{} \Big)_+^0\!
\lambda(\dd^{3N}\!q)
\geq && 
\label{SUPERaddRUELLEintLOG}\\
&&\hskip-7truecm \ln \!\int\!\!\!
\Big({\veps} - \big\langle\uli{\Delta}^{(1)}_{X^{(n)}}\,,\uli{\Delta}^{(1)}_{X^{(n)}}\big\rangle_{} \Big)_+^0\!
\lambda(\dd^{3n}\!q) +
\nonumber\\
&&\hskip-6.5truecm 
\ln \!\int\!\!\!
\Big(
	{\veps} - \big\langle\uli{\Delta}^{(1)}_{X^{(N-n)}}\,,\uli{\Delta}^{(1)}_{X^{(N-n)}}\big\rangle_{} 
\Big)_+^0\! \lambda(\dd^{3(N-n)}\!q),
\nonumber
\end{eqnarray}
which proves super-additivity of 
$N\!\mapsto\! \ln \!\int\!
\bigl({\veps} - \big\langle\uli{\Delta}^{(1)}_{X^{(N)}}\,,\uli{\Delta}^{(1)}_{X^{(N)}}\big\rangle_{} \bigr)_+^0
\lambda(\dd^{3N}\!q)$ 
without further ado.	
	Furthermore, since $(\cdots)_+^0$ is either 1 or 0, we conclude that 
$\ln \!\int\!
\bigl({\veps} - \big\langle\uli{\Delta}^{(1)}_{X^{(N)}}\,,\uli{\Delta}^{(1)}_{X^{(N)}}\big\rangle_{} \bigr)_+^0
\lambda(\dd^{3N}\!q) \leq 0$.
	This upper bound and super-additivity now yield that the following limit exists, 
\begin{equation}
\lim_{N\to\infty} {\textstyle\frac{1}{N}}
\ln \!\int\!\!\!
\Big(\veps - \big\langle\uli{\Delta}^{(1)}_{X^{(N)}}\,,\uli{\Delta}^{(1)}_{X^{(N)}}\big\rangle_{} \Big)_+^0\!
\lambda(\dd^{3N}\!q)
= 
\tilde{s}_{_\Lambda,{}_I}(\veps);
\label{TILsIaux}
\end{equation}
moreover, $\tilde{s}_{_\Lambda,{}_I}(\veps)\leq 0$ is monotonic increasing, since l.h.s.\Ref{TILsIaux} is.

	Next we would like to prove continuity of $\tilde{s}_{_\Lambda,{}_I}(\veps)$ as function of $\veps$ 
and then conclude the proof as at the end of the proof of Theorem \ref{thm:ENTROPYperPARTICLElimit}, 
but so far a proof of continuity of $\tilde{s}_{_\Lambda,{}_I}(\veps)$ has eluded us.
	Fortunately we can bypass this obstacle because $\tilde{s}_{_\Lambda,{}_I}(\veps)$ is a monotonic 
increasing function of $\veps$.
	We define
\begin{equation}
\tilde{s}_{_\Lambda,{}_I}(\veps^+)
=
\inf_{x>1} \tilde{s}_{_\Lambda,{}_I}(x\veps)
\label{TILsIauxPLUS}
\end{equation}
and show that 
\begin{equation}
\lim_{N\to\infty} {\textstyle\frac{1}{N}}
\ln \!\int\!\!\!
\left(1 - {\textstyle{\frac{1}{\veps N^2}}}{I}_\Lambda^{(N)}(\qV_1,...,\qV_N) \right)_+^{0}\!
\lambda(\dd^{3N}\!q)
= 
\tilde{s}_{_\Lambda,{}_I}(\veps^+),
\label{OLsI}
\end{equation}
which proves Proposition \ref{prop:RUELLEconfigENTROPYperPARTICLElimitSPECIAL}, with
$\oli{s}_{_\Lambda,{}_I}(\veps)=\tilde{s}_{_\Lambda,{}_I}(\veps^+)$.

	To accomplish this, we recall \Ref{IasBform} and \Ref{UbilinearFORM} and rewrite \Ref{RUELLEconfigREWRITE} as
\begin{equation}
\Xi_{I^{(N)}_\Lambda}(N^2\veps) 
= 
\int\!\!\Big(\veps - 
\big\langle\uli{\Delta}^{(1)}_{X^{(N)}}\,,\,\uli{\Delta}^{(1)}_{X^{(N)}}\big\rangle_{}
+
{\textstyle{\frac{1}{N}}}\big\langle\uli{\Delta}^{(1)}_{X^{(N)}}\big\rangle_{}
\Big)_+^{0}\!\lambda(\dd^{3N}\!q),
\label{RUELLEconfigREreWRITE}
\end{equation}
where we also introduced the abbreviation
\begin{equation}
\big\langle\uli{\Delta}^{(1)}_{X^{(N)}}\big\rangle_{}
=
\int\!{\tfrhalf} U_\Lambda({\qV,\qV}) \uli\Delta^{(1)}_{X^{(N)}}(\qV)\dd^3q .
\label{UdiagTERMabbreviate}
\end{equation}
	Since now $\veps_g>0$,
there exist constants $\uli{B},\oli{B}$ satisfying $0\!<\!\uli{B}\!<\!\oli{B}\!<\!\infty$ so that
\begin{equation}
\uli{B}\leq \big\langle\uli{\Delta}^{(1)}_{X^{(N)}}\big\rangle_{}\leq \oli{B}.
\label{UdiagTERMbounds}
\end{equation}
	But then, for all $N>N_U(\veps)$ big enough, we have
\begin{equation}
{\textstyle\frac{1}{N}} \ln \Xi_{I^{(N)}_\Lambda}(N^2\veps) 
\geq
\tilde{s}_{_\Lambda,{}_I}\big(\veps +N^{-1}\uli{B}\big) +o(1)
\geq
\tilde{s}_{_\Lambda,{}_I}\big(\veps^+\big) +o(1)
\label{lnXiLOWERbound}
\end{equation}
where $o(1)\to 0$ as $N\to\infty$.
	So
\begin{equation}
\liminf_{N\to\infty} {\textstyle\frac{1}{N}} \ln \Xi_{I^{(N)}_\Lambda}(N^2\veps) 
\geq
\tilde{s}_{_\Lambda,{}_I}\big(\veps^+\big).
\label{limINFlnXiLOWERbound}
\end{equation}
	On the other hand, for all $N>N_U(\veps)$ we also have that
\begin{equation}
{\textstyle\frac{1}{N}} \ln \Xi_{I^{(N)}_\Lambda}(N^2\veps) 
\leq
\tilde{s}_{_\Lambda,{}_I}\big(\veps +N^{-1}\oli{B}\big) +o(1),
\label{lnXiUPPERbound}
\end{equation}
and so
\begin{equation}
\limsup_{N\to\infty} {\textstyle\frac{1}{N}} \ln \Xi_{I^{(N)}_\Lambda}(N^2\veps) 
\leq
\tilde{s}_{_\Lambda,{}_I}\big(\veps^+\big).
\label{limSUPlnXiUPPERbound}
\end{equation}
The estimates \Ref{limINFlnXiLOWERbound} and \Ref{limSUPlnXiUPPERbound} prove \Ref{OLsI}.

	So $\oli{s}_{_\Lambda,{}_I}(\veps)=\tilde{s}_{_\Lambda,{}_I}(\veps^+)$.
	Of course, $\oli{s}_{_\Lambda,{}_I}(\veps)=\tilde{s}_{_\Lambda,{}_I}(\veps)$ at all $\veps$ which are 
points of continuity of $\tilde{s}_{_\Lambda,{}_I}(\veps)$, and the two functions share their points of discontinuity.
	At such points $\oli{s}_{_\Lambda,{}_I}(\veps)$ is right-continuous and may or may not agree with
$\tilde{s}_{_\Lambda,{}_I}(\veps)$.

Proposition \ref{prop:RUELLEconfigENTROPYperPARTICLElimitSPECIAL} is proved.
\qed

	We are now ready to state our second upgrade of our Theorem \ref{thm:ENTROPYperPARTICLElimit}. 
\setcounter{theorem}{0}
\begin{theorem}
\label{thm:ENTROPYperPARTICLElimitUPup}$^{\hskip-.2truecm ++}$
	Under the hypotheses of Proposition \ref{prop:RUELLEconfigENTROPYperPARTICLElimitSPECIAL},
Theorem \ref{thm:ENTROPYperPARTICLElimit} holds and 
the system-specific Boltzmann entropy per particle $s_{_\Lambda}(\veps)$ given in \Ref{MCentropyASYMPexpansionI}
satisfies the variational principle
\begin{equation}
s_{_\Lambda}(\veps) 
=
\sup_{0\leq x\leq 1}\Bigl( s_{_\Lambda,{}_K}(x\veps) 
+  \oli{s}_{_\Lambda,{}_I}([1-x]\veps)\Bigr).
\label{VPs}
\end{equation}
\end{theorem}
\setcounter{theorem}{3}

\medskip\noindent
{\textit{Proof of Theorem \ref{thm:ENTROPYperPARTICLElimitUPup}$^{++}$:}} 

	Integration by parts yields, for any $\ell>0$ and $\veps>\veps_g$, 
\begin{equation}
\int\!\!\! 
\left(1 - {\textstyle{\frac{1}{\veps N^2}}}{I}_\Lambda^{(N)} \right)_+^\ell\!
\lambda(\dd^{3N}\!q)
= 
\int_0^1 
\int\!\!\! 
\left(1 - {\textstyle{\frac{1}{[1-x]\veps N^2}}}{I}_\Lambda^{(N)}\right)_+^0\!
\lambda(\dd^{3N}\!q)\dd{x^\ell},
\label{NsIpartINT}
\end{equation}
where we suppressed the arguments $(\qV_1,...,\qV_N)$ from ${I}_\Lambda^{(N)}(\qV_1,...,\qV_N)$.
	Setting $\ell={\frac{3N}{2}-1}$, recalling \Ref{RUELLEconfigREWRITE},
and using that $N^{-1}\ln\big( {\frac{3N}{2}-1}\big)\to 0$, we find
\begin{equation}
s_{_\Lambda,{}_I}(\veps)
=
\lim_{N\to\infty} {\textstyle{\frac{1}{N}}} \ln \!
\int_0^1 
 \Xi_{I^{(N)}_\Lambda}(N^2[1-x]\veps)\;x^{\frac{3N}{2}-2}\,\dd{x}.
\label{sInewLIM}
\end{equation}
	Proposition \ref{prop:RUELLEconfigENTROPYperPARTICLElimitSPECIAL} and Laplace's method 
(cf. sect. II.7 in \cite{EllisBOOK}) now yield
\begin{equation}
s_{_\Lambda,{}_I}(\veps) 
=
\sup_{0\leq x\leq 1}\Bigl( {\textstyle{\frac{3}{2}}} \ln x + \oli{s}_{_\Lambda,{}_I}([1-x]\veps)\Bigr);
\label{VPsI}
\end{equation}
note that \Ref{VPsI} implies that $\veps\mapsto s_{_\Lambda,{}_I}(\veps)$ is continuous even when 
$\oli{s}_{_\Lambda,{}_I}(\veps)$ is not. 
	Recalling next the definition \Ref{sK} of $s_{_\Lambda,{}_K}(\veps)$ as well as \Ref{sKsI} of
Theorem \ref{thm:ENTROPYperPARTICLElimit}$^+$, we see that Theorem \ref{thm:ENTROPYperPARTICLElimit}$^{++}$ 
is proved.
\qed

	We end this subsection by pointing out that our method of proving 
Theorem \ref{thm:ENTROPYperPARTICLElimit}$^{++}$ not only avoids the regularization of Dirac's $\delta$ measure,
we also tackled  the map $\cE\mapsto S(\cE)$ directly rather than its inverse $\cS\mapsto E(\cS)$ \cite{RuelleBOOK}.
	The strategy to tackle $\cS\mapsto E(\cS)$ is due to Griffiths \cite{Griffiths}.

%%%%%%%%%%%%%%%%%%%%%%%%%%%%%%%%%%%%%%%%%%%%%%%%%%%%%%%%%%%%%%%%
%%%%%%%%%%%%%%%%%%%%%%%%%%%%%%%%%%%%%%%%%%%%%%%%%%%%%%%%%%%%%%%%
\subsection{Proof of Theorem \ref{thm:maxENTROPYvp}} 
%%%%%%%%%%%%%%%%%%%%%%%%%%%%%%%%%%%%%%%%%%%%%%%%%%%%%%%%%%%%%%%%
%%%%%%%%%%%%%%%%%%%%%%%%%%%%%%%%%%%%%%%%%%%%%%%%%%%%%%%%%%%%%%%%

	Since formula \Ref{MCentropyINTEGRATEDinPasympREV} holds also under the assumptions $(H1)$--$(H5)$ 
on the interactions, and since it is well-known that the system-specific Boltzmann entropy per particle of the 
perfect gas \Ref{sK} minimizes Boltzmann's  $H$ functional under the constraint of prescribing the value of the 
kinetic Hamiltonian, it suffices  to study the \emph{interaction entropy of Boltzmann's ergodic ensemble}, 
\begin{equation}
S_{{I}^{(N)}_\Lambda}(\cE)
=
\ln \!\int\!
\big(1 -  {\textstyle{\frac{1}{\cE}}}{I}_\Lambda^{(N)}(\qV_1,...,\qV_N) \big)_+{\!\!\!\!}^{\frac{3N}{2}-1}
\lambda(\dd^{3N}\!q).
\label{INTentropyN}
\end{equation}
	Note that \Ref{INTentropyN} is non-positive, and under hypotheses $(H1)$--$(H5)$ we also have
\begin{eqnarray}
\hskip-1truecm
&& \left[\!\int\!
\Big(1 -  \invveps {\textstyle{\frac{1}{N^2}}}{I}_\Lambda^{(N)}(\qV_1,...,\qV_N) \Big)_+{\!\!\!\!}^{\frac{3N}{2}-1}
\lambda(\dd^{3N}\!q)\right]^{{\frac{2}{3N-2}}}
\geq
\nonumber\\
&& \left[\!\int\!\!\!
\Big(1 -  \invveps {\textstyle{\frac{1}{N^2}}}{I}_\Lambda^{(N)}(\qV_1,...,\qV_N) \Big){\!}^{\frac{3N}{2}-1}
\!\prod_{1\leq k\leq N}\chi_{B_\delta[k]}\lambda(\dd^3q_k)
\right]^{{\frac{2}{3N-2}}}
\geq
\nonumber\\
&&
|C_\delta|^{{\frac{2}{3-2/N}}}
\!\int\!\!\!
\Big(1 -  \invveps {\textstyle{\frac{1}{N^2}}}{I}_\Lambda^{(N)}(\qV_1,...,\qV_N) \Big)
\!\prod_{1\leq k\leq N}
\textstyle{\frac{\chi_{B_\delta[k]}}{\int_{B_\delta[k]\cap\Lambda}\lambda(\dd^3{q})}}\lambda(\dd^3q_k)
\geq
\nonumber\\
&&
|C_\delta|^{{\frac{2}{3}}}
\left(1- \tfrhalf\left(1+\textstyle{\frac{\veps_g}{\veps}}\right)\right)
> 0,
\label{lowerBOUNDexpSI}
\end{eqnarray}
where again $ C_\delta$ is given in \Ref{Cdelta}, 
but now with $\delta(\veps)$ independent of $k$ and $N$ chosen so  that 
$N^{-2}{I}_\Lambda^{(N)}(\qV_1,...,\qV_N)\leq \tilde\veps_g(N) + (\veps- \veps_g)/2$ 
when the $\qV_k$ vary in $B_\delta(\qV_{g,k})\cap\oli{\Lambda}$,
where $(\qV_{g,1},...,\qV_{g,N})$ is a ground state configuration for ${I}_\Lambda^{(N)}(\qV_1,...,\qV_N)$
with a fat neighborhood, which exists by $(H2)\&(H3)$.
	We also used that 
$\tilde\veps_g(N)=\min N^{-2}{I}_\Lambda^{(N)}(\qV_1,...,\qV_N) \leq\veps_g$ (see Appendix A).
	So
\begin{equation}
{\textstyle{\frac{2}{3N-2}}} S_{{I}^{(N)}_\Lambda}(\cE) 
\geq 
\ln \left(|C_\delta|^{{\frac{2}{3}}}
\left(1- \tfrhalf\left(1+\textstyle{\frac{\veps_g}{\veps}}\right)\right)
\right)>-\infty
\label{lowerBOUNDc}
\end{equation}
for all $N>1$.
	The estimate \Ref{lowerBOUNDc} guarantees the existence of limit points of the (negative) interaction entropy
per particle as $N\to\infty$.
	We want to show that the interaction entropy per particle actually has a limit and characterize the limit
by the variational principle stated in Theorem \ref{thm:maxENTROPYvp}.

	We begin by characterizing \Ref{INTentropyN} by its own maximum entropy principle.
	We introduce the \textit{quasi-interaction energy of} $\varrho^{(N)} \in \Psp^s(\Lambda^N)$, defined by
\begin{equation}
\pzcQ_{\,\,I/\veps}^{(N)}\left(\varrho^{(N)}\right) 
= 
{\textstyle{\frac{3N-2}{2}}}\!
\int\!\ln \!\left(1 -  {\textstyle{\frac{1}{\veps N^2}}}{I}_\Lambda^{(N)}(\qV_1,...,\qV_N)\right)_{\!+}
\!\varrho^{(N)}(\dd^{3N}\!q)
\label{quasiCONFIGenergy}
\end{equation}
whenever $\supp\varrho^{(N)}\subset\supp \big(\veps - N^{-2}{I}_\Lambda^{(N)}\big)_+$; 
else we set $\pzcQ_{\,\,I/\veps}^{(N)}\left(\varrho^{(N)}\right) = -\infty$.
	The \emph{entropy of} $\varrho^{(N)}$ \emph{relative to}
$\varrho_{ap}^{(N)}\in \Psp^s(\Lambda^N)$ is defined as usual\footnote{Our physicists' 
		sign convention of relative entropy is opposite to the probabilists' one.}
by
\begin{equation}
\pzcR^{(N)}\left(\varrho^{(N)}|\varrho_{ap}^{(N)}\right) 
=
 - \int \ln \left(\frac{d\varrho^{(N)}}{d\varrho_{ap}^{(N)}} \right) \varrho^{(N)}(\dd^{3N}q)
\label{relativeENTROPY}
\end{equation}
if $\varrho^{(N)}$ is absolutely continuous w.r.t. the a-priori measure $\varrho_{ap}^{(N)}$, and provided the integral 
in \Ref{relativeENTROPY} exists. 
	In all other cases, $\pzcR^{(N)}\big(\varrho^{(N)}|\varrho_{ap}^{(N)}\big) = -\infty$.
	Finally, we define what we call the \textit{interaction entropy of} $\varrho^{(N)}$ by
\begin{equation}
\pzcS_{I/\veps}^{(N)}(\varrho^{(N)}) 
\equiv
\pzcR^{(N)}\left(\varrho^{(N)}|\lambda\right) + \pzcQ_{\,\,I/\veps}^{(N)}\left(\varrho^{(N)}\right).
\label{ACTUALentropyNBODY}
\end{equation}
	We are now ready to state our variational principle.
\begin{proposition}
\label{prop:VPfinN}
	For $\veps >\veps_g\geq 0$, the interaction entropy functional \Ref{ACTUALentropyNBODY} achieves 
its supremum.
	The maximizer is the unique probability measure 
\begin{equation}
\varrho_{{N^2\veps}}^{(N)}(\dd^{3N}\!q)
=
\frac{\quad
\left(1-{\textstyle{\frac{1}{\veps N^2}}}I_\Lambda^{(N)}(\qV_1,...,\qV_N) \right)_+{\!\!\!\!}^{\frac{3N}{2}-1}
\dd^{3N}\!q}
{\displaystyle{\int}\!
\left(1-
{\textstyle{\frac{1}{\veps N^2}}}I_\Lambda^{(N)}(\tilde\qV_1,...,\tilde\qV_N)\right)_+{\!\!\!\!}^{\frac{3N}{2}-1} 
\dd^{3N}\!\tilde{q}} 
\in (\Psp^s\cap\Lsp^\infty)(\Lambda^N);
\label{micLNmeasure}
\end{equation}
\vskip-.2truecm
\noindent
thus
\begin{equation}
\max_{\varrho^{(N)}\in \Psp^s(\Lambda^N)} \pzcS_{I/\veps}^{(N)}\bigl(\varrho^{(N)}\bigr) =
 \pzcS_{I/\veps}^{(N)}\bigl(\varrho_{{N^2\veps}}^{(N)}\bigr).
\label{vpFE}
\end{equation}
\vskip-.1truecm
\noindent
	Moreover,
\begin{equation}
\pzcS_{I/\veps}^{(N)}\bigl(\varrho_{{N^2\veps}}^{(N)}\bigr) 
= 
S_{{I}^{(N)}_\Lambda}(N^2\veps).
\label{vpFEzwei}
\end{equation}
\end{proposition}
\vskip-.2truecm

\medskip\noindent
{\textit{Proof of Proposition \ref{prop:VPfinN}:}} 

	Under our hypotheses on $I_\Lambda^{(N)}$ the measure $\varrho_{{N^2\veps}}^{(N)}$ is absolutely continuous 
w.r.t. $\lambda$ and bounded whenever $\veps>\veps_g$, so the standard convexity 
argument due to Boltzmann \cite{Boltzmann},  cf. \cite{RuelleBOOK,EllisBOOK}, applies and shows that 
$\pzcS_{I/\veps}^{(N)}\bigl(\varrho^{(N)}\bigr)-\pzcS_{I/\veps}^{(N)}\bigl(\varrho_{{N^2\veps}}^{(N)}\bigr)\leq 0$, 
with equality~holding if and only if $\varrho^{(N)} = \varrho_{{N^2\veps}}^{(N)}$.
	Identity \Ref{vpFEzwei} is verified by explicit calculation. 
\qed

	Since ultimately we are interested in the limit $N\to\infty$ of our finite-$N$ results, we recall the
formalism of probabilities on infinite sequences $\Lambda^\Nset$, as encountered already in section 3 for
$(\Lambda\times\Rset^3)^\Nset$.
	Thus, by $\Psp^s(\Lambda^\Nset)$ we denote the permutation-symmetric probability measures 
on the set of infinite exchangeable sequences in $\Lambda$.
	Let $\{\varrhon\}_{n\in\Nset}$ denote the sequence of marginals of any $\varrho\in{\Psp}^s(\Lambda^\Nset)$. 
	The de Finetti \cite{deFinetti} -- Dynkin \cite{Dynkin} -- Hewitt-Savage \cite{HewittSavage} 
decomposition theorem for $\Psp^s(\Lambda^\Nset)$ states that every $\varrho\in \Psp^s(\Lambda^\Nset)$ is uniquely 
presentable as a linear convex superposition of infinite product measures, i.e., for each 
$\varrho \in \Psp^s(\Lambda^\Nset)$ there exists a unique probability measure 
$\varsigma(d\rho|\varrho)$ on $\Psp(\Lambda)$, such that for each $n\in \Nset$,
\begin{equation}
\varrhon (\dd^{3n}q) =
	\int_{\Psp(\Lambda)}\rho^{\otimes n}(\dd^{3}q_1\cdots\dd^{3}q_n)\,\varsigma(\dd\rho|\varrho),
\label{deFinettiDECOMPconfig}
\end{equation}
where $\varrhon$ is the $n$-th marginal measure of $\varrho$, and where
$\rho^{\otimes n}(\dd^{3}q_1\cdots\dd^{3}q_n)\equiv  \rho(\dd^{3}q_1)\times \cdots\times\rho(\dd^{3}q_n)$.
	Also, \Ref{deFinettiDECOMPconfig} expresses the extreme point decomposition of the convex set 
$\Psp^s(\Lambda^\Nset)$, see \cite{HewittSavage}.

	Next we would like to formulate the $N=\infty$ analogue of \Ref{ACTUALentropyNBODY}, but the naive 
manipulation of the formulas is not recommended.
	The functional $\pzcQ_{\,\,I/\veps}^{(N)}$ is well-defined by \Ref{quasiCONFIGenergy} and its accompanying
text for all $N\in\Nset$; however, since our conditions on ${I}_\Lambda^{(N)}(\qV_1,...,\qV_N)$ allow it to be unbounded above when two positions $\qV_k$ and $\qV_l$ approach each other (for example: Coulomb interactions), 
we find that $\pzcQ_{I/\veps}^{(N)}\big(\rho^{\otimes n}\big)=-\infty$ for all product measures $\rho^{\otimes n}$,
but these are exactly the $N$-point marginals of the extreme points of our set of exchangeable measures on the
infinite Cartesian product $\Lambda^\Nset$.
	This obstacle can be circumvented by noting that the finite-$N$
quasi-interaction energy defined in \Ref{quasiCONFIGenergy} and the line ensuing \Ref{quasiCONFIGenergy} is
the monotone limit of a family of concave functionals in which the integrand function 
$\ln(1-x)_+$ (with $\ln 0=-\infty$ understood) is replaced 
by $\ln(1-x)\chi_{\{x<1-\alpha\}}+[\ln\alpha +(1-\alpha -x)/\alpha]\chi_{\{x\geq 1-\alpha\}}$;
thus
\begin{eqnarray}
{}^\alpha\pzcQ_{\,\,I/\veps}^{(N)}\left(\varrho^{(N)}\right) 
\!\!\!\!\!\!\!\!
&&= 
{\textstyle{\frac{3N-2}{2}}}
\int\!\Bigl(\ln \! \left(1 -  {\textstyle{\frac{1}{\veps N^2}}}{I}_\Lambda^{(N)}\right)\!
\chi_{\left\{I^{(N)}_\Lambda < \veps N^2(1-\alpha)\right\}}
\Bigr.
\label{ALPHAquasiCONFIFenergy}\\
&&\quad 
+
\Bigl[\ln\alpha + {\textstyle\frac{1}{\alpha}} 
\left(1 - {\textstyle{\frac{1}{\veps N^2}}}{I}_\Lambda^{(N)} -\alpha\right)\!\Bigr]
\!\chi_{\left\{I^{(N)}_\Lambda\geq\veps N^2(1-\alpha)\right\}}\Bigr) \varrho^{(N)}(\dd^{3N}\!q),
\nonumber
\end{eqnarray}
where we omitted the argument $(\qV_1,...,\qV_N)$ from $I_\Lambda^{(N)}$, for brevity, and
\begin{equation}
\pzcQ_{\,\,I/\veps}^{(N)}\left(\varrho^{(N)}\right) 
= 
\lim_{\alpha\downarrow 0} {}^\alpha\pzcQ_{\,\,I/\veps}^{(N)}\left(\varrho^{(N)} \right).
\label{quasiCONFIGenergyREdef}
\end{equation}

	We also define ${}^\alpha\pzcS_{I/\veps}^{(N)}\big(\varrho^{(N)}\big)$ precisely like 
$\pzcS_{I/\veps}^{(N)}\big(\varrho^{(N)}\big)$ except that 
$\pzcQ_{\,\,I/\veps}^{(N)}\big(\varrho^{(N)}\big)$ is replaced by
${}^\alpha\pzcQ_{\,\,I/\veps}^{(N)}\big(\varrho^{(N)}\big)$.
	We have
${}^\alpha\pzcS_{I/\veps}^{(N)}\big(\rho^{\otimes n}\big)>-\infty$ for
all $\rho \in (\Psp\cap\Lsp^1\ln\Lsp^1)(\Lambda)$, and 
$\lim_{\alpha\downarrow 0}{}^\alpha\pzcS_{I/\veps}^{(N)}\big(\rho^{\otimes n}\big)=-\infty$
whenever $I/\veps\not \leq 1$.
	By ${}^\alpha\varrho_{{N^2\veps}}^{(N)}$ we denote the unique maximizer of 
${}^\alpha\pzcS_{I/\veps}^{(N)}\big(\varrho^{(N)}\big)$, easily proven to exist as done for
$\pzcS_{I/\veps}^{(N)}\big(\varrho^{(N)}\big)$.
	Equally easily we find
$\lim_{\alpha\downarrow 0} {}^\alpha \pzcS_{I/\veps}^{(N)}\big( {}^\alpha\varrho_{{N^2\veps}}^{(N)}\big) 
	= \pzcS_{I/\veps}^{(N)}\big(\varrho_{{N^2\veps}}^{(N)}\big)$.

	We are now ready to formulate the $N=\infty$ analogue of \Ref{ACTUALentropyNBODY}.

	To define the \emph{mean quasi-interaction energy of} $\varrho\in \Psp^s(\Lambda^\Nset)$,
we introduce the subset $\Psp^s_{U^2_\Lambda}(\Lambda^\Nset)\subset\Psp^s(\Lambda^\Nset)$ 
for which the expected value of $U^2_\Lambda$ is finite; i.e. 
$\int U^2_\Lambda(\qV,\qV')\, {}^2\!\varrho(\dd^3q\dd^3q')<\infty$, where
${}^2\!\varrho(\dd^3q\dd^3q')$ is the second marginal measure of $\varrho\in\Psp^s_{U^2_\Lambda}(\Lambda^\Nset)$.
	Also, by $\Psp_{U^2_\Lambda}(\Lambda)$ we denote the subset of $\Psp(\Lambda)$ which consists of 
Lebesgue-absolutely continuous probability measures $\rho$ for which 
$\int U^2_\Lambda(\qV,\qV')\rho^{\otimes 2}(\dd^3q\dd^3q')<\infty$,
which implies $\left\langle\rho^{}\,,\,\rho_{}\right\rangle < \infty$;
here we recycled the bilinear form notation \Ref{UbilinearFORM} for lower semi-continuous 
(rather than continuous) $U_\Lambda$.
	If $\varrho\in \Psp^s_{U^2_\Lambda}(\Lambda^\Nset)$, then the decomposition measure 
$\varsigma(\dd\rho|\varrho)$ is concentrated on $\Psp_{U^2_\Lambda}(\Lambda)$;
this can be shown by adapting arguments from \cite{HewittSavage}; cf. also \cite{MesserSpohn}.
	The \emph{mean quasi-interaction energy of} $\varrho\in \Psp^s_{U^2_\Lambda}(\Lambda^\Nset)$ 
is defined as
\begin{equation}
\uli\pzcQ_{\,\,I/\veps}(\varrho)
\equiv
\lim_{\alpha\downarrow 0} 
\lim_{n\to\infty} 
\ {\textstyle{\frac{1}{n}}} {}^\alpha\pzcQ^{(n)}_{\,\,I/\veps}\big(\varrhon\big).
\label{meanQuasiINTENERGY}
\end{equation}

	We show that $\uli\pzcQ_{\,\,I/\veps}(\varrho)$ is well-defined.
	By the linearity of 
$\varrhon\mapsto  {}^\alpha\pzcQ^{(n)}_{\,\,I/\veps}\big(\varrhon\big)$, 
the presentation \Ref{deFinettiDECOMPconfig} yields
\begin{equation}
{}^\alpha\pzcQ^{(n)}_{\,\,I/\veps}\big(\varrhon\big)
=
\int {}^\alpha\pzcQ^{(n)}_{\,\,I/\veps}(\rho^{\otimes n})\,\varsigma(\dd\rho|\varrho),
\label{QnVIArep}
\end{equation}
and on $\Psp^s_{U^2_\Lambda}(\Lambda^\Nset)$ the conventional law of large numbers for $U$ statistics applies (see \cite{Hoeffding})
and yields
\begin{eqnarray}
\lim_{n\to\infty} 
{\textstyle\frac{1}{n}}
 {}^\alpha\pzcQ^{(n)}_{\,\,I/\veps}(\rho^{\otimes n})
\!\!\!\!\!\!\!\!&&
=
{\textstyle{\frac{3}{2}}} \left[
\ln\bigl[1 - {\textstyle{\frac{1}{\veps}}} \big\langle\rho\,,\,\rho\big\rangle_{}\bigr]
	\chi_{_{\left\{\left\langle\rho^{}\,,\,\rho_{}\right\rangle < \veps(1-\alpha)\right\}}}
\right.
\label{meanQalphaREP}\\
&&
+
\left.
\left(\ln \alpha +{\textstyle{\frac{1}{\alpha}}}
\bigl[1 -\alpha- {\textstyle{\frac{1}{\veps}}} \big\langle\rho\,,\,\rho\big\rangle_{}\bigr]\right)
	\chi_{_{\left\{\left\langle\rho^{}\,,\,\rho_{}\right\rangle \geq \veps(1-\alpha)\right\}}}
\right].
\nonumber
\end{eqnarray}
	Clearly, when $\alpha\downarrow 0$ in \Ref{meanQalphaREP} the ``value'' $-\infty$ is assigned
to all $\rho$ for which $\left\langle\rho^{}\,,\,\rho_{}\right\rangle \geq \veps$; 
the $\alpha\downarrow 0$ limit is finite when $\left\langle\rho^{}\,,\,\rho_{}\right\rangle < \veps$.
	We conclude with:
\begin{lemma} 
\label{lem:meanINTENTROPYisAFFINE}
	The mean quasi-interaction energy \Ref{meanQuasiINTENERGY} is well-defined and affine linear.
	For $\varrho\in \Psp^s_{U^2_\Lambda}(\Lambda^\Nset)$ having decomposition measure 
$\varsigma(\dd\rho|\varrho)$ supported entirely by $\rho$ for which 
$\left\langle\rho^{}\,,\,\rho_{}\right\rangle < \veps$, we have
 \Ref{meanQuasiINTENERGY} given by
\begin{equation}
\uli\pzcQ_{\,\,I/\veps}(\varrho)
=
\int \pzcQ_{\,\,I/\veps}(\rho)\,\varsigma(\dd\rho|\varrho),
\label{meanQrep}
\end{equation}
where
\begin{equation}
\pzcQ_{\,\,I/\veps}(\rho)
\equiv
{\textstyle{\frac{3}{2}}}\ln\bigl[1 - {\textstyle{\frac{1}{\veps}}} \big\langle\rho\,,\,\rho\big\rangle_{}\bigr];
\label{Qdef}
\end{equation}
otherwise, $\uli\pzcQ_{\,\,I/\veps}(\varrho)=-\infty$.
\end{lemma} 

	The $N=\infty$ analogue of \Ref{relativeENTROPY} is the well-known \emph{mean (relative) entropy}
of $\varrho\in {\Psp}^s(\Lambda^\Nset)$, which is well-defined as limit
\begin{equation}
\uli\pzcR(\varrho)
\equiv
\lim_{n\to\infty} \ {\textstyle{\frac{1}{n}}} \pzcR^{(n)}\big(\varrhon|\lambda\big).
\label{meanENTROPY}
\end{equation}
	Here, $\pzcR^{(n)}\big(\varrhon|\lambda\big)$, $n \in\{0,1,...\}$, is the relative entropy of 
$\varrhon$, as defined in \Ref{relativeENTROPY}; we also set 
$\pzcR^{(-k)}\big({}^{-k}\varrho|\lambda\big)\equiv 0$ for all $k\in\Nset$.
	The limit \Ref{meanENTROPY} exists or is $-\infty$.
	This is a consequence of the next lemma, which holds for $\varrho\in \Psp^s(\Lambda^N)$ or
$\varrho\in \Psp^s(\Lambda^\Nset)$.
	If $\varrho=\varrho^{(N)}$, it is understood that $k\leq N$ in $\varrho^{(N)}_k$. 

\begin{lemma} 
\label{lem:ENTROPYproperties}
	Relative entropy $n\mapsto\pzcR^{(n)}\big(\varrhon|\lambda\big)$ has the following properties:

(A) {\emph{Non-positivity}}: For all $n$,
\begin{equation}
	\pzcR^{(n)}\big(\varrhon|\lambda\big)\leq 0;
\end{equation}

(B) {\emph{Monotonic decrease}}: If $n>m$ then
\begin{equation}
\pzcR^{(n)}\big(\varrhon|\lambda\big)
\leq 
\pzcR^{(m)}\big(\varrhom|\lambda\big);
\end{equation} 

(C) {\emph{Strong sub-additivity}}: For $m,\,n \leq \ell$, and $k={\ell - m - n}$,
\begin{equation}
\hskip-.2truecm
\pzcR^{(\ell)}\big({}^\ell\varrho|\lambda\big)
 \leq 
\pzcR^{(m)}\big(\varrhom|\lambda\big)
+
\pzcR^{(n)}\big(\varrhon|\lambda\big)
+
\pzcR^{(k)}\big(\varrhok|\lambda\big)
-
\pzcR^{(-k)}\big({}^{-k}\varrho|\lambda\big).
\end{equation} 
\end{lemma}

	The proof of Lemma \ref{lem:ENTROPYproperties} is a straightforward
adaptation from a proof by Robinson and Ruelle \cite{RobinsonRuelle} (section 2, proof of proposition 1)
for the standard-thermodynamic-limit problem to the Vlasov limit, studied here, cf. \cite{KieCPAM}.

	The next lemma also has an elementary proof which likewise is an adaption
from \cite{RobinsonRuelle}, proof of their proposition 3, cf. \cite{KieCPAM}.
\begin{lemma} 
\label{lem:meanENTROPYisAFFINE}
	The mean entropy functional \Ref{meanENTROPY} is affine linear.
\end{lemma} 

	Lemma \ref{lem:meanENTROPYisAFFINE} in conjunction with the 
de Finetti \cite{deFinetti} -- Dynkin \cite{Dynkin} -- Hewitt-Savage \cite{HewittSavage} 
decomposition theorem for $\Psp^s(\Lambda^\Nset)$ yields a key formula for the mean entropy 
which does not hold for the finite-$N$ entropy.
	Namely, as a consequence of Lemma \ref{lem:meanENTROPYisAFFINE}, the extremal decomposition of $\varrho$ 
yields
\begin{equation}
 \uli\pzcR(\varrho) = \int \pzcR(\rho|\lambda)\,\varsigma(\dd\rho|\varrho),
\label{RaffineREP}
\end{equation}
where we also set $\pzcR(\rho|\lambda)\equiv\pzcR^{(1)}(\rho|\lambda)$.

	Lemma \ref{lem:ENTROPYupperSEMIconti}, 
also proved by adaption of a corresponding proof in \cite{RobinsonRuelle}, 
proposition 4, ends the listing of properties of mean relative entropy \Ref{meanENTROPY}.
\begin{lemma} 
\label{lem:ENTROPYupperSEMIconti}
	The mean entropy functional is weakly upper semi-continuous.
\end{lemma} 

	Finally we define the \emph{mean interaction entropy of} $\varrho\in \Psp^s(\Lambda^\Nset)$,
\begin{equation}
\uli{\pzcS}_{I/\veps}(\varrho)
\equiv
\uli\pzcR(\varrho) + \uli\pzcQ_{\,\,I/\veps}(\varrho).
\label{meanAdefine}
\end{equation}
	By \Ref{RaffineREP} and \Ref{meanQrep} we have 
\begin{equation}
\uli{\pzcS}_{I/\veps}(\varrho)
=
\int_{\Psp(\Lambda)} \pzcS_{_{I/\veps}}(\rho) \,\varsigma(\dd\rho|\varrho),
\label{Arep}
\end{equation}
where we introduced the functional
\begin{equation}
\pzcS_{_{I/\veps}}(\rho)
\equiv
\pzcR(\rho|\lambda) +  \pzcQ_{\,\,I/\veps}(\rho),
\label{ACTUALentropyONEBODY}
\end{equation}
which is well-defined and finite whenever $\rho \in (\Psp\cap \Lsp^1\ln\Lsp^1)(\Lambda)$ and 
$\big\langle\rho\,,\,\rho\big\rangle <\veps$; else we have $\pzcS_{_{I/\veps}}(\rho)=-\infty$.
	Note that $\pzcS_{_{I/\veps}}(\rho)\leq 0$, for $\pzcR\bigl(\varrho|\lambda\bigr) \leq 0$ and
$\pzcQ_{\,\,I/\veps}(\rho)\leq 0$, the latter because $U_\Lambda\geq 0$ by hypothesis.

	Because of \Ref{Arep} the problem of maximizing $\uli{\pzcS}_{I/\veps}(\varrho)$
reduces to maximizing $\pzcS_{_{I/\veps}}(\rho)$ given in \Ref{ACTUALentropyONEBODY}.
\begin{proposition}
\label{prop:ACTUALentropyFUNCofRHO} 
$\pzcS_{_{I/\veps}}(\rho)$ is weakly upper semi-continuous for $\veps >\veps_g\geq 0$ and 
takes its finite non-positive maximum at a solution of the fixed point equation 
\begin{equation}
\rho(\qV)
=
\frac{
 \exp\left(-\vartheta^{-1}_\veps(\rho) \int_\Lambda U_\Lambda(\qV,\tilde{\qV})\rho(\tilde{\qV})\dd^3\tilde{q} 
    \right)}
          {\int_\Lambda
 \exp\left(-\vartheta^{-1}_\veps(\rho) \int_\Lambda U_\Lambda(\hat{\qV},\tilde{\qV})\rho(\tilde{\qV})\dd^3\tilde{q}
    \right)
	\dd{\hat{q}}},
\label{fixPOINTEQrhoUagain}
\end{equation}
where
\begin{equation}
\vartheta_\veps(\rho)
=
{\textstyle{\frac{2}{3}}}
\bigl(1 - {\textstyle{\frac{1}{\veps}}} \big\langle\rho\,,\,\rho\big\rangle_{}\bigr)
\veps >0.
\label{THETArewrite}
\end{equation}
\end{proposition}
\medskip\noindent
{\textit{Proof of Proposition \ref{prop:ACTUALentropyFUNCofRHO}:}} 

	Since relative entropy $\pzcR(\rho|\lambda)$ is weakly upper semi-continuous 
(\cite{ReedSimon}, Suppl. to IV.5; \cite{EllisBOOK}, chpt.VIII), and since the functional 
$\pzcQ_{\,\,I/\veps}(\rho)$ is weakly upper semi-continuous
as a consequence of hypothesis $(H2)$ and the positivity of $U_\Lambda$, so is $\pzcS_{_{I/\veps}}(\rho)$.
	Since $\oli{\Lambda}$ is compact, $\pzcS_{_{I/\veps}}(\rho)$ now takes its maximum, which is
non-positive because $\pzcS_{_{I/\veps}}(\rho)\leq 0$, and finite (i.e. $>-\infty)$ because of the 
following.
	Let $k\mapsto \rho_{(k)}$ in $(\Psp\cap\Csp^\infty_0)(\Lambda)$ be a minimizing sequence for 
$\big\langle\rho\,,\,\rho\big\rangle$.
	Since $\veps >\veps_g\geq 0$, by $(H3)$ there is a $K$ such that 
$\veps_g< \big\langle\rho_{(k)}\,,\,\rho_{(k)}\big\rangle<\veps$ for all $k\geq K$. 
	Then 
$\max_\rho \pzcS_{_{I/\veps}}(\rho)\geq \pzcS_{_{I/\veps}}(\rho_{(K)}) = 
\pzcR(\rho_{(K)}|\lambda)+ \pzcQ_{\,\,I/\veps}(\rho_{(K)})>-\infty$.

	Let $\qV\mapsto\rho_\veps(\qV)$ denote any maximizer for $\pzcS_{_{I/\veps}}(\rho)$.
	Suppose $\big\langle\rho_\veps\,,\,\rho_\veps\big\rangle\geq\veps$.
	Then $\pzcQ_{\,\,I/\veps}(\rho_\veps)=-\infty$, and because $\pzcR(\rho_\veps|\lambda)\leq 0$ 
then also $\pzcS_{_{I/\veps}}(\rho_\veps)=-\infty$.
	Therefore $\big\langle\rho_\veps\,,\,\rho_\veps\big\rangle <\veps$ strictly, and since $\veps>0$,
this proves \Ref{THETArewrite}.

	The standard variational argument now shows that the maximizer satisfies the Euler-Lagrange equation for 
$\pzcS_{_{I/\veps}}(\rho)$, which is \Ref{fixPOINTEQrhoUagain}.
\qed
\begin{corollary} 
\label{cor:maxAisAmax}
	The functional $\uli{\pzcS}_{I/\veps}(\varrho)$ given in \Ref{meanAdefine} achieves its supremum.
	If $\varrho_\veps$ is a maximizer of $\uli{\pzcS}_{I/\veps}(\varrho)$, then the support of its 
decomposition measure $\varsigma(\dd\rho|\varrho_\veps)$ is the set of maximizers
$\{\rho_\veps\}$ of the functional $\pzcS_{_{I/\veps}}(\rho)$ given in \Ref{ACTUALentropyONEBODY}.
\end{corollary} 
\medskip\noindent
{\textit{Proof of Corollary \ref{cor:maxAisAmax}:}} 

	Abstractly, by Lemma \ref{lem:ENTROPYupperSEMIconti} and the linearity of the mean quasi-interaction 
energy functional,
the mean interaction entropy functional $\uli{\pzcS}_{I/\veps}(\varrho)$ given in \Ref{meanAdefine} is weakly 
upper semi-continuous, and so achieves its supremum over the compact set of permutation symmetric probabilities 
$\Psp^s_{U^2_\Lambda}(\Lambda^\Nset)$.

	Alternatively, by \Ref{Arep} and two obvious estimates, we have right away that
\begin{equation}
 \pzcS_{_{I/\veps}}(\rho_\veps) 
=
 \uli{\pzcS}_{I/\veps}(\rho_\veps^\Nset) 
\leq
 \sup_\varrho \uli{\pzcS}_{I/\veps}(\varrho) 
\leq \max_\rho \pzcS_{_{I/\veps}}(\rho)
=
 \pzcS_{_{I/\veps}}(\rho_\veps),
\label{supAlessAsup}
\end{equation}
so $\sup_\varrho\uli{\pzcS}_{I/\veps}(\varrho) = \max_\varrho \uli{\pzcS}_{I/\veps}(\varrho)
=\uli{\pzcS}_{I/\veps}(\rho_\veps^\Nset)$.
	Now let $\varrho_\veps$ maximize $\uli{\pzcS}_{I/\veps}(\varrho)$ and suppose that 
$\supp\varsigma(\dd\rho|\varrho_\veps)$ is not a subset of the maximizers $\{\rho_\veps\}$ of 
$\pzcS_{_{I/\veps}}(\rho)$.
	Then 
\begin{equation}
\uli{\pzcS}_{I/\veps}(\varrho_\veps)
=
\int_{\Psp(\Lambda)} \pzcS_{_{I/\veps}}(\rho) \,\varsigma(\dd\rho|\varrho_\veps)
<
\max_\rho \pzcS_{_{I/\veps}}(\rho)
=
 \uli{\pzcS}_{I/\veps}(\rho_\veps^\Nset),
\label{supAnotAsup}
\end{equation}
so $\varrho_\veps$ is not a maximizer --- a contradiction to the supposition.
\qed

	We now relate the sequence of maximizers $\{\varrho_{{N^2\veps}}^{(N)}\}_{N\in\Nset}$ 
of $\{\pzcS_{I/\veps}^{(N)}\}$ to the set of maximizers $\{\rho_\veps\}$ of $\pzcS_{_{I/\veps}}$.
	We begin with the maxima of $\pzcS_{I/\veps}^{(N)}\big(\varrho^{(N)}\big)$ and $\pzcS_{_{I/\veps}}(\rho)$.

\begin{proposition}
\label{prop:ACTUALentropyLIM}
	We have
\begin{equation}
\lim_{N\to\infty} {\textstyle{\frac{1}{N}}}\pzcS_{I/\veps}^{(N)}\big(\varrho_{{N^2\veps}}^{(N)}\big) 
=
\pzcS_{_{I/\veps}}(\rho_\veps).
\end{equation}
\end{proposition}
\medskip\noindent
{\textit{Proof of Proposition \ref{prop:ACTUALentropyLIM}:}} 

	For all $\alpha\in(0,1)$, we have 
\begin{equation}
 {}^\alpha \pzcS_{I/\veps}^{(N)}\big( {}^\alpha\varrho_{{N^2\veps}}^{(N)}\big) 
\geq  
 {}^\alpha\pzcS_{I/\veps}^{(N)}\big(\rho_\veps^{\otimes n}\big).
\end{equation}
	We compute
\begin{equation}
{}^\alpha\pzcS_{I/\veps}^{(N)}\big(\rho_\veps^{\otimes n}\big) 
= 
N \pzcR^{(1)}\big(\rho_\veps|\lambda\big) + {}^\alpha\pzcQ_{\,\,I/\veps}^{(N)}\big(\rho_\veps^{\otimes n}\big) .
\end{equation}
	Since $\big\langle\rho_\veps\,,\,\rho_\veps\big\rangle_{}<\veps$, when $\alpha\in(0,1)$ is
sufficiently small we have by $(H4)$ and Proposition \ref{prop:ACTUALentropyFUNCofRHO} that
\begin{equation}
\lim_{N\to\infty} {\textstyle{\frac{1}{N}}} {}^\alpha\pzcQ_{\,\,I/\veps}^{(N)}\big(\rho_\veps^{\otimes n}\big) 
= 
{\textstyle{\frac{3}{2}}}\ln\bigl[1 - 
{\textstyle{\frac{1}{\veps}}} \big\langle\rho_\veps\,,\,\rho_\veps\big\rangle_{}\bigr].
\end{equation}
	Hence, for all sufficiently small $\alpha\in(0,1)$, 
\begin{equation}
\lim_{N\to\infty} {\textstyle{\frac{1}{N}}}{}^\alpha\pzcS_{I/\veps}^{(N)}\big(\rho_\veps^{\otimes n}\big) 
= 
\pzcS_{_{I/\veps}}\big(\rho_\veps\big) .
\end{equation}
	Thus
\begin{equation}
\liminf_{N\to\infty} {\textstyle{\frac{1}{N}}}{}^\alpha\pzcS_{I/\veps}^{(N)}
\big({}^\alpha\varrho_{{N^2\veps}}^{(N)}\big) 
\geq 
\pzcS_{_{I/\veps}}(\rho_\veps)
\label{liminfAofNalpha}
\end{equation}
for all sufficiently small $\alpha\in(0,1)$, and this yields the first desired estimate
\begin{equation}
\liminf_{N\to\infty} {\textstyle{\frac{1}{N}}}\pzcS_{I/\veps}^{(N)}
\big(\varrho_{{N^2\veps}}^{(N)}\big) 
\geq 
\pzcS_{_{I/\veps}}(\rho_\veps).
\label{liminfAofN}
\end{equation}

	Now consider \Ref{micLNmeasure} as extended to a probability on $\Lambda^\Nset$.
	Since $\Lambda$ is bounded, $\oli{\Lambda}$ is compact, and 
then the sequence $\{\varrho_{{N^2\veps}}^{(N)}\}_{N\in\Nset}$ is weakly compact, so 
\begin{equation}
\lim_{N\to\infty} \varrhon_{{\dot{N}}^2\veps}^{(\dot{N}[N])} = \dotvarrhon_{\veps}
 \in\Psp^s(\oli\Lambda^{n})\quad \forall n\in \Nset\,,
\label{VLASOVlimERGODEconfigMARG}
\end{equation}
after extraction of a subsequence $\{\varrho^{(\dot{N}[N])}\}_{N\in\Nset}$; note that 
the $\{\dotvarrhon_{\veps}\}_{n\in\Nset}$ form a compatible sequence of marginals.
	Furthermore, we have $\int_{\partial\Lambda}{}^1\!{\dot\varrho}_{\veps}(\dd^3{q})=0$, 
or else $\uli\pzcR(\dot\varrho_{\veps})=-\infty$, a contradiction; 
so $\dotvarrhon_{\veps} \in\Psp^s(\Lambda^{n})$.

	Following \cite{MesserSpohn,KieCPAM} we now use sub-additivity of relative entropy
(property $(C)$ in Lemma \ref{lem:ENTROPYproperties}) and then negativity of relative entropy
(property $(A)$ in Lemma \ref{lem:ENTROPYproperties}) (valid also with $\varrho_{{{\dot{N}}^2\veps}}^{(\dot{N})}$
in place of $\varrho$), and obtain
\begin{eqnarray} 
\pzcR^{(\dot{N})}\big(\varrho_{{{\dot{N}}^2\veps}}^{(\dot{N})}|\lambda\big) 
\!\!&\leq&\!\!
\left\lfloor{\textstyle{\frac{\dot{N}}{n}}}\right\rfloor
\pzcR^{(n)}\big(\varrhon_{{{\dot{N}}^2\veps}}^{(\dot{N})}|\lambda\big) 
+
\pzcR^{(m)}\big(\varrhom_{{{\dot{N}}^2\veps}}^{(\dot{N})}|\lambda\big) 
\nonumber \\
\!\!&\leq&\!\!
\left\lfloor{\textstyle{\frac{\dot{N}}{n}}}\right\rfloor
\pzcR^{(n)}\big(\varrhon_{{{\dot{N}}^2\veps}}^{(\dot{N})}|\lambda\big) 
\label{subADDestimate}
\end{eqnarray} 
where $\lfloor{a/b}\rfloor$ is the integer part of $a/b$, and where $m<n$.
	Upper semi-continuity  for the relative entropy gives 
\begin{equation}
\limsup_{N\to\infty} \pzcR^{(n)}\big(\varrhon_{{{\dot{N}}^2\veps}}^{(\dot{N}[N])}|\lambda\big) 
\leq
 \pzcR^{(n)}(\dotvarrhon_\veps|\lambda),
\label{semiCONTestimate}
\end{equation}
while $\frac{1}{\dot{N}}\left\lfloor\frac{\dot{N}}{n}\right\rfloor\to{\frac{1}{n}}$.
	Hence, dividing \Ref{subADDestimate} by $\dot{N}[N]$ and letting $N\to\infty$ gives
\begin{equation}
\limsup_{N\to\infty} 
{\textstyle{\frac{1}{\dot{N}}}}\pzcR^{(\dot{N})}\big(\varrho_{{{\dot{N}}^2\veps}}^{(\dot{N})}|\lambda\big) 
\leq
{\textstyle{\frac{1}{n}}} \pzcR^{(n)}(\dotvarrhon_\veps|\lambda) \quad \forall n\in\Nset,
\end{equation}
and now taking the supremum over $n$ (equivalently: the limit $n\to\infty$) we get
\begin{equation}
\limsup_{N\to\infty} 
{\textstyle{\frac{1}{\dot{N}}}}\pzcR^{(\dot{N})}\big(\varrho_{{\dot{N}}^2\veps}^{(\dot{N})}|\lambda\big) 
\leq
\uli\pzcR(\dot\varrho_{\veps}).
\label{limsupRofN}
\end{equation}
	Lastly, using \Ref{RaffineREP} in \Ref{limsupRofN} yields
\begin{equation}
\limsup_{N\to\infty} 
{\textstyle{\frac{1}{\dot{N}}}}\pzcR^{(\dot{N})}\big(\varrho_{{N^2\veps}}^{(\dot{N})}|\lambda\big) 
\leq
\int \pzcR(\rho|\lambda) \,\varsigma(\dd\rho|\dot\varrho_\veps)
\label{limsupRofNrep}
\end{equation}
where $\varsigma(\dd\rho|\dot\varrho_\veps)$ be the Hewitt--Savage decomposition measure for
$\dot\varrho_\veps$. 
	For each $\rho\in\supp \varsigma(\dd\rho|\dot\varrho_\veps)$
we can choose a family of $\varrho^{(\dot{N})}[\rho]\in\Psp^s(\Lambda^{\dot{N}})$ satisfying 
\begin{equation}
\lim_{N\to\infty} \varrhon^{(\dot{N})}[\rho]
=
\rho^{\otimes n}
\label{finiteNdecompoLIMIT}
\end{equation}
for each $n\in\Nset$, such that for each $\dot{N}[N]$, with $N\in\Nset$, we have
\begin{equation}
 \varrho^{(\dot{N})}_{{\dot{N}}^2\veps} = \int \varrho^{(\dot{N})}[\rho] \,\varsigma(\dd\rho|\dot\varrho_\veps).
\label{finiteNdecompo}
\end{equation}
	In contrast to the de Finetti-Dynkin-Hewitt-Savage decomposition, this finite $N$
decomposition is not unique, but this is immaterial. 
	We remark that in the physically (presumably) most important situations, namely
when $\supp \varsigma(\dd\rho|\varrho_\veps)$ is either a finite set or a continuous
group orbit of a compact group, then a decomposition \Ref{finiteNdecompo} satisfying \Ref{finiteNdecompoLIMIT}
can easily be constructed explicitly, as shown in Appendix B.
 
	By \Ref{finiteNdecompo}, the linearity of the map 
$\varrho^{(\dot{N})}\mapsto {}^\alpha\pzcQ^{(\dot{N})}_{\,\,I/\veps}\big(\varrho^{(\dot{N})}\big)$ gives
\begin{equation}
{}^\alpha\pzcQ^{(\dot{N})}_{\,\,I/\veps}\big(\varrho_{{N^2\veps}}^{(\dot{N})}\big) 
=
\int {}^\alpha\pzcQ^{(\dot{N})}_{\,\,I/\veps}(\varrho^{(\dot{N})}[\rho])
 \,\varsigma(\dd\rho|\dot\varrho_\veps),
\label{QofNrep}
\end{equation}
and by the concavity of the map 
$I\mapsto{}^\alpha\pzcQ^{(\dot{N})}_{\,\,I/\veps}\big(\varrho^{(\dot{N})}\big)$,
Jensen's inequality gives
\begin{eqnarray}
 {}^\alpha\pzcQ^{(\dot{N})}_{\,\,I/\veps}(\varrho^{(\dot{N})}[\rho])
\!\!\!\!\!\!\!\!&&
\leq 
{\textstyle{\frac{3\dot{N}-2}{2}}} \left[
\ln\bigl[1 - {\textstyle{\frac{1}{\veps}}}\cU^{(N)}(\varrho^{(\dot{N})})
\bigr]
	\chi_{_{\left\{\cU^{(N)}(\varrho^{_{(\dot{N})}}) < \veps(1-\alpha)\right\}}}
\right.
\label{upperQalphaESTIM}\\
&&\quad
+
\left.
\left(\ln \alpha +{\textstyle{\frac{1}{\alpha}}}
\bigl[1-\alpha-{\textstyle{\frac{1}{\veps}}}\cU^{(N)}(\varrho^{(\dot{N})}) 
\bigr]\right)
	\chi_{_{\left\{\cU(\varrho^{_{(\dot{N})}}) \geq \veps(1-\alpha)\right\}}}
\right],
\nonumber
\end{eqnarray}
\vskip-.2truecm
\noindent
where
\begin{equation}
\cU^{(N)}(\varrho^{(\dot{N})}) 
=
\left(1-\dot{N}^{-1}\right)\!\int\tfrhalf U_\Lambda(\check{\qV},\hat{\qV})\,{}^2\!\varrho^{(\dot{N})}[\rho](\dd^3\check{q}\dd^3\hat{q}).
\end{equation}
	The weak lower semi-continuity of $U_\Lambda$ now gives
\begin{equation}
\liminf_{N\to\infty} \int {\tfrhalf}U_\Lambda\,{}^2\!\rho^{(\dot{N})}[\rho] \dd^6q
\geq 
\big\langle\rho,\,\rho\big\rangle,
\label{liminfUfunc}
\end{equation}
and since $\dot{N}^{-1}\to 0$, we find for each convergent subsequence of measures that
\begin{eqnarray}
\limsup_{N\to\infty} 
{\textstyle{\frac{1}{\dot{N}}}}
{}^\alpha\pzcQ^{(\dot{N})}_{\,\,I/\veps}\big(\varrho^{(\dot{N})}[\rho]\big) 
\!\!\!\!\!\!\!\!&&\leq
{\textstyle{\frac{3}{2}}} \left[
\ln\bigl[1 - {\textstyle{\frac{1}{\veps}}} \big\langle\rho\,,\,\rho\big\rangle_{}\bigr]
	\chi_{_{\left\{\left\langle\rho^{}\,,\,\rho_{}\right\rangle < \veps(1-\alpha)\right\}}}
\right.
\label{limsupQofNrepALPHA}\\
&&\quad
+
\left.
\left(\ln \alpha +{\textstyle{\frac{1}{\alpha}}}
\bigl[1 -\alpha- {\textstyle{\frac{1}{\veps}}} \big\langle\rho\,,\,\rho\big\rangle_{}\bigr]\right)
	\chi_{_{\left\{\left\langle\rho^{}\,,\,\rho_{}\right\rangle \geq \veps(1-\alpha)\right\}}}
\right].
\nonumber
\end{eqnarray}
for each $\alpha\in(0,1)$.
	Now suppose that $\big\langle\rho\,,\,\rho\big\rangle_{}\geq\veps$; then 
r.h.s.\Ref{limsupQofNrepALPHA}$\downarrow-\infty$ as $\alpha\downarrow 0$, in which case by
\Ref{limsupQofNrepALPHA} and \Ref{QofNrep} also ${}^\alpha\pzcQ^{(\dot{N})}_{\,\,I/\veps}\big(\varrho_{{N^2\veps}}^{(\dot{N})}\big)\downarrow-\infty$
as $\alpha\downarrow 0$, and by \Ref{quasiCONFIGenergyREdef} and \Ref{ACTUALentropyNBODY} and
property (A) in Lemma \ref{lem:ENTROPYproperties}, and then
Proposition \ref{prop:VPfinN}, this contradicts the lower bound \Ref{lowerBOUNDc}.
 	Therefore, $\big\langle\rho\,,\,\rho\big\rangle_{}<\veps$ for every
$\rho\in\supp\varsigma(\dd\rho|\dot\varrho_\veps)$, and so, and recalling \Ref{Qdef}, we conclude that
\begin{equation}
\hskip-.35truecm
\limsup_{N\to\infty} 
{\textstyle{\frac{1}{\dot{N}}}}\pzcQ^{(\dot{N})}_{\,\,I/\veps}\big(\varrho_{{N^2\veps}}^{(\dot{N})}\big) 
\leq 
{\textstyle{\frac{3}{2}}}\int\!\ln \bigl[1 - {\textstyle{\frac{1}{\veps}}} \big\langle\rho\,,\,\rho\big\rangle_{}\bigr]
 \varsigma(\dd\rho|\dot\varrho_\veps)
=
\!\int\!\! \pzcQ_{\,\,I/\veps}(\rho) \,\varsigma(\dd\rho|\dot\varrho_\veps).
\label{limsupQofNrep}
\end{equation}

	The estimates \Ref{limsupQofNrep} and \Ref{limsupRofNrep} and two obvious estimates now give
\begin{eqnarray}
\limsup_{N\to\infty} 
{\textstyle{\frac{1}{\dot{N}}}}\pzcS^{(\dot{N})}_{I/\veps}\big(\varrho_{{{\dot{N}}^2\veps}}^{(\dot{N})}\big) 
\!\!&\leq&\!\!
\limsup_{N\to\infty} 
{\textstyle{\frac{1}{\dot{N}}}}\pzcR^{(\dot{N})}\big(\varrho_{{{\dot{N}}^2\veps}}^{(\dot{N})}|\lambda\big) 
+
\limsup_{N\to\infty} 
{\textstyle{\frac{1}{\dot{N}}}}\pzcQ^{(\dot{N})}_{\,\,I/\veps}\big(\varrho_{{{\dot{N}}^2\veps}}^{(\dot{N})}\big) 
\nonumber\\
\!\!&\leq&\!\!
\int \pzcR(\rho|\lambda) \,\varsigma(\dd\rho|\dot\varrho_\veps) +
\int \pzcQ_{\,\,I/\veps}(\rho) \,\varsigma(\dd\rho|\dot\varrho_\veps)
\nonumber\\
\!\!&=&\!\!
\int \pzcS_{I/\veps}(\rho) \,\varsigma(\dd\rho|\dot\varrho_\veps)
\nonumber\\
\!\!&\leq&\!\!
\max_\rho \pzcS_{I/\veps}(\rho),
\label{limsupAofNrep}
\end{eqnarray}
and since this holds for each limit point $\dot\varrho_\veps$ we can drop the dot to get
\begin{equation}
\limsup_{N\to\infty} {\textstyle{\frac{1}{N}}}\pzcS_{I/\veps}^{(N)}\big(\varrho_{{N^2\veps}}^{(N)}\big) 
\leq \pzcS_{_{I/\veps}}(\rho_\veps).
\label{limsupAofN}
\end{equation}

	By \Ref{liminfAofN} and \Ref{limsupAofN}, Proposition \ref{prop:ACTUALentropyLIM} is proved.
\qed

	By Propositions \ref{prop:VPfinN}, \ref{prop:ACTUALentropyFUNCofRHO}, and \ref{prop:ACTUALentropyLIM},
the interaction entropy per particle for Boltzmann's ergodic ensemble converges as follows,
\begin{equation}
s_{_\Lambda,{}_I}(\veps) 
	\equiv 
\lim_{N\to\infty} {\textstyle{\frac{1}{N}}} S_{{I}^{(N)}_\Lambda}(N^2\veps)
	= 
\pzcS_{_{I/\veps}}(\rho_\veps),
\label{limSIperN}
\end{equation}
and $s_{_\Lambda,{}_I}(\veps)$ is characterized by its own variational principle expressed in 
Proposition \ref{prop:ACTUALentropyFUNCofRHO}.
	Moreover, by  formula \Ref{MCentropyINTEGRATEDinPasympREV}, which holds under 
assumptions $(H1)$--$(H5)$, the expansion \Ref{MCentropyASYMPTOTICS} now follows, with 
\begin{equation}
s_{_\Lambda}(\veps) 
=
s_{_\Lambda,{}_K}(\veps) 
+ 
s_{_\Lambda,{}_I}(\veps) ,
\label{sKsIagain}
\end{equation}
where $s_{_\Lambda,{}_K}(\veps)$ is given in \Ref{sK}, and $s_{_\Lambda,{}_I}(\veps)$ in \Ref{limSIperN}
and Proposition \ref{prop:ACTUALentropyFUNCofRHO}.
	By Proposition \ref{prop:ACTUALentropyFUNCofRHO}, any maximizer $\rho_\veps$ of 
$\pzcS_{_{I/\veps}}(\rho)$ satisfies \Ref{fixPOINTEQrhoU} and \Ref{ThetaEasFNCTLrhoU}.

	Lastly, one readily verifies that the just described $s_{_\Lambda}(\veps)$ equals the 
negative minimum of Boltzmann's $H$ functional over the set of trial densities
$\Asp_\veps = \{f \in (\Psp\cap\Lsp^1\cap\Lsp^1\ln\Lsp^1)(\Rset^3\!\times\!\Lambda): \pzcE(f) = \veps \}$.
	This is done by explicitly carrying out the standard variational argument for $H(f)$, taking
the constraints into account with the help of Lagrange multipliers which are then eliminated with the help
of the very functionals of $\rho_\veps$ displayed in Theorem \ref{thm:maxENTROPYvp}.

	This completes the proof of Theorem \ref{thm:maxENTROPYvp}.
\qed

%%%%%%%%%%%%%%%%%%%%%%%%%%%%%%%%%%%%%%%%%%%%%%%%%%%%%%%%%%%%%%%%
%%%%%%%%%%%%%%%%%%%%%%%%%%%%%%%%%%%%%%%%%%%%%%%%%%%%%%%%%%%%%%%%
\subsection{Proof of Theorem 3} 
%%%%%%%%%%%%%%%%%%%%%%%%%%%%%%%%%%%%%%%%%%%%%%%%%%%%%%%%%%%%%%%%
%%%%%%%%%%%%%%%%%%%%%%%%%%%%%%%%%%%%%%%%%%%%%%%%%%%%%%%%%%%%%%%%
	We begin with the observation that \Ref{micLNmeasure} is clearly the $N$-th configurational marginal 
measure of \Ref{BOLTZMANNsERGODE}, i.e. \Ref{micLNmeasure} is \Ref{BOLTZMANNsERGODE} with the
Hamiltonian given by \Ref{HAMrescaled}, integrated over all the $\pV$ variables in \Ref{HAMrescaled}.
	Put differently, \Ref{micLNmeasure} is the joint $N$-point distribution on configuration space $\Lambda^N$ 
of an $N$-body system with Hamiltonian \Ref{HAMrescaled} chosen w.r.t. the a-priori measure 
\Ref{BOLTZMANNsERGODE} on $(\Rset^{3}\!\times\!\Lambda)^{N}$.
	Hence, our proof of Theorem \ref{thm:maxENTROPYvp} also proves the following
weaker version of Theorem~\ref{thm:deFinettiERGODE}.

\setcounter{theorem}{2}
\begin{theorem}
\label{thm:deFinettiERGODEconfig}$^{\!\!\!\!-}$
	Under the same assumptions as in Theorem \ref{thm:maxENTROPYvp}, consider \Ref{BOLTZMANNsERGODE}
for the Hamiltonian \Ref{HAMrescaled} as extended to a probability on $(\Rset^3\!\times\!\Lambda)^\Nset$.
	Then the sequence $\{\varrho_{{N^2\veps}}^{(N)}\}_{N\in\Nset}$ of its configuration space marginals,
obtained by integrating over all the $\pV$ variables in \Ref{HAMrescaled} and given in \Ref{micLNmeasure}, 
is weakly compact in $\Psp^s(\oli\Lambda^\Nset)$, so one can extract a subsequence 
$\{\varrho_{{{\dot{N}}^2\veps}}^{(\dot{N}[N])}\}_{N\in\Nset}$ such that
\begin{equation}
\lim_{N\to\infty} \varrho_{{{\dot{N}}^2\veps}}^{(\dot{N}[N])} = \dot\varrho_\veps \in\Psp^s(\Lambda^\Nset)\,,
\label{VLASOVlimERGODEconfig}
\end{equation}
in the sense that
\begin{equation}
\lim_{N\to\infty} \varrhon_{{{\dot{N}}^2\veps}}^{(\dot{N}[N])} 
= 	
\int_{\Psp(\Lambda)} \prod_{1\leq k\leq n}\rho(\qV_k) \dd^{3}q_k\, \varsigma(\dd\rho|\dot\varrho_\veps) 
\quad\forall n\in\Nset\,.
\label{VLASOVlimERGODEconfigMARGn}
\end{equation}
	The decomposition measure $\varsigma(\dd\rho|\dot\varrho_\veps)$ 
of each such limit point $\dot\varrho_\veps$ is supported on the subset of 
$\Psp(\Lambda)$ which consists of the probability measures ${\rho_\veps}(\qV)\dd^{3}q$
which maximize the functional $\pzcS_{_{I/\veps}}(\rho)$.
\end{theorem}

	Since each limit point $\dot\varrho_\veps$ of \Ref{micLNmeasure} is a convex linear superposition
of infinite product measures on $\Lambda^\Nset$ consisting of ``Boltzmann factors'' $\rho_\veps(\qV)$
on $\Lambda$, satisfying \Ref{fixPOINTEQrhoU} with \Ref{ThetaEasFNCTLrhoU} and maximizing the 
interaction entropy functional $\pzcS_{_{I/\veps}}(\rho)$, and since each such 
Boltzmann factor is associated with a unique ``Maxwellian'' $\sigma_\veps(\pV)$ on $\Rset^3$ through
\Ref{sigmaOFrho}, each such Boltzmann factor thereby defines a unique Maxwell--Boltzmann distribution 
$\sigma_\veps(\pV)\rho_\veps(\qV)$ on $\Rset^3\!\times\!\Lambda$ given by the product of this Boltzmann factor 
with its associated Maxwellian. 
	So the very decomposition measure $\varsigma(\dd\rho|\dot\varrho_\veps)$ of each limit point 
$\dot\varrho_\veps$ on $\Lambda^\Nset$ allows us to define a unique probability measure  $\dot\mu_\veps$ 
on $(\Rset^3\!\times\!\Lambda)^\Nset$, viz. 
\begin{equation}
	\dotmun_{\veps} (\dd^{3n}p\dd^{3n}q) 
=
	\int_{\Psp(\Lambda)} 
\prod_{1\leq k\leq n}
\sigma(\rho)(\pV_k)\rho(\qV_k) 
\dd^{3}p_k\dd^{3}q_k
\, \varsigma(\dd\rho|\dot\varrho_\veps) 
\quad
\forall n\in \Nset,
\label{deFinettiDECOMPprime}
\end{equation} 			
and this measure $\varsigma(\dd\rho|\dot\varrho_\veps)$ on ${\Psp(\Lambda)}$
can be mapped into a unique measure $\nu(\dd\tau|\dot\mu_\veps)$ on ${\Psp(\Rset^3\!\times\!\Lambda)}$
which is concentrated on those $\tau\in {\Psp(\Rset^3\!\times\!\Lambda)}$ which are of the form
$\tau(\dd^{3}p\dd^{3}q) = \sigma(\rho)(\pV)\rho(\qV)\dd^{3}p\dd^{3}q$, with $\sigma(\rho)$ given by
\Ref{sigmaOFrho} and $\rho$ satisfying \Ref{fixPOINTEQrhoU} with \Ref{ThetaEasFNCTLrhoU} 
and maximizing $\pzcS_{_{I/\veps}}(\rho)$, thus
\begin{equation}
\dotmun_{\veps} (\dd^{3n}p\dd^{3n}q) 
=
	\int_{\Psp(\Rset^3\!\times\!\Lambda)} 
\prod_{1\leq k\leq n}
 \tau(\dd^{3}p_k\dd^{3}q_k)
\, \nu(\dd\tau|\dot\mu_\veps) 
\qquad
\forall n\in \Nset.
\label{ooomph}
\end{equation}
	The corresponding infinite product measures $\prod_{1\leq k\leq \infty} \tau(\dd^{3}p_k\dd^{3}q_k)$
on $(\Rset^3\!\times\!\Lambda)^\Nset$ are extreme points of
$\Psp^s((\Rset^3\!\times\!\Lambda)^\Nset)$, and so \Ref{ooomph} is the extremal representation of $\dot\mu_\veps$.
	So, having Theorem \ref{thm:deFinettiERGODEconfig}$^{-}$ and its consequence \Ref{ooomph},
all we need to do to finish the proof of Theorem \ref{thm:deFinettiERGODE} is to show that each such 
defined $\dot\mu_\veps$ is indeed a limit point of \Ref{BOLTZMANNsERGODE} under the stated hypotheses. 	

	To see this, we use that by Theorem \ref{thm:deFinettiERGODEconfig}$^{-}$ we already know that
\Ref{VLASOVlimERGODEconfigMARGn} holds, and we know the support of $\varsigma(\dd\rho|\dot\varrho_\veps)$.
	Writing $\varrhon_{{{\dot{N}}^2\veps}}^{(\dot{N}[N])}$ explicitly gives
\begin{equation}
\varrhon_{{{\dot{N}}^2\veps}}^{(\dot{N})}(\dd^{3n}\!q)
=
\frac{
      \displaystyle{\int}\!
\left(1-{\textstyle{\frac{1}{\veps {\dot{N}}^2}}}
I_\Lambda^{(\dot{N})}(\qV_1,...,\qV_{\dot{N}})
\right)_+{\!\!\!\!}^{\frac{3\dot{N}}{2}-1}
\dd^{3(\dot{N}-n)}\!q}
     {\displaystyle{\int}\!
\left(1-
{\textstyle{\frac{1}{\veps {\dot{N}}^2}}}I_\Lambda^{(\dot{N})}(\tilde\qV_1,...,\tilde\qV_{\dot{N}})
\right)_+{\!\!\!\!}^{\frac{3\dot{N}}{2}-1} 
\dd^{3\dot{N}}\!\tilde{q}} 
\dd^{3n}\!q,
\label{micLNmeasureMARGn}
\end{equation}
where the integral in the numerator runs over the variables $\qV_{n+1}$ to $\qV_{\dot{N}}$.
	For any $\dot{N}$ and $1\leq n<\dot{N}$ we now write 
\begin{equation}
\hskip-.5truecm
I_\Lambda^{({\dot{N}})}(\qV_1,...,\qV_{{\dot{N}}})
\!=\!
I_\Lambda^{({n})}(\qV_1,...,\qV_{{n}})
\!+\!
I_\Lambda^{({n|\dot{N}})}(\qV_1,...,\qV_{{\dot{N}}})
\!+\!
I_\Lambda^{({\dot{N}-n})}(\qV_{n+1},...,\qV_{{\dot{N}}})
\label{Isplitting}
\end{equation}
which defines $I_\Lambda^{({n|\dot{N}})}(\qV_1,...,\qV_{{\dot{N}}})$.
	Henceforth we omit the arguments from the $I$s to keep the formulas within sight; 
by \Ref{Isplitting} the superscripts convey which variables are used.
	With the help of \Ref{Isplitting} we rewrite the integrands thusly,\footnote{We are using
		that $(fg)_+ = f_+g_+ + f_-g_-$ for two arbitrary functions $f$ and $g$, and that in
		our case $f$ cannot be strictly negative if $g$ is, giving $(fg)_+ = f_+g_+$; in our case, 
		$f$ and $g$ are the respective expressions between the two pairs
		of big parentheses at r.h.s.\Ref{plusISplusplus}.}
\begin{equation}
\left(1-{\textstyle{\frac{1}{\veps {\dot{N}}^2}}}I_\Lambda^{(\dot{N})}\right)_+
= 
\left(1- \frac{I_\Lambda^{(n)}+I_\Lambda^{(n|\dot{N})}}
		{{\dot{N}}^{2}\big(\veps-{\dot{N}}^{-2} I_\Lambda^{(\dot{N}-n)}\big)}\right)_{\!\!+}\!\!
\left(1-{\textstyle{\frac{1}{\veps {\dot{N}}^2}}}I_\Lambda^{(\dot{N}-n)}\right)_+.
\label{plusISplusplus}
\end{equation}
	Now $\bigl(1-(\veps {\dot{N}}^2)^{-1}I_\Lambda^{(\dot{N})}\bigr)_+$ vanishes in an open
neighborhood of configurations 
\vskip-.1truecm
\noindent
$(\qV_1,...,\qV_n)_\infty$ for which $I_\Lambda^{(\dot{N})}= \infty$, so that
$I_\Lambda^{(n)}< \infty$ on the support of \Ref{plusISplusplus}.
	And for any configuration $(\qV_1,...,\qV_n)$ for which $I_\Lambda^{(n)}< \infty$, we have 
${\dot{N}}^{-2}I_\Lambda^{(n)}\to 0$ as $\dot{N}[N]\to\infty$. 
	Moreover, by our Theorem \ref{thm:deFinettiERGODEconfig}$^{-}$ and its explication
\Ref{VLASOVlimERGODEconfigMARGn}, we have that 
$\left(1-{\textstyle{\frac{1}{\veps {\dot{N}}^2}}}I_\Lambda^{(\dot{N}-n)}
 \right)_+{\!\!\!\!}^{\frac{3\dot{N}}{2}-1}\dd^{3(\dot{N}-n)}\!q$, 
interpreted as a measure on the convex set of probability measures $\Psp(\Lambda)$ with support in the 
set of empirical one-point ``densities'' with $\dot{N}-n$ atoms,\footnote{If $U_\Lambda$ is bounded 
		continuous on $\Lambda^2$, then we already know that we can rewrite $I_\Lambda^{(N)}$ as 
		a sum of a bilinear and a linear form on $\Psp(\Lambda)$ evaluated at a normalized empirical 
		one-point ``density'' with $N$ atoms; see \Ref{IasBform}.
		If $U_\Lambda$ is only lower semi-continuous this particular identification ceases to make 
		sense, but happily we can \emph{always} interpret $I_\Lambda^{(N)}$ as a linear form on the 
		convex set of probability measures $\Psp(\Lambda^{2})$ (cf. \Ref{upperQalphaESTIM} and
		 \Ref{liminfUfunc}), evaluated at a normalized
		empirical two-point ``density'' with $N$ atoms \Ref{normalEMPmeasTWO}, and we note that
		any empirical two-point ``density'' with $N$ atoms \Ref{normalEMPmeasTWO} is uniquely 
		determined by its associated empirical one-point ``density'' with $N$ atoms 
		\Ref{normalEMPmeasONE}.}
converges (up to normalization) to 
$\varsigma(\dd\rho|\dot\varrho_\veps)$, and for any $\rho$ in the support
of $\varsigma(\dd\rho|\dot\varrho_\veps)$ we have that
${\dot{N}}^{-2}I_\Lambda^{(\dot{N}-n)}\to\langle\rho,\rho\rangle$ while
${\dot{N}}^{-1}I_\Lambda^{(n|\dot{N})}
\to \sum_{1\leq k\leq n} \int_\Lambda U_\Lambda({\qV_k},\tilde{\qV})\rho(\tilde{\qV})\dd^3{q}$
when $\dot{N}[N]\to\infty$.
	So for any $\rho$ in the support of the decomposition measure
$\varsigma(\dd\rho|\dot\varrho_\veps)$ we have 
\begin{equation}
\hskip-.5truecm
\left(\!\!1-\frac{I_\Lambda^{(\dot{n})}+I_\Lambda^{(n|\dot{N})}}
	{{\dot{N}}^{2}\big(\veps-{\dot{N}}^{-2}I_\Lambda^{(\dot{N}-n)}\big)}\right)_+^{\!\!\!\!\frac{3\dot{N}}{2}-1}
\!\!\!\!\!\!\to
\prod_{1\leq k\leq n} \!\!
\exp\!\left(\!-\frac{1}{\vartheta_\veps(\rho)}\!\int_\Lambda U_\Lambda(\qV_k,\tilde{\qV})\rho(\tilde{\qV})
\dd^3\tilde{q}\!
\right)
\label{powTOexp}
\end{equation}
with $\frac{3}{2}\vartheta_\veps(\rho)= \veps - \langle\rho,\rho\rangle$.
	After this preparation, we now explicitly compute the marginal $\mun^{(\dot{N})}_{{\dot{N}}^2\veps}$ and find
\begin{equation}
\mun_{{{\dot{N}}^2\veps}}^{(\dot{N})}(\dd^{3n}\!p\dd^{3n}\!q)
=
\frac{\displaystyle{\int}\!
\left(
	1-{\textstyle{\frac{1}{\veps {\dot{N}}^2}}}\left(K^{(n|N)} +I_\Lambda^{(\dot{N})}\right)
\right)_+{\!\!\!\!}^{\frac{3(\dot{N}-n)}{2}-1}\dd^{3(\dot{N}-n)}\!q}
     {\displaystyle{\int}\!
\left(1-
	{\textstyle{\frac{1}{\veps {\dot{N}}^2}}}\left(K^{(n|N)}+I_\Lambda^{(\dot{N})}\right)
\right)_+{\!\!\!\!}^{\frac{3(\dot{N}-n)}{2}-1} \dd^{3\dot{N}}\!\tilde{q}\,\dd^{3n}\!\tilde{p}} 
\dd^{3n}\!p\,\dd^{3n}\!q,
\label{micNmeasureMARGn}
\end{equation}	%(\qV_1,...,\qV_{\dot{N}}) (\tilde\qV_1,...,\tilde\qV_{\dot{N}})
where $K^{(n|N)}(\pV_1,...,\pV_{n})  = N\sum_{1\leq k \leq n} \frac{1}{2}|\pV_k|^2$.
	Using \Ref{Isplitting} we factor the integrands as in \Ref{plusISplusplus}, though
now we get 
\begin{equation}
\left(\!1-\frac{K^{(n|N)} + I_\Lambda^{(\dot{N})}}{\veps {\dot{N}}^2}\!\right)_{\!\!+}
\!\!=\!
\left(\!\!1- \frac{K^{(n|N)} + I_\Lambda^{(n)}+I_\Lambda^{(n|\dot{N})}}
		{{\dot{N}}^{2}\big(\veps-{\dot{N}}^{-2} I_\Lambda^{(\dot{N}-n)}\big)}\right)_{\!\!+}\!\!\!
\left(1-{\textstyle{\frac{1}{\veps {\dot{N}}^2}}}I_\Lambda^{(\dot{N}-n)}\right)_+
\label{plusISplusplusAGAIN}
\end{equation}
and by following essentially verbatim the arguments which lead from \Ref{plusISplusplus} to \Ref{powTOexp},
we now find that for any $\rho\in\supp\varsigma(\dd\rho|\dot\varrho_\veps)$,
\begin{equation}
\hskip-.5truecm
\left(\!\!1-\frac{K^{(n|N)} +I_\Lambda^{(\dot{n})}+I_\Lambda^{(n|\dot{N})}}
    {{\dot{N}}^{2}\big(\veps-{\dot{N}}^{-2}I_\Lambda^{(\dot{N}-n)}\big)}\right)_+^{\!\!\!\!\frac{3(\dot{N}-n)}{2}-1}
\!\!\!\!\!\!\!\!\!\to
\prod_{1\leq k\leq n}\!\! \exp\!\left(\!-\frac{\frac{1}{2}|\pV_k|^2 +
\int_\Lambda U_\Lambda(\qV_k,\tilde{\qV})\rho(\tilde{\qV})\dd^3\tilde{q}}
{\vartheta_\veps(\rho)}\!\right)
\label{powTOexpFINAL}
\end{equation}

	Our Theorem~\ref{thm:deFinettiERGODE} is proved.
\qed
%%%%%%%%%%%%%%%%%%%%%%%%%%%%%%%%%%%%%%%%%%%%%%%%%%%%%%%%%%%%%%%%%%%%%%%%%%%%%%%%%%%%%%%%%
%%%%%%%%%%%%%%%%%%%%%%%%%%%%%%%%%%%%%%%%%%%%%%%%%%%%%%%%%%%%%%%%%%%%%%%%%%%%%%%%%%%%%%%%%
%%%%%%%%%%%%%%%%%%%%%%%%%%%%%%%%%%%%%%%%%%%%%%%%%%%%%%%%%%%%%%%%%%%%%%%%%%%%%%%%%%%%%%%%%
\section{Spin-offs of our results} 
%%%%%%%%%%%%%%%%%%%%%%%%%%%%%%%%%%%%%%%%%%%%%%%%%%%%%%%%%%%%%%%%%%%%%%%%%%%%%%%%%%%%%%%%%
%%%%%%%%%%%%%%%%%%%%%%%%%%%%%%%%%%%%%%%%%%%%%%%%%%%%%%%%%%%%%%%%%%%%%%%%%%%%%%%%%%%%%%%%%
%%%%%%%%%%%%%%%%%%%%%%%%%%%%%%%%%%%%%%%%%%%%%%%%%%%%%%%%%%%%%%%%%%%%%%%%%%%%%%%%%%%%%%%%%
	In this section we list a number of corollaries of our results. 
%%%%%%%%%%%%%%%%%%%%%%%%%%%%%%%%%%%%%%%%%%%%%%%%%%%%%%%%%%%%%%%%%%%%%%%%%%%%%%%%%%%%%%%%%
%%%%%%%%%%%%%%%%%%%%%%%%%%%%%%%%%%%%%%%%%%%%%%%%%%%%%%%%%%%%%%%%%%%%%%%%%%%%%%%%%%%%%%%%%
\subsection{A weak law of large numbers / ergodic theorem} 
%%%%%%%%%%%%%%%%%%%%%%%%%%%%%%%%%%%%%%%%%%%%%%%%%%%%%%%%%%%%%%%%%%%%%%%%%%%%%%%%%%%%%%%%%
%%%%%%%%%%%%%%%%%%%%%%%%%%%%%%%%%%%%%%%%%%%%%%%%%%%%%%%%%%%%%%%%%%%%%%%%%%%%%%%%%%%%%%%%%
	Whenever $\pzcHB(f)$ has a unique minimizer $f_\veps$ over $\Asp_\veps$,
then necessarily all limit points in \Ref{VLASOVlimptERGODE} coincide, i.e. any $\dot\mu_\veps=\mu_\veps$.
	By the weak compactness of $\Psp^s(\oli\Lambda^\Nset)$ (in product topology) we then in fact do have 
weak convergence,
\begin{equation}
\lim_{N\to\infty} \mun^{(N)}_{N^2\veps} (\dd^{3n}p\dd^{3n}q)
= 
\mun_\veps(\dd^{3n}p\dd^{3n}q) \in\Psp^s((\Rset^3\!\times\!\Lambda)^n)\quad \forall n\in \Nset\,.
\label{VLASOVlimERGODE}
\end{equation}
	Since in this case the decomposition measure $\nu(d\tau|\mu_\veps)$ is a singleton, 
the limit $\mu_\veps= \{\mun_\veps\}_{n\in\Nset}$ is of the form
\begin{equation}
\mun_\veps (\dd^{3n}p\dd^{3n}q) 
=
\prod_{1\leq k\leq n}f_\veps(\pV_k,\qV_k)\dd^{3}p_k\dd^{3}q_k 
\label{fac}
\end{equation}
with $f_\veps(\pV,\qV)=\sigma_\veps(\pV)\rho_\veps(\qV)$ as defined in Theorem \ref{thm:maxENTROPYvp}.
	As discussed in \cite{SpohnBOOK}, the factorization property \Ref{fac} is equivalent to a weak law 
of large numbers --- or to an ergodic theorem, depending on ones point of view.
	Since the single particle momentum $\bP$ and position $\bQ$ of an individual $N$-body system picked from 
Boltzmann's Ergode \Ref{BOLTZMANNsERGODE}, with Hamiltonian \Ref{HAMrescaled},
are random variables, any bounded continuous single-particle
test function $\theta$ on $\Rset^3\!\times\!\Lambda$ defines a new random variable 
$\bTheta=\theta(\bP,\bQ)$, and so does its sample mean over a single $N$-body system,
\begin{equation}
\bigl\langle \bTheta \bigr\rangle_N 	
\equiv 
\frac{1}{N} \sum_{j=1}^{N}\theta(\bP_{j},\bQ_{j})\, .
\label{EMPaverage}
\end{equation}
	Theorem \ref{thm:deFinettiERGODE} in the special case \Ref{fac} implies that, for all such $\theta$,
\begin{equation}
\lim _{ N \to \infty}\bigl\langle \bTheta \bigr\rangle_N 	
= 
\int_{\Rset^3\!\times\!\Lambda} \theta(\pV,\qV) f_\veps(\pV,\qV)\dd^{3}p\dd^{3}q\, ,
\label{SPOHNlln}
\end{equation}
in probability.
	The generalization to $n$-body test functions holds as well.
%%%%%%%%%%%%%%%%%%%%%%%%%%%%%%%%%%%%%%%%%%%%%%%%%%%%%%%%%%%%%%%%%%%%%%%%%%%%%%%%%%%%%%%%%
%%%%%%%%%%%%%%%%%%%%%%%%%%%%%%%%%%%%%%%%%%%%%%%%%%%%%%%%%%%%%%%%%%%%%%%%%%%%%%%%%%%%%%%%%
\subsection{\!The\! Vlasov\! limit\! for\! other\! thermodynamic\! potentials} 
%%%%%%%%%%%%%%%%%%%%%%%%%%%%%%%%%%%%%%%%%%%%%%%%%%%%%%%%%%%%%%%%%%%%%%%%%%%%%%%%%%%%%%%%%
%%%%%%%%%%%%%%%%%%%%%%%%%%%%%%%%%%%%%%%%%%%%%%%%%%%%%%%%%%%%%%%%%%%%%%%%%%%%%%%%%%%%%%%%%
	A second corollary, or actually a whole family of corollaries, is the existence of the Vlasov limit
for the thermodynamic potentials of the canonical and grandcanonical ensembles under the same hypotheses. 
	We only discuss the Vlasov limit for the thermodynamic potential of the canonical ensemble.
 
	Thus, taking the Laplace transform of \Ref{STRUKTURfunktion}, i.e. multiplying
by $e^{-\beta \cE}$ and integrating over $\cE$, yields what is known
as the \emph{canonical partition function},
\begin{equation}
Z_{H^{(N)}_\Lambda}(\beta) 
= 
{\textstyle{\frac{1}{N!}}}
\int \exp\left(-\beta H^{(N)}_\Lambda(X^{(N)})\right)\dd^{6N}\!X.
\label{PARTITIONfunktion}
\end{equation}
	The Hamiltonian $H^{(N)}_\Lambda(X^{(N)})$ is given in \Ref{HAMrescaled}.
	Clearly \Ref{PARTITIONfunktion} factors as follows,
\begin{equation}
Z_{H^{(N)}_\Lambda}(\beta) 
= 
Z_{K^{(N)}}(\beta) Z_{I^{(N)}_\Lambda}(\beta) 
\label{ZisZKtimesZI}
\end{equation}
where
\begin{equation}
Z_{I^{(N)}_\Lambda}(\beta) 
= 
\!\int\!\!\exp\left(-\beta I_\Lambda^{(N)}(\qV_1,...,\qV_N)\right)\!\lambda(\dd^{3N}\!q)
\label{CANONconfigINT}
\end{equation}
is the \emph{canonical configurational integral}, with $\lambda(\dd^3{q}) = |\Lambda|^{-1}\dd^3q$ 
the normalized Lebesgue measure introduced in section \ref{sec:proofs}, and $\lambda(\dd^{3N}\!q)$ its 
$N$-fold product, and
\begin{equation}
Z_{K^{(N)}}(\beta) 
= 
{\textstyle{\frac{|\Lambda|^N}{N!}}}\int \exp\left(-\beta K^{(N)}(\pV_1,...,\pV_N)\right)\dd^{3N}\!p
\label{PARTITIONfunktionPG}
\end{equation}
is the canonical partition function of a spatially uniform perfect gas in $\Lambda$, a Gaussian on the Cartesian
product of the $\pV$ spaces, which evaluates to 
\begin{equation}
Z_{K^{(N)}}(\beta) 
= 
{\textstyle{\frac{|\Lambda|^N}{N!}}}
\big(2\pi \vartheta\big)^{3N/2};
\label{PARTITIONfunktionPGevaluated}
\end{equation}
here, we introduced $N\vartheta= \beta^{-1}$, with $\vartheta$ independent of $N$, not to be confused with
$\vartheta_\veps$ which is a functional of $\rho$.
	Since $\beta^{-1}$ receives the meaning of a temperature of a heat bath 
(up to the absorbed factor $\kB$), it needs to grow $\propto N$ to compensate for the growth of
the system's energy $\cE\propto N^2$. 
	Taking the logarithm of \Ref{PARTITIONfunktion} gives what we call
the \emph{canonical thermodynamic potential} (canonical $T$-potential, for short)\footnote{Multiplying 
		the canonical $T$-potential by the temperature of the heat bath yields the negative of 
		what is usually called the canonical free energy, which in the thermodynamic limit yields 
		the Helmholtz free energy of the physical systems.}
$\Phi_{H^{(N)}_\Lambda}(\beta)$.
	Using \Ref{ZisZKtimesZI} and \Ref{PARTITIONfunktionPGevaluated} 
as well as $\beta = \frNT$ yields the asymptotic expansion
\begin{eqnarray}
\Phi_{H^{(N)}_\Lambda}(\frNT) 
\!\!\! &=&\!\!\! 
- N\ln N + N \ln \left(e|\Lambda|(2\pi \vartheta)^{3/2}\right) + O(\ln N)
\nonumber\\
&&\!\!\! + \ln Z_{I^{(N)}_\Lambda}(\frNT). 
\label{FREEenergyASYMP}
\end{eqnarray}
	Again, the $N\ln N$ term is due to Gibbs' $N!$ and purely combinatorial in origin.
	In the absence of interactions (save the confinement to $\Lambda$)
\Ref{FREEenergyASYMP} reduces to
\begin{equation}
\Phi_{K^{(N)}}(\frNT) 
=
- N\ln N + N \ln \left(e|\Lambda|(2\pi \vartheta)^{3/2}\right) + O(\ln N),
\label{FREEenergyASYMPpg}
\end{equation}
the asymptotic expansion of the canonical $T$-potential of the spatially uniform perfect gas.
	The coefficient of the $O(N)$ term in \Ref{FREEenergyASYMPpg} is the system-specific
Helmholtz $T$-potential per particle of the uniform perfect gas in $\Lambda$, denoted by
\begin{equation}
\phi_{_\Lambda,{}_K}(\vartheta) = \ln \left(e|\Lambda|(2\pi \vartheta)^{3/2}\right).
\label{freeENERGYperPARTICLEpg}
\end{equation}
	The system-specific interaction Helmholtz $T$-potential per particle is defined by
\begin{equation}
\phi_{_\Lambda,{}_I}(\vartheta) = \lim_{N\to\infty} \frN \ln Z_{I^{(N)}_\Lambda}(\frNT).
\label{freeENERGYperPARTICLEinter}
\end{equation}
	The limit \Ref {freeENERGYperPARTICLEinter} exists for Hamiltonians satisfying $(H1)$--$(H5)$, as 
follows by corollary from Theorem \ref{thm:maxENTROPYvp}; if $(H2)$ is replaced by bounded continuity of 
the interaction, as explained earlier, then we can also infer the existence of the limit 
\Ref{freeENERGYperPARTICLEinter} from our Theorem \ref{thm:ENTROPYperPARTICLElimit}.
	The argument is quite standard, cf. \cite{RuelleBOOK}. 
	Namely, note that  
\begin{equation}
\frN \ln Z_{H^{(N)}_\Lambda}(\beta) 
= 
\frN \ln \int e^{-\beta\cE + S(\cE)}\dd\cE
\label{ZasLAPtransf}
\end{equation}
where $S(\cE)$ is shorthand for $S_{H^{(N)}_\Lambda}(\cE)$.
	Setting $\cE=N^2\veps$ and $\beta = \frNT$ and expanding $S(\cE)$ using \Ref{MCentropyASYMPTOTICS}
(if $U_\Lambda$ is bounded continuous on $\Lambda^2$ we can alternately use \Ref{MCentropyASYMPexpansionI}),
we find
\begin{equation}
\frN \ln Z_{H^{(N)}_\Lambda}(\frNT) 
= 
-\ln N + \ln \left[\int e^{N\left(-\vartheta^{-1}\veps + s_\Lambda(\veps)\right) + o(N)}\dd\veps\right]{}^{\!\!\tfrN}
+
O\big({\textstyle{\frac{\ln N}{N}}}\big).
\label{ZasLAPtransfEXPANDED}
\end{equation}
	Clearly, $\norm{g}_N\to\norm{g}_\infty$ as $N\to\infty$, and so the following asymptotic expansion
for the canonical $T$-potential results,
\begin{equation}
\Phi_{H^{(N)}_\Lambda}(\frNT) 
=
-N\ln N + N \phi_{_\Lambda}(\vartheta)  + o(N)
\label{freeENERGYasymptotics}
\end{equation}
with 
\begin{equation}
\phi_{_\Lambda}(\vartheta) 
\equiv 
\sup_{\veps>\veps_g}\big(-\vartheta^{-1}\veps + s_{_\Lambda}(\veps)\big).
\label{LEGENDREtransformEtoT}
\end{equation}
	By \Ref{FREEenergyASYMP} and \Ref{freeENERGYasymptotics}, we also have 
$N^{-1} \ln Z_{I^{(N)}_\Lambda}(\frNT)\stackrel{N\to\infty}{\longrightarrow} \phi_{_\Lambda,{}_I}(\vartheta)$, with 
\begin{equation}
\phi_{_\Lambda,{}_I} (\vartheta) = \phi_{_\Lambda}(\vartheta) - \phi_{_\Lambda,{}_K}(\vartheta) .
\label{fIisfminfK}
\end{equation}
	This concludes our demonstration that the Vlasov limit for the system-specific Helmholtz $T$-potential 
per particle follows from our theorems about the Vlasov limit of the system-specific Boltzmann entropy per particle.

	Next we notice that also the familiar ``minimum free energy principle'' for 
$-\phi_{_\Lambda}(\vartheta)$ follows from combining the Legendre--Fenchel transform \Ref{LEGENDREtransformEtoT}
with our ``maximum entropy principle'' in Theorem \ref{thm:maxENTROPYvp}. 
	Thus, for the system-specific Helmholtz $T$-potential per particle we find the variational principle
\begin{equation}
- \vartheta\phi_{_\Lambda}(\vartheta) = \inf_{f\in\Asp} \pzcF_\vartheta(f),
\label{specTPOTisMINUSenergyMINUsHfunc}
\end{equation}
with $\Asp = \{f \in (\Psp_{U_\Lambda}\cap\Lsp^1\cap\Lsp^1\ln\Lsp^1)(\Rset^3\!\times\!\Lambda)\}$
the admissible trial densities, and 
\begin{equation}
\pzcF_\vartheta(f) = \pzcE(f)+\vartheta \pzcHB(f),
\label{FREEenergyFUNC}
\end{equation}
the \emph{Helmholtz free energy functional of} $f$, where
$\pzcHB(f)$ is Boltzmann's $H$ function of $f$, given in \Ref{HfuncOFf}, and
$\pzcE(f)$ is the energy functional given in \Ref{eOFfU}.
	It also follows directly from our results that $\pzcF_\vartheta(f)$ takes its infimum over the set $\Asp$, 
and that any minimizer $f_\vartheta$ of $\pzcF_\vartheta(f)$ over $\Asp$ is of the form
\begin{equation}
{f_\vartheta}(\pV,\qV) 
= 
\sigma_\vartheta(\pV)\rho_\vartheta(\qV),
\label{fMaxwellBoltzmannCAN}
\end{equation}
where 
\begin{equation}
\sigma_\vartheta(\pV)
=
\left({\textstyle{2\pi\vartheta}}\right)^{-\frac{3}{2}}
\exp\bigl(-\vartheta^{-1}{\textstyle{\frac{1}{2}}} \left|\pV\right|^2\bigr),
\label{sigmaVARTHETA}
\end{equation}	
while $\rho_\vartheta(\qV)$ now solves the following fixed point equation on $\qV$ space,
\begin{equation}
\rho_\vartheta(\qV)
=
\frac{
		\exp\left(
-\frac{1}{\vartheta}\int_\Lambda U_\Lambda(\qV,\tilde{\qV})\rho_\vartheta(\tilde{\qV})\dd^3\tilde{q} 
			\right)}
   {\int_\Lambda\exp\left(
-\frac{1}{\vartheta}\int_\Lambda U_\Lambda(\hat{\qV},\tilde{\qV})\rho_\vartheta(\tilde{\qV})\dd^3\tilde{q}
			\right)
\dd{\hat{q}}}
\label{fixPOINTEQrhoUcan}
\end{equation}
with $\vartheta>0$ prescribed.

	We remark that the various \emph{possible} relationships between the set of maximizers of the 
maximum entropy variational principle and the set of minimizers of the minimum free energy variational 
principle have been discussed in great detail in \cite{EllisETalA,EllisETalB}.
	Note that this can be (and was) done without proving that the maximum entropy variational principle
characterizes the limit points of Boltzmann's Ergode  \Ref{BOLTZMANNsERGODE} proper.

	We also remark that the existence of the system-specific Helmholtz $T$-potential per particle in
the Vlasov limit for the canonical ensemble was shown previously by various techniques.
	Sub-additivity arguments, such as those used to prove Theorem \ref{thm:ENTROPYperPARTICLElimit},
are used in \cite{KieCPAM}. 
	The very strategy which we applied to prove Theorems \ref{thm:maxENTROPYvp} and \ref{thm:deFinettiERGODE}, 
which not only yields the variational principle for the system-specific Boltzmann entropy but also identifies the 
limit points of the sequence of ergodic ensemble measures as convex linear superpositions of infinite products of 
the optimizers for this maximum entropy principle, was originally applied in \cite{MesserSpohn} to the canonical 
ensemble for Lipschitz continuous interactions $I_\Lambda^{(N)}$; 
subsequently in \cite{KieCPAM} and in \cite{clmpCMPcan} this approach to the canonical ensemble was 
generalized to less regular interactions including the ones studied here; and in \cite{KieSpoCMP} the 
limit $N\to\infty$ of $N^{-1} \ln Z_{I^{(N)}_\Lambda}(1/\vartheta)$ was obtained by adapting this strategy 
(note the different $N$ scaling of $\beta$). 
	We emphasize that none of these canonical results implies the existence of the Vlasov limit for
the system-specific Boltzmann entropy per particle, nor captures the limit points of the ergodic ensemble measures,
unless it is a priori known that the ensembles are (convexly) equivalent, i.e. unless it is known that 
$\veps\mapsto s_{_\Lambda}(\veps)$ is concave (more on that in section 7).
	Our results, by contrast, hold irrespective of whether $\veps\mapsto s_{_\Lambda}(\veps)$ is concave or not.
%	
%%%%%%%%%%%%%%%%%%%%%%%%%%%%%%%%%%%%%%%%%%%%%%%%%%%%%%%%%%%%%%%%%%%%%%%%%%%%%%%%%%%%%%%%%
%%%%%%%%%%%%%%%%%%%%%%%%%%%%%%%%%%%%%%%%%%%%%%%%%%%%%%%%%%%%%%%%%%%%%%%%%%%%%%%%%%%%%%%%%
\subsection{The Vlasov limit for subergodic ensembles} 
%%%%%%%%%%%%%%%%%%%%%%%%%%%%%%%%%%%%%%%%%%%%%%%%%%%%%%%%%%%%%%%%%%%%%%%%%%%%%%%%%%%%%%%%%
%%%%%%%%%%%%%%%%%%%%%%%%%%%%%%%%%%%%%%%%%%%%%%%%%%%%%%%%%%%%%%%%%%%%%%%%%%%%%%%%%%%%%%%%%
	Another spin-off, or in this case rather a variation on the theme of our microcanonical results 
is the straightforward generalization of our Theorems to subensembles whose invariant measures
are concentrated on sub-manifolds of $\{H=\cE\}$ determined by further isolating integrals of 
the Hamiltonian \Ref{HAMrescaled}, such as angular momentum if the domain $\Lambda$ is rotationally 
symmetric, or the Lynden-Bells' invariant \cite{LBLBa,LBLBb} which occurs in a generalization of the 
Calogero--Moser model to particles moving in $\Rset^3$ confined by a quadratic potential.
	Hypothesis $(H4)$ does not hold for these interactions, but can be replaced by a weaker one
at the expense of some extra work.
	In those cases the entropy maximizer factors into a product of a \emph{locally (at $\qV$) shifted Maxwellian}
on $\pV$ space and a purely space-dependent \emph{Boltzmann factor}.
	The shifted Maxwellian which generalizes \Ref{fMaxwellBoltzmannMC} to include
angular momentum is known as a ``rotating Maxwellian;'' in the case of the Lynden-Bells' 
Hamiltonian one finds a ``rotating-dilating Maxwellian.'' 
	An announcement of these results was made in \cite{KieASSISI}; details will appear
in \cite{KieLanINprep}.

%%%%%%%%%%%%%%%%%%%%%%%%%%%%%%%%%%%%%%%%%%%%%%%%%%%%%%%%%%%%%%%%%%%%%%%%%%%%
%%%%%%%%%%%%%%%%%%%%%%%%%%%%%%%%%%%%%%%%%%%%%%%%%%%%%%%%%%%%%%%%%%%%%%%%%%%%
%%%%%%%%%%%%%%%%%%%%%%%%%%%%%%%%%%%%%%%%%%%%%%%%%%%%%%%%%%%%%%%%%%%%%%%%%%%%
\section{Unfinished business} 
%%%%%%%%%%%%%%%%%%%%%%%%%%%%%%%%%%%%%%%%%%%%%%%%%%%%%%%%%%%%%%%%%%%%%%%%%%%%
%%%%%%%%%%%%%%%%%%%%%%%%%%%%%%%%%%%%%%%%%%%%%%%%%%%%%%%%%%%%%%%%%%%%%%%%%%%%
%%%%%%%%%%%%%%%%%%%%%%%%%%%%%%%%%%%%%%%%%%%%%%%%%%%%%%%%%%%%%%%%%%%%%%%%%%%%
	In this last section of our paper we point out some open problems related to the ones treated here. 
%%%%%%%%%%%%%%%%%%%%%%%%%%%%%%%%%%%%%%%%%%%%%%%%%%%%%%%%%%%%%%%%%%%%%%%%%%%%
%%%%%%%%%%%%%%%%%%%%%%%%%%%%%%%%%%%%%%%%%%%%%%%%%%%%%%%%%%%%%%%%%%%%%%%%%%%%
\subsection{The maximum interaction entropy principle}
%%%%%%%%%%%%%%%%%%%%%%%%%%%%%%%%%%%%%%%%%%%%%%%%%%%%%%%%%%%%%%%%%%%%%%%%%%%%
%%%%%%%%%%%%%%%%%%%%%%%%%%%%%%%%%%%%%%%%%%%%%%%%%%%%%%%%%%%%%%%%%%%%%%%%%%%%
	To the best of the author's knowledge, the maximum interaction entropy principle formulated in
Proposition \ref{prop:ACTUALentropyFUNCofRHO} is new. 
	As made clear in Theorem  \ref{thm:maxENTROPYvp} it offers a way to directly evaluate the usual variational
principle of maximum entropy with energy constraint.
	By contrast, the standard approach to evaluate this constrained maximum entropy principle has 
been rather indirect.
	Namely, a Lagrange parameter (basically $\vartheta$) is introduced for the energy constraint,
yielding the corresponding fix point equation \Ref{fixPOINTEQrhoUcan} for the stationary points of the
free energy functional.
	After finding all solution families (not just the minimizers of the free energy functional), 
a parameter representation of energy and entropy along the various solution families of \Ref{fixPOINTEQrhoUcan}
results, among which the one with highest entropy for given energy has then to be selected. 
	Clearly our new variational approach appears to be more economical than that.

	One of the simplest tasks would be to prove the existence of a unique solution to \Ref{fixPOINTEQrhoU}
at sufficiently high energies $\veps$.
	For Coulomb interactions a unique solution is expected for all energies, while for (regularized) Newton
interactions multiplicity of solutions is expected for sufficiently low energies. 
	This is suggested by the detailed numerical evaluations of the standard principle of maximum
entropy with constraints for related equations, cf. \cite{StahlETal,Chavanis}.
%%%%%%%%%%%%%%%%%%%%%%%%%%%%%%%%%%%%%%%%%%%%%%%%%%%%%%%%%%%%%%%%%%%%%%%%%%%%
%%%%%%%%%%%%%%%%%%%%%%%%%%%%%%%%%%%%%%%%%%%%%%%%%%%%%%%%%%%%%%%%%%%%%%%%%%%%
\subsection{Convergence of the ergodic ensemble measures}
%%%%%%%%%%%%%%%%%%%%%%%%%%%%%%%%%%%%%%%%%%%%%%%%%%%%%%%%%%%%%%%%%%%%%%%%%%%%
%%%%%%%%%%%%%%%%%%%%%%%%%%%%%%%%%%%%%%%%%%%%%%%%%%%%%%%%%%%%%%%%%%%%%%%%%%%%
	We already pointed out in subsection 6.1 that the sequence of ergodic ensemble measures converges 
whenever a unique optimizer exists for the maximum interaction entropy variational principle in 
Theorem \ref{thm:maxENTROPYvp} and Proposition \ref{prop:ACTUALentropyFUNCofRHO}. 
	We don't see any reason why the sequence of ergodic ensemble measures should not converge when 
the entropy maximizer is not unique, and so we expect that the mere existence of limit points concluded 
in this paper by using weak compactness can actually be upgraded to the existence of a limit.
%%%%%%%%%%%%%%%%%%%%%%%%%%%%%%%%%%%%%%%%%%%%%%%%%%%%%%%%%%%%%%%%%%%%%%%%%%%%
%%%%%%%%%%%%%%%%%%%%%%%%%%%%%%%%%%%%%%%%%%%%%%%%%%%%%%%%%%%%%%%%%%%%%%%%%%%%
\subsection{Characterization of the de Finetti--Dynkin measure}
%%%%%%%%%%%%%%%%%%%%%%%%%%%%%%%%%%%%%%%%%%%%%%%%%%%%%%%%%%%%%%%%%%%%%%%%%%%%
%%%%%%%%%%%%%%%%%%%%%%%%%%%%%%%%%%%%%%%%%%%%%%%%%%%%%%%%%%%%%%%%%%%%%%%%%%%%
	As also noted in subsection 6.1, the decomposition measure $\nu(d\tau|\mu_\veps)$ is a 
singleton whenever a unique optimizer exists for the maximum interaction entropy variational principle in 
Theorem \ref{thm:maxENTROPYvp}.
	In more general situations we have little information on the decomposition measure $\nu(d\tau|\mu_\veps)$,
beyond knowing that it reduces to $\varsigma(\dd\rho|\varrho_\veps)$ and that $\varsigma(\dd\rho|\varrho_\veps)$ 
is supported on the maximizers of the maximum interaction entropy principle formulated in
Proposition \ref{prop:ACTUALentropyFUNCofRHO}.
	Of course, we already mentioned earlier that experience with explicitly studied physical systems 
suggests that $\supp \varsigma(\dd\rho|\varrho_\veps)$ is either a finite set or a continuous
group orbit of a compact group, but a general proof or disproof seems not available.
	More is known for the canonical ensemble \cite{KusuokaTamura}, and  their approach should apply 
to the microcanonical ensemble to determine $\nu(d\tau|\mu_\veps)$. 
%%%%%%%%%%%%%%%%%%%%%%%%%%%%%%%%%%%%%%%%%%%%%%%%%%%%%%%%%%%%%%%%%%%%%%%%%%%%
%%%%%%%%%%%%%%%%%%%%%%%%%%%%%%%%%%%%%%%%%%%%%%%%%%%%%%%%%%%%%%%%%%%%%%%%%%%%
\subsection{Large deviation principles}
%%%%%%%%%%%%%%%%%%%%%%%%%%%%%%%%%%%%%%%%%%%%%%%%%%%%%%%%%%%%%%%%%%%%%%%%%%%%
%%%%%%%%%%%%%%%%%%%%%%%%%%%%%%%%%%%%%%%%%%%%%%%%%%%%%%%%%%%%%%%%%%%%%%%%%%%%
	Whenever $\pzcHB(f)$ has a unique minimizer $f_\veps$ over $\Asp_\veps$, then
Theorems \ref{thm:maxENTROPYvp} and \ref{thm:deFinettiERGODE} imply that
\begin{equation}
\Prob\left(\dKR\left(\uli\Delta^{(n)}_{\bX^{(N)}},f^{\otimes n}_{\veps}\right) > \delta\right) \withNto 0
\quad \forall \delta>0,
\label{physPROBnormalEMPmeasDISTfeqTOzeroAGAIN}
\end{equation}
where ``Prob'' refers to the ensemble measure \Ref{BOLTZMANNsERGODE} with Hamiltonian
\Ref{HAMrescaled}.
	It is desirable to improve \Ref{physPROBnormalEMPmeasDISTfeqTOzeroAGAIN} to a large deviation principle, 
a rigorous variation on the theme of Einstein's fluctuation formula.
	Heuristically we expect
\begin{equation}
\Prob\left(\dKR\left(\uli\Delta^{(n)}_{\bX^{(N)}},f^{\otimes n}_{\veps}\right) > \delta\right) 
\asymp \sup_{f\in\Asp_\veps^\delta} e^{-N(\pzcHB(f)-\pzcHB(f_\veps))} \quad \forall \delta>0,
\label{LDP}
\end{equation}
where 
$\Asp_\veps^\delta 
	= \{f \in (\Psp\cap\Lsp^1\cap\Lsp^1\ln\Lsp^1)(\Rset^3\!\times\!\Lambda): \pzcE(f) 
	= \veps \} \backslash \widetilde{B}_\delta(f_\veps)$.
	In \cite{EyinkSpohn,EllisETalA,EllisETalB} such a feat was accomplished for the regularized 
microcanonical ensembles at the level of the 1-point functions.
	The recent article \cite{EichelsbacherSchmock} establishes some nice large deviation principles
for the $n$-point functions in a strong topology which allows one to handle some singular interactions. 
	 We expect that the conjectured large deviation principle can be proved along their lines.

	We also refer to Lanford's article \cite{Lanford} and the books by Varadhan \cite{Varadhan} and
Ellis \cite{EllisBOOK} for mathematical background on large deviation principles and their applications
to statistical mechanics, and to \cite{Touchette} for a more recent review.
%%%%%%%%%%%%%%%%%%%%%%%%%%%%%%%%%%%%%%%%%%%%%%%%%%%%%%%%%%%%%%%%%%%%%%%%%%%%
%%%%%%%%%%%%%%%%%%%%%%%%%%%%%%%%%%%%%%%%%%%%%%%%%%%%%%%%%%%%%%%%%%%%%%%%%%%%
\subsection{Vlasov limit for the canonical ensemble measures}
%%%%%%%%%%%%%%%%%%%%%%%%%%%%%%%%%%%%%%%%%%%%%%%%%%%%%%%%%%%%%%%%%%%%%%%%%%%%
%%%%%%%%%%%%%%%%%%%%%%%%%%%%%%%%%%%%%%%%%%%%%%%%%%%%%%%%%%%%%%%%%%%%%%%%%%%%
	Using the very strategy used in this paper to prove our Theorems \ref{thm:maxENTROPYvp} and 
\ref{thm:deFinettiERGODE}, the Vlasov limit for the {canonical ensemble measures} associated with
\Ref{PARTITIONfunktion} was established in \cite{MesserSpohn,clmpCMPcan,KieCPAM} under various hypotheses
on the interactions, covering our $(H1)$--$(H5)$.
	This raises the question of whether one can conclude the convergence of the {canonical ensemble measures} 
associated with \Ref{PARTITIONfunktion} from the convergence of the microcanonical ensemble measures (or, if 
convergence cannot be shown, the analog for the limit points). 
	Put differently, we ask to extend the conclusions reached at the level of the thermodynamic functions 
to the level of the measures.
	In \cite{EllisETalA,EllisETalB} such a feat was accomplished for the canonical ensemble measures in terms
of regularized microcanonical ensemble measures, using large deviation principle techniques, and issues of
equivalence of ensembles were addressed.

%%%%%%%%%%%%%%%%%%%%%%%%%%%%%%%%%%%%%%%%%%%%%%%%%%%%%%%%%%%%%%%%%%%%%%%%%%%%
%%%%%%%%%%%%%%%%%%%%%%%%%%%%%%%%%%%%%%%%%%%%%%%%%%%%%%%%%%%%%%%%%%%%%%%%%%%%
\subsection{Interactions without lower bound} 
%%%%%%%%%%%%%%%%%%%%%%%%%%%%%%%%%%%%%%%%%%%%%%%%%%%%%%%%%%%%%%%%%%%%%%%%%%%%
%%%%%%%%%%%%%%%%%%%%%%%%%%%%%%%%%%%%%%%%%%%%%%%%%%%%%%%%%%%%%%%%%%%%%%%%%%%%
	By hypothesis $(H2)$ we allow the pair interactions to diverge when two particles approach each other 
infinitely closely.
	However, $W_\Lambda(\qV,\tilde{\qV})$ is only allowed to diverge to $+\infty$,
which happens with the repulsive Coulomb interactions when $\qV\to\tilde{\qV}$.
	Divergence of $W_\Lambda(\qV,\tilde{\qV})$ to $-\infty$ is excluded from our analysis, because
our postulates imply that $I^{(N)}_\Lambda$ is bounded below by $\cE_g(N)>-\infty$.
	In particular, the $-\infty$ singularity of the attractive Newton interactions in $\Rset^3$ will have to be 
regularized. 

	The canonical ensemble and regularized microcanonical ensembles have been controlled under weaker 
hypotheses, allowing in particular the interactions to diverge logarithmically to $-\infty$,
 see \cite{clmpCMPcan,KieCPAM} for the canonical and \cite{clmpCMPmic,KieLebLMP,KiePHYSICA} for the 
regularized microcanonical ensembles.
	It should be possible to adapt the technical arguments in these papers to establish the 
Vlasov limit for \Ref{BOLTZMANNsERGODE} for negative logarithmically singular interactions.
%%%%%%%%%%%%%%%%%%%%%%%%%%%%%%%%%%%%%%%%%%%%%%%%%%%%%%%%%%%%%%%%%%%%%%%%%%%%
%%%%%%%%%%%%%%%%%%%%%%%%%%%%%%%%%%%%%%%%%%%%%%%%%%%%%%%%%%%%%%%%%%%%%%%%%%%%
\subsection{Unbounded domains}
%%%%%%%%%%%%%%%%%%%%%%%%%%%%%%%%%%%%%%%%%%%%%%%%%%%%%%%%%%%%%%%%%%%%%%%%%%%%
%%%%%%%%%%%%%%%%%%%%%%%%%%%%%%%%%%%%%%%%%%%%%%%%%%%%%%%%%%%%%%%%%%%%%%%%%%%%
	In \cite{KieSpoCMP} and \cite{ChaKieDMJ}, unbounded $\Lambda$ where allowed for the canonical ensemble, and
our microcanonical theorems should similarly be extendible to unbounded domains under a suitable confinement 
hypothesis which replaces hypothesis $(H5)$, presumably
\begin{equation}
(H5') \qquad\qquad {\mbox{\textit{Confinement}:}}
	\ e^{- U_\Lambda(\qV,\tilde{\qV})} \in \Lsp^1(\Lambda\!\times\!\Lambda).
\end{equation}
	Incidentally, $(H5')$ not only imposes on behavior of $U_\Lambda$ as any of its two arguments is sent to 
infinity, it also restricts the manner in which $U_\Lambda$ can diverge to $-\infty$, e.g. when its two 
arguments approach each other infinitely closely, allowing logarithmic divergence.
%%%%%%%%%%%%%%%%%%%%%%%%%%%%%%%%%%%%%%%%%%%%%%%%%%%%%%%%%%%%%%%%%%%%%%%%%%%%
%%%%%%%%%%%%%%%%%%%%%%%%%%%%%%%%%%%%%%%%%%%%%%%%%%%%%%%%%%%%%%%%%%%%%%%%%%%%
\subsection{Ergodic ensembles of quasi-particles}
%%%%%%%%%%%%%%%%%%%%%%%%%%%%%%%%%%%%%%%%%%%%%%%%%%%%%%%%%%%%%%%%%%%%%%%%%%%%
%%%%%%%%%%%%%%%%%%%%%%%%%%%%%%%%%%%%%%%%%%%%%%%%%%%%%%%%%%%%%%%%%%%%%%%%%%%%
	Our analysis does not cover ergodic ensembles of quasi-particle systems like point vortices moving 
in two dimensions whose Kirchhoff Hamiltonian is of the type \Ref{HAM} without the sum of $|\pV|^2$ terms.
	The ergodic point vortex ensemble measures are of the type
\begin{equation}
\mu^{(N)}_\cE(\dd^{2N}\!X)
= 
\big(N!{\Omega^\prime_{I^{(N)}_\Lambda}(\cE)}\big)^{-1}\,\delta\big(\cE-I^{(N)}_\Lambda(X^{(N)})\big)\dd^{2N}\!X\,,
\label{ONSAGERsERGODE}
\end{equation}
\vskip-.2truecm
\noindent
where $X^{(N)}: =(\qV_1,...,\qV_N)\in \Lambda^N$, where now $\Lambda\subset\Rset^2$, and $\dd^{2N}\!X$ is 
$2N$-dimensional Lebesgue measure, and the pair interactions now feature positive logarithmic singularities
(for a single specie of point vortices).
	Onsager \cite{Onsager} observed that for such systems a critical $\cE$ value exists
such that the map $\cE\mapsto S(\cE)$ is \emph{decreasing} when $\cE>\cE_{crit}$, giving rise to negative
ensemble temperatures.
	Regularized microcanonical measures for such  vortex Hamiltonians have been analyzed in \cite{clmpCMPmic}
under an equivalence assumption to the canonical ensemble, and in \cite{KieLebLMP,KiePHYSICA} without such an 
equivalence assumption.\footnote{The authors of \cite{clmpCMPmic}
		use the primitive $\Omega_{I^{(N)}_\Lambda}(\cE)$ of $\Omega^\prime_{I^{(N)}_\Lambda}(\cE)$ 
		(i.e. \Ref{STRUKTURfunktion} with $H\equiv I$) 
		\vskip-.1truecm
		\noindent
		to define a quasi-microcanonical ensemble entropy when $\cE<\!\cE_{crit}$, and for $\cE\!>\cE_{crit}$
		they use $\Omega_{I^{(N)}_\Lambda}(\infty)-\Omega_{I^{(N)}_\Lambda}(\cE)$.
		In \cite{KieLebLMP,KiePHYSICA} a Gaussian approximation to $\delta(I-\cE)$ is used.
                We also mention \cite{EyinkSpohn} where the approximation 
		$\Omega_{I^{(N)}_\Lambda}(\cE)-\Omega_{I^{(N)}_\Lambda}(\cE-\triangle\cE)$is used; these
                authors also regularize the logarithmic singularity of the interactions.}
	It is desirable to find a way to handle the proper ergodic ensemble for point vortex and other 
quasi-particle systems for which the sum of squares of kinematical momenta is absent from their Hamiltonian,
but clearly this will require the introduction of new technical ideas.
	Incidentally, this last sentence applies verbatim also to other scalings than Vlasov scaling, in 
particular to the conventional thermodynamic limit scaling explained in the introduction.

	There is one exception to what we just wrote: precisely at the critical energy $\cE_{crit}$ of
a point vortex system it is \emph{a priori} known that all the $n$-point measures have densities given by 
$(1/|\Lambda|)^{\otimes n}$.
	Taking advantage of this fact, O'Neil and collaborators \cite{oneilredner,campbelloneil} found 
that for a neutral \emph{two-species} system the vicinity of $\cE_{crit}\propto N\ln N$ can be analyzed 
directly using $\delta(I-\cE)$; it turns out to be a small-entropy regime where $S$, not $S/N$, converges 
to a  limit when  $N\to\infty$, with $\cE- CN\ln N \propto N$.
	Interestingly enough, this scaling falls in between the conventional thermodynamic limit and the 
Vlasov scaling.
	To the author's knowledge, so far these are the only results for point vortices obtained for 
$\delta(I-\cE)$ proper, i.e. without regularization of the Dirac measure.

\medskip

%%%%%%%%%%%%%%%%%%%%%%%%%%%%%%%%%%%%%%%%%%%%%%%%%%%%%%%%%%%%%%
%%%%%%%%%%%%%%%%%%%%%%%%%%%%%%%%%%%%%%%%%%%%%%%%%%%%%%%%%%%%%%
\noindent
{\textbf{Acknowledgment}:} The author thanks Carlo Lancellotti for his careful reading of the
manuscript and for his comments. 
This paper was written with support from the NSF under grant DMS-0807705.
Any opinions expressed in this paper are entirely those of the author and not necessarily those of the NSF. 
%%%%%%%%%%%%%%%%%%%%%%%%%%%%%%%%%%%%%%%%%%%%%%%%%%%%%%%%%%%%%%
%%%%%%%%%%%%%%%%%%%%%%%%%%%%%%%%%%%%%%%%%%%%%%%%%%%%%%%%%%%%%%

\newpage
\appendix

\section{Monotonicity of the ground state energy}

	In this appendix we will prove two monotonic convergence results about the ground state energy
which are used in the setup of our construction of the Vlasov limit $N\to\infty$.
	The results and their proofs are rather elementary and presumably known, and quite likely 
to be found in the vast literature on $U$ statistics; however, my (certainly incomplete) perusal of
the pertinent literature has not yet met with success.\footnote{In fact, I originally did not 
		expect monotonicity results of the type proved here to hold at all.
		I was prompted to conjecture the results, and then to prove them, by analyzing the numerical 
		results of the computations of the (conjectured) ground state energies $\cE_g(N)$ for Thomson's 
		problem \cite{Thomson} reported in \cite{AWRTSDW,PerezGetal}, which -- divided by either $N^2$ or 
		$N(N-1)$ -- arranged themselves monotonically increasing when plotted vs. $N$. 
		An interesting spin-off of the monotonicity of the pair-specific Thomson energies is 
		a necessary criterion for minimality which can be used as a test for the empirical 
		numerical experiments.
		After the present paper was submitted I successfully carried out such a test; see
		\cite{KieClassEnullPAPER}.}
	
	Here is our first proposition.
\begin{proposition}
\label{prop:UperNsqMinusN}
	Let $\Lambda\subset\Rset^{\rm D}$ be a bounded and connected domain.
	Assume the following hypotheses regarding $U_\Lambda(\qV,\tilde{\qV})$:

\smallskip\noindent
\begin{eqnarray}
&&(H1)\quad   {\mbox{\textit{Symmetry}:}}
		\ U_\Lambda(\check{\qV},\hat{\qV})=U_\Lambda(\hat{\qV},\check{\qV}) 
\nonumber\\
&&(H2)\quad   {\mbox{\textit{Lower\ Semi-Continuity}:}}
	\ U_\Lambda(\check{\qV},\hat{\qV})\ {\rm is\ l.s.c.\ on\ }  \oli{\Lambda}\!\times\!\oli{\Lambda} 
\nonumber\\
&&
(H3)\quad  {\mbox{\textit{Sublevel\ Set\ Regularity}:}}\,
	\lambda^{\otimes 2}\Big( \Big\{
U_\Lambda(\check{\qV},\hat{\qV})- \min U_\Lambda < \eps
\Big\}\Big) 
>0 
\nonumber\\
&&
(H4)\quad  {\mbox{\textit{Local\ Square\ Integrability}:}}
	\ U_\Lambda(\qV,\,\cdot\,)\in \Lsp^2\left(B_r(\qV)\cap\Lambda\right)\ \forall\ \qV\in{\Lambda}
\nonumber
\end{eqnarray}
where $\lambda$ is normalized Lebesgue measure for $\Lambda$.
	For $N\geq 2$ define the pair-specific ground state energy by 
\begin{equation} 
\veps_g(N)
\equiv
\min_{\{\qV_1,\dots,\qV_N\}}
{\textstyle{\frac{1}{N(N-1)}}} \sum\sum_{\hskip-.7truecm 1\leq i < j\leq N} {U}_\Lambda(\qV_i,\qV_j).
\label{pairSPECpotENERGY}
\end{equation}	
	Then the sequence $N\mapsto \veps_g(N)$ so defined  
is monotonic increasing and converges to $\veps_g<\infty$ defined by
\begin{equation}
\veps_g = \min_{\rho\in\Psp(\oli{\Lambda})}
\iint{\tfrhalf}
U_\Lambda(\qV,\tilde{\qV})\rho(\qV)\rho(\tilde{\qV})\dd^{\rm D}q\dd^{\rm D}\tilde{q}.
\label{EPSnull}
\end{equation}
\end{proposition}

	Note that $\veps_g$ as defined in \Ref{EPSnull} coincides with $\veps_g$ as defined in \Ref{Enull}
when ${\rm D}=3$ and $U_\Lambda$ is decomposed into the earlier stipulated sum of $V_\Lambda$ and $W_\Lambda$.

\medskip\noindent
{\textit{Proof of Proposition \ref{prop:UperNsqMinusN}:}} 

	We begin with the mandatory observation that under hypotheses $(H1)$ and $(H2)$ the
pair-specific ground state energy $\veps_g(N)$ defined in \Ref{pairSPECpotENERGY} is well-defined;
i.e. $\veps_g(N)\in\Rset$ (note that $(H3)\&(H4)$ are immaterial here).

	We next prove the monotonicity of $N\mapsto \veps_g(N)$, with $N\geq2$. 
	Elementary (combinatorial) identities and the single inequality that the minimum of a sum is
not less than the sum of the minima shows that $\veps_g(N+1)\geq \veps_g(N)$, viz.
\begin{eqnarray} 
\veps_g(N+1)
\!\!\!\!\!&& =
\min_{\ \{\qV_1,\dots,\qV_{N+1}\}}
{\textstyle{\frac{1}{(N+1)N}}} \sum\sum_{\hskip-.7truecm 1\leq i < j\leq N+1} {U}_\Lambda(\qV_i,\qV_j)
\nonumber\\
\!\!\!\!\!&&  =
\min_{\ \{\qV_1,\dots,\qV_{N+1}\}}
{\textstyle{\frac{1}{(N+1)N}}} \sum_{1\leq k \leq N+1}
\Biggl[
{\textstyle{\frac{1}{N-1}}}
\sum\sum_{\hskip-.7truecm \genfrac{}{}{0pt}{}{1\leq i < j\leq N+1}{ i\neq k\neq j} } 
{U}_\Lambda(\qV_i,\qV_j)
\Biggr]
\nonumber\\
\!\!\!\!\!&& \geq
{\textstyle{\frac{1}{(N+1)N}}} \sum_{1\leq k \leq N+1}
\Biggl[
\min_{\{\qV_1,\dots,\qV_{N+1}\}\backslash\{\qV_k\}}
{\textstyle{\frac{1}{N-1}}}
\sum\sum_{\hskip-.7truecm \genfrac{}{}{0pt}{}{1\leq i < j\leq N+1}{ i\neq k\neq j} } {U}_\Lambda(\qV_i,\qV_j)
\Biggr]
\nonumber\\
\!\!\!\!\!&& =
{\textstyle{\frac{1}{(N+1)N}}}(N+1) 
\Biggl[
\min_{\{\qV_1,\dots,\qV_{N}\}}
{\textstyle{\frac{1}{N-1}}}
\sum\sum_{\hskip-.7truecm {1\leq i < j\leq N}} {U}_\Lambda(\qV_i,\qV_j)
\Biggr]
\nonumber\\
\!\!\!\!\!&&=
\min_{\{\qV_1,\dots,\qV_{N}\}}
{\textstyle{\frac{1}{N(N-1)}}}
\sum\sum_{\hskip-.7truecm {1\leq i < j\leq N}} {U}_\Lambda(\qV_i,\qV_j)
\nonumber\\
\!\!\!\!\!&&=\
\veps_g(N),
\label{pairSPECpotENERGYmono}
\end{eqnarray}
and the proof of monotonicity of $N\mapsto\veps_g(N)$ is complete.

	Next, to prove convergence to $\veps_g$ given by \Ref{EPSnull} we begin by noting that
under hypotheses $(H1)$, $(H2)$ and $(H4)$, the ground state energy $\veps_g$ defined in \Ref{EPSnull}
is well-defined; actually, for this issue we can even relax $(H4)$ to the weaker $\Lsp^1_{loc}(\Lambda)$ 
condition which is implied by $(H4)$.
	We now use the density 
of empirical $N$-point measures in the weakly compact set of all probability measures on $\oli{\Lambda}^2$,
and the existence (by $(H2)\&(H3)$) of a minimizing sequence $\in\Csp^0_b(\Lambda)$ for 
$\langle\rho,\rho\rangle$, to prove convergence $\veps_g(N)\nearrow\veps_g$.
	We let $\uli{\Delta}^{(2)}_{X^{(N)}_g}$ denote the 2-point measure in $\oli{\Lambda}^2$ for 
a ground state $X^{(N)}_g=(0_1,\qV_1;...;0_N,\qV_N)_g$ of $N$ points in $\oli{\Lambda}$
(which need not be unique), and let
$\uli{\Delta}^{(2)}_{X^{({N})}}$ be any other 2-point measure on $\oli{\Lambda}^2$ with $N$ support points.

	We define the linear functional ${}^2\!\rho\mapsto \cU({}^2\!\rho)$ by
\begin{equation}
\cU({}^2\!\rho) 
=
\iint\tfrhalf U_\Lambda(\check{\qV},\hat{\qV})\,{}^2\!\rho(\dd^{\rm D}\check{q}\dd^{\rm D}\hat{q}).
\end{equation}
	Note that for product measures ${}^2\!\rho=\rho^{\otimes 2}$ we have
\begin{equation}
\cU(\rho^{\otimes 2}) 
=
\langle\rho,\,\rho\rangle.
\end{equation}
	Note furthermore that the functional 
${}^2\!\rho\mapsto \cU({}^2\!\rho)$ is generally \emph{not} continuous,
because we have only (weak) lower semi-continuity of $U_\Lambda$.
	In particular, while any continuous change in the supporting points of the 2-point measure 
$\uli{\Delta}^{(2)}_{X^{({N})}}$ on $\oli{\Lambda}^2$ results in a weakly continuous change of the 
2-point measure, the functional $\cU$ evaluated at these 2-point measures, i.e. $\cU(\uli{\Delta}^{(2)}_{X^{({N})}})$, 
generally changes discontinuously.
	However, we do have
\begin{equation}
\veps_g(N) = \cU(\uli{\Delta}^{(2)}_{X^{({N})}_g}) \leq \cU(\uli{\Delta}^{(2)}_{X^{({N})}}).
\label{UofDELTAgUPbound}
\end{equation}
	Now let $\{\rho_n\}_{n\in\Nset}$ be a minimizing sequence in $(\Psp\cap\Csp^0_b)(\oli\Lambda)$
for $\langle\rho,\rho\rangle = \cU(\rho^{\otimes 2})$; note that it is not necessary to postulate also that 
$\rho_n\to\rho$ for any actual minimizer $\rho$, as this will follow automatically from the proof.
	Then, by $(H3)$, for any $\eps>0$ we can find an $n_\eps$ such that 
$\cU(\rho_n^{\otimes 2})\leq \veps_g+\eps$ whenever $n\geq n_\eps$.
	So pick any $\eps>0$, let $n=n_\eps$, and let $\{\qV_k\}_{k\in\Nset}$ be i.i.d. with
a-priori measure $\rho_{n_\eps}\in (\Psp\cap\Csp^0_b)(\oli\Lambda)$ for each $\qV_k$.
	Then by $(H4)$ the weak law of large numbers for $U$ statistics (of order 2) holds \cite{Hoeffding},
and so, in probability,
\begin{equation}
\cU(\uli{\Delta}^{(2)}_{X^{({N})}})
\stackrel{{N}\to\infty}{\longrightarrow} 
\big\langle\rho_{n_\eps},\,\rho_{n_\eps}\big\rangle
\leq
\veps_g +\eps
\label{limUofDELTA}
\end{equation}
for each $\eps>0$.
	By \Ref{limUofDELTA} and \Ref{UofDELTAgUPbound} we have 
\begin{equation}
\limsup_{N\to\infty} \veps_g(N) = \limsup_{N\to\infty} \cU(\uli{\Delta}^{(2)}_{X_g^{({N})}})
\leq 
\veps_g.
\label{limsupUofDELTA}
\end{equation}

	On the other hand, by the compactness of $\oli{\Lambda}$ and the weak$^*$ 
compactness of $\Psp(\oli{\Lambda})$ we can extract a ${}^*$-weakly convergent subsequence 
$\uli{\Delta}^{(2)}_{X^{(\dot{N})}_g}\to {}^2\!\dot\rho \in\Psp(\oli{\Lambda}^2)$.
	Moreover, since any convergent sequence of $n$-point measures $\uli{\Delta}^{(n)}_{X^{(\dot{N})}_g}$
necessarily converges to an $n$-fold product measure, we have ${}^2\!\dot\rho = \dot\rho^{\otimes 2}$.
	Now the weak lower semi-continuity of $\cU$ gives
\begin{equation}
\liminf_{\dot{N}\to\infty} \cU(\uli{\Delta}^{(2)}_{X_g^{(\dot{N})}})
\geq 
\big\langle\dot\rho,\,\dot\rho\big\rangle
\geq 
\veps_g.
\label{liminfUofDELTA}
\end{equation}

	Estimates \Ref{limsupUofDELTA} and \Ref{liminfUofDELTA} prove convergence $\veps_g(N)\to\veps_g$.

	Convergence and the earlier proved monotonicity of $N\mapsto\veps_g(N)$
completes the proof of Proposition \ref{prop:UperNsqMinusN}.
\qed

	We notice that our proof of Proposition \ref{prop:UperNsqMinusN} yields as a ``byproduct''
that $\big\langle\dot\rho,\,\dot\rho\big\rangle= \veps_g$.
	Thus we have the following noteworthy corollary:

\begin{corollary}
 \label{coro:weakLIMofDELTAg}
	Any limit point $\dot\rho^{\otimes 2}$ of the sequence of ground state 2-point measures
$\{\uli{\Delta}^{(2)}_{X_g^{({N})}}\}_{N\in\Nset}$ minimizes the bilinear form 
$\cU(\rho^{\otimes 2})=\langle\rho,\rho\rangle$.
\end{corollary}
	
Here is our second proposition.
\begin{proposition}
\label{prop:UperNsq}
	Assume the hypotheses on $U_\Lambda(\qV,\tilde{\qV})$ stated in the previous proposition, and 
in addition assume that $U_\Lambda\geq 0$.
	Then the quasi pair-specific ground state energy, defined by
\begin{equation} 
\tilde\veps_g(N)
\equiv
\min_{\qV_1,\dots,\qV_N}
{\textstyle{\frac{1}{N^2}}} \sum\sum_{\hskip-.7truecm 1\leq i < j\leq N} {U}_\Lambda(\qV_i,\qV_j),
\label{potHAMwithUU}
\end{equation}	
is a strictly increasing function of $N$ which converges to $\veps_g$ defined in \Ref{EPSnull}.
\end{proposition}

\medskip\noindent
{\textit{Proof of Proposition \ref{prop:UperNsq}:}} 

	First of all, $\tilde\veps_g(N)$ is as well-defined as $\veps_g(N)$.

	Next, inspection of the monotonicity part of the proof of proposition \ref{prop:UperNsqMinusN} reveals that 
the same steps as in \Ref{pairSPECpotENERGYmono} now yield
\begin{equation} 
\tilde\veps_g(N+1)
\geq
{\textstyle{\frac{N^2}{(N+1)(N-1)}}} \,
\tilde\veps_g(N),
\label{QUASIpairSPECpotENERGYmono}
\end{equation}
and $\tilde\veps_g(N)\geq 0$ because of the here assumed positivity of $U_\Lambda$.
	The strict monotonicity of $N\mapsto\tilde\veps_g(N)$ now follows because
\begin{equation} 
{(N+1)(N-1)} < N^2.
\end{equation}

	Lastly, since $1-N^{-1}\to 1$, the limit of $\tilde\veps_g(N)$ coincides with that of $\veps_g(N)$.

	This concludes the proof of Proposition \ref{prop:UperNsq}.
\qed

\section{Decomposition of the finite $N$ measures}

	Let $\varrho_\veps\in\Psp^s(\Lambda^\Nset)$ be the weak limit of
$\{\varrho_{N^2\veps}^{(N)}\in\Psp^s(\Lambda^N)\}_{N\in\Nset}$,
and let $\varsigma(\dd\rho|\varrho_\veps)$ be its unique de Finetti-Dynkin-Hewitt-Savage decomposition measure.
	(If $\{\varrho_{N^2\veps}^{(N)}\}_{N\in\Nset}$ has several limit points, as accounted for in the main
text, the following considerations are valid for the associated converging subsequences of finite $N$ measures.)
	We now show that if $\supp \varsigma(\dd\rho|\varrho_\veps)$ is either a finite set or a continuous
group orbit of a compact group, then for each $\rho\in\supp \varsigma(\dd\rho|\varrho_\veps)$
we can explicitly construct a family of $\varrho^{({N})}[\rho]\in\Psp^s(\Lambda^{{N}})$ satisfying 
\begin{equation}
\lim_{N\to\infty} \varrhon^{({N})}[\rho]
=
\rho^{\otimes n}
\end{equation}
for each $n\in\Nset$, such that for each $N\in\Nset$,
\begin{equation}
 \varrho^{({N})}_{N^2\veps} = \int \varrho^{({N})}[\rho] \,\varsigma(\dd\rho|\varrho_\veps).
\label{finiteNdecompoAPP}
\end{equation}

\subsection{The support of $\varsigma(\dd\rho|\varrho_\veps)$ is a finite set}

	In the simplest case $\varsigma(\dd\rho|\varrho_\veps)$ is a singleton, so that 
$\varrho_\veps = \rho_\veps^{\otimes\Nset}$, i.e. 
\begin{equation}
\lim_{N\to\infty} \varrhon_{N^2\veps}^{({N})}
=
\rho_\veps^{\otimes n} \quad \forall\quad n\in\Nset.
\end{equation}
	In this case 
$\int\varrho^{({N})}[\rho]\,\varsigma(\dd\rho|\varrho_\veps)=\varrho^{(N)}[\rho_\veps] =\varrho_{N^2\veps}^{(N)}$, 
and we are done.

	Next, assume that $\varsigma(\dd\rho|\varrho_\veps)$ is an arithmetic mean of two singletons, viz.
\begin{equation}
 \varsigma(\dd\rho|\varrho_\veps) = \nu_1\delta_{\rho_1}(\dd\rho) +\nu_2\delta_{\rho_2}(\dd\rho)
\end{equation}
with $0<\nu_1 = 1- \nu_2<1$, and let $\dKR(\rho_1,\rho_2)=D>0$ be the usual Kantorovich-Rubinstein
distance between $\rho_1$ and $\rho_2$.
	Let $B_{D/2}(\rho_k)$ be the $KR$-open ball in $\Psp(\Lambda)$ which is centered 
at $\rho_k$ and has radius $D/2$.
	Now decompose $\Lambda^N= \Lambda^N_1\cup\Lambda^N_2$, where $\Lambda_1\cap\Lambda_2=\emptyset$ and
$\varrho^{(N)}_{N^2\veps}(\Lambda_k^N)=\nu_k$, such that 
$\Lambda_k^N$ contains all points for which  ${}^1\uli\Delta^{(N)}\in B_{D/2}(\rho_k)$; when $N$ is too small
there may be no such points, but by the weak density  in $\Psp(\Lambda)$ of the empirical one-point measures 
the set of such points $\in\Lambda^N$ has positive $\varrho^{(N)}_{N^2\veps}$ measure when $N$ is large enough.
	In fact, since by hypothesis the weak limit of
$\{\varrho_{N^2\veps}^{(N)}\in\Psp^s(\Lambda^N)\}_{N\in\Nset}$ is given by 
$\varrho_\veps=\nu_1\rho_1^{\otimes\Nset} +\nu_2\rho_2^{\otimes \Nset}\in\Psp^s(\Lambda^\Nset)$, it follows that when $N\nearrow\infty$ then the probability w.r.t. $\varrho_{N^2\veps}^{(N)}$ that
${}^1\uli\Delta^{(N)}\in B_{D/2}(\rho_k)$ approaches $\nu_k$.
	So if we define
\begin{equation}
 \varrho^{({N})}[\rho_k] = \nu_k^{-1}\varrho^{(N)}_{N^2\veps}\chi_{\Lambda_k^N}
\end{equation}
and recall that $\varrho^{(N)}_{N^2\veps}(\Lambda_k^N)=\nu_k$, it follows that
\begin{equation}
\lim_{N\to\infty} \varrhon^{({N})}[\rho_k]
=
\rho^{\otimes n}_k
\end{equation}
for each $n\in\Nset$ and $k=1$ or $2$, and such that for each $N\in\Nset$,
\begin{equation}
 \varrho^{({N})}_{N^2\veps} = \nu_1 \varrho^{({N})}[\rho_1] + \nu_2 \varrho^{({N})}[\rho_2],
\end{equation}
which is \Ref{finiteNdecompoAPP} in the case that $\varsigma$ is the arithmetic mean of two singletons.

	The general case of $\supp\varsigma$ being a finite set is treated similarly in an obvious manner, with
$D$ now the minimum of the set of distances between any pair $(\rho_k,\rho_l)$ picked from the support of $\varsigma$.
	
\subsection{The support of $\varsigma(\dd\rho|\varrho_\veps)$ is a continuous group orbit}

	For simplicity we assume that we are dealing with a one-parameter continuous group $G$ acting on
the base space, like $SO(2)$ acting on $\Lambda$; the generalization to more complicated situations 
(e.g. $SO(3)$ acting on $\Lambda$) is straightforward.
	In this case we can pick any particular $\rho_0\in\supp\varsigma$ and obtain every other (say)
$\rho_\theta\in\supp\varsigma$ by acting with a group element $g_\theta\in G$ thusly, 
$\rho_\theta = \rho_0\circ g_\theta$.
	The de Finetti etc. decomposition of $\varrho_\veps$ can then be written as an integral w.r.t. Haar
measure over the group $G$ of the infinite product measures $\rho_\theta^{\otimes \Nset}$. 
	The corresponding finite $N$ presentation is simply obtained by change of variables for
$\varrho_{N^2\veps}^{(N)}$ through factoring out the group $G$, which gives each $\varrho^{(N)}[\rho_\theta]$
uniquely.
\newpage

%%%%%%%%%%%%%%%%%%%%%%%%%%%%%%%%%%%%%%%%%%%%%%%%%%%%%%%%%%%%%%%%%%%%%%%%%%%%%%%%%%%%%%%%%%%
%%%%%%%%%%%%%%%%%%%%%%%%%%%%%%%%%%%%%%%%%%%%%%%%%%%%%%%%%%%%%%%%%%%%%%%%%%%%%%%%%%%%%%%%%%%


\begin{thebibliography}{999999.}
\small
\bibitem[Aetal97]{AWRTSDW}
Altschuler, E.L., 
Williams, T.J., 
Ratner, E.R., 
Tipton, R., 
Stong, R., 
Dowla, F., 
and 
Wooten, F.: 
	``Possible global minimum lattice conﬁgurations for Thomson’s problem of charges on the sphere,''
	\textit{Phys. Rev. Lett.} \textbf{78}, 2681--2685 (1997).
\vskip-.3truecm 
\noindent
\bibitem[Bol96]{Boltzmann} 
Boltzmann, L.,
	\textit{Vorlesungen \"uber Gastheorie},
	J.A. Barth, Leipzig (1896); English translation:
	``Lectures on Gas theory'' (S.G. Brush, transl.),
	Univ. California Press, Berkeley (1964).
\vskip-.3truecm 
\noindent
\bibitem[CLMP92]{clmpCMPcan}
Caglioti, E.,
Lions, P. L., 
Marchioro, C., 
and 
Pulvirenti, M., 
	``A special class of stationary flows for two-dimensional
        Euler equations: A statistical mechanics description,'' 
        \textit{Commun. Math. Phys.} \textbf{143}, 501--525 (1992).
\vskip-.3truecm 
\noindent
\bibitem[CLMP95]{clmpCMPmic}
Caglioti, E., 
Lions, P. L., 
Marchioro, C., 
and 
Pulvirenti, M., 
	``A special class of stationary flows for two-dimensional
        Euler equations: A statistical mechanics description. II,'' 
        \textit{Comm. Math. Phys.} \textbf{174}, 229--260 (1995).
\vskip-.3truecm 
\noindent
\bibitem[CON91]{campbelloneil}
Campbell, L. J.,  
and 
O'Neil, K., 
        ``Statistics of two-dimensional point vortices and high energy vortex states,''
        \textit{J. Stat. Phys.} \textbf{65}, 495--529 (1991).
\vskip-.3truecm 
\noindent
\bibitem[ChKi00]{ChaKieDMJ}
Chanillo, S., 
and 
Kiessling, M. K.-H.,
	``Surfaces with prescribed Gauss curvature,''
	\textit{Duke Math. J.} \textbf{105}, 309--353 (2000).
\vskip-.3truecm 
\noindent
\bibitem[Cha02]{Chavanis}
Chavanis, 
P.H., 
	``Phase transitions in self-gravitating systems. Self-gravitating fermions and hard spheres models,'' 
	\textit{Phys. Rev. E} \textbf{65}, 056123 ff. (2002).
\vskip-.3truecm 
\noindent
\bibitem[CETT05]{EllisETalB}
Costeniuc, M., 
Ellis, R.S., 
Touchette, H.,
and
Turkington, B.,
	``The generalized canonical ensemble and its universal equivalence with the microcanonical ensemble,'' 
	\textit{J. Stat. Phys.} \textbf{119}, 1283--1329 (2005).
\vskip-.3truecm 
\noindent
\bibitem[Dyn53]{Dynkin}
Dynkin, E.B.,
	``Klassy ekvivalentnyh slu$\check{c}$a$\check{i}$nyh veli$\check{c}$in,'' 
	\textit{Uspeki Mat. Nauk.} \textbf{6}, 125--134 (1953).
\vskip-.3truecm 
\noindent
\bibitem[EiSch02]{EichelsbacherSchmock}
Eichelsbacher, P.,
and 
Schmock, U.,
	``Large deviations of $U$-empirical measures in strong topologies and applications,''
	\textit{Ann. Inst. Henri Poincar\'e -- PR} \textbf{38}, 779--797 (2002).
\vskip-.3truecm 
\noindent
\bibitem[Ell85]{EllisBOOK}
Ellis, R.S.,
	\textit{Entropy, large deviations, and statistical mechanics}, 
	Springer-Verlag, New York (1985).
\vskip-.3truecm 
\noindent
\bibitem[EHT00]{EllisETalA}
Ellis, R.S., 
Haven, K.,
and
Turkington, B.,
	``Large deviation principles and complete equivalence and nonequivalence results for pure and 
		mixed ensembles,''
	\textit{J. Stat. Phys.} \textbf{101}, 999--1064 (2000).
\vskip-.3truecm 
\noindent
\bibitem[EySp93]{EyinkSpohn}
Eyink, G., 
and 
Spohn, H., 
        ``Negative temperature states and large-scale, 
        long-lived vortices in two-dimensional turbulence,''
        \textit{J. Stat. Phys.} \textbf{70}, 833--886 (1993).
\vskip-.3truecm 
\noindent
\bibitem[deF37]{deFinetti}
de Finetti, B.,
		``La pr\'evision: ses lois logiques, ses sources subjectives,''
	\textit{Annales Inst. Henri Poincar\'e} \textbf{7}, 1--68 (1937).
\vskip-.3truecm 
\noindent
\bibitem[Gib02]{Gibbs} 
Gibbs, J.W.,
	\textit{Elementary Principles in Statistical Mechanics},
	Yale Univ. Press, New Haven (1902); reprinted by Dover,
	New York (1960).
\vskip-.3truecm
\noindent
\bibitem[Gri65]{Griffiths} 
Griffiths, R.B.,
	``Microcanonical ensemble in quantum statistical mechanics,''
	\textit{J. Math. Phys.} \textbf{6}, 1447--1461 (1965).
\vskip-.3truecm 
\noindent
\bibitem[HeSa55]{HewittSavage}
Hewitt, E., 
and 
Savage, L. J., 
		``Symmetric measures on Cartesian products,'' 
        \textit{Trans. Amer. Math. Soc.} \textbf{80}, 470--501 (1955).
\vskip-.3truecm 
\noindent
\bibitem[Hoe48]{Hoeffding}
Hoeffding, W.
	``A class of statistics with asymptotically normal distributions''
	\textit{Annals Statist.} \textbf{19}, 293--325 (1948);
	Reprinted (except for sect. 9.e-9.h)  in: 
	\textit{Breakthroughs in Statistics}, Vol. I, p.308-334 (Kotz, S., Johnson, N.L.; eds.)
	Springer-Verlag, New York  (1992).
\vskip-.3truecm \noindent
\bibitem[Kie93]{KieCPAM} 
Kiessling,  M. K.-H., 
		``Statistical mechanics of classical particles with logarithmic interactions,''
        \textit{Commun. Pure Appl. Math.} \textbf{47}, 27--56 (1993).
\vskip-.3truecm \noindent
\bibitem[Kie00]{KiePHYSICA} 
Kiessling, M.K.-H., 
	``Statistical mechanics approach to some problems in conformal geometry''
	\textit{Physica} \textbf{A 297}, 353-368 (2000).
\vskip-.3truecm \noindent
\bibitem[Kie08]{KieASSISI} 
Kiessling, M.K.-H., 
	        ``Statistical equilibrium dynamics,''
	pp.91-108 in \textit{AIP Conf. Proc.} \textbf{97}, 
	A. Campa, A. Giansanti, G. Morigi, and F. Sylos Labini (eds.), American Inst. Phys. (2008).
\vskip-.3truecm \noindent
\bibitem[Kie09a]{KieRUELLEpaper}
Kiessling, M.K.-H., 
		``On Ruelle's construction of the thermodynamic limit for the classical microcanonical entropy,''
\textit{J. Stat. Phys.} \textbf{134}, 19--25 (2009).
\vskip-.3truecm \noindent
\bibitem[Kie09b]{KieClassEnullPAPER}
Kiessling, M.K.-H., 
		``A note on classical ground state energies,''
\textit{J. Stat. Phys.} \textbf{136}, 275--284 (2009).
\vskip-.3truecm \noindent
\bibitem[KiLa09]{KieLanINprep}
Kiessling, M.K.-H., 
and 
Lancellotti, C.,
		(in preparation) (2009).
\vskip-.3truecm \noindent
\bibitem[KiLe97]{KieLebLMP}
Kiessling, M.K.-H., 
and 
Lebowitz, J.L.,
		``The microcanonical point vortex ensemble: Beyond equivalence,''
	\textit{Lett. Math. Phys.} \textbf{42}, 43--56 (1997).
\vskip-.3truecm \noindent
\bibitem[KiSp99]{KieSpoCMP}
Kiessling, M.K.-H., 
and 
Spohn, H.,
	``A note on the eigenvalue density of random matrices,''
	\textit{Commun. Math. Phys.} \textbf{199}, 683--695 (1999).		
\vskip-.3truecm \noindent
\bibitem[KuTa84]{KusuokaTamura}
Kusuoka, S., 
and 
Tamura, Y., 
		``Gibbs measures for mean field potentials,''
	\textit{J. Fac. Sci. Univ. Tokyo, Sec. IA, Math.} \textbf{31}, 223--245 (1984).
\vskip-.3truecm
\noindent
\bibitem[Lan73]{Lanford}
Lanford, O.E.,III.,
		``Entropy and equilibrium states in classical statistical physics,''
	pp.1--107 in \cite{BattelleSeattleRecontres} (1973).
\vskip-.3truecm
\noindent
\bibitem[Len73]{BattelleSeattleRecontres}
Lenard, A. (ed.), 
	\textit{Statistical mechanics and mathematical problems}, 
	Conf. Proc. of the Battelle Seattle Recontres 1971, \textit{Lect.  Notes Phys.} \textbf{20}
	(J. Ehlers et al., eds.), Springer (1973).
\vskip-.3truecm
\noindent
\bibitem[ML79]{MartinLoef}
Martin-L\"of, A.,
	\textit{Statistical mechanics and the foundations of thermodynamics},
	in \textit{Lect. Notes Phys.} \textbf{101} (J. Ehlers et al., eds.), Springer (1979).
\vskip-.3truecm
\noindent
\bibitem[MeSp82]{MesserSpohn}
Messer, J., 
and 
Spohn, H., 
		``Statistical mechanics of the isothermal Lane-Emden equation,'' 
        \textit{J. Stat. Phys.} \textbf{29}, 561--578 (1982).
\vskip-.3truecm \noindent
\bibitem[LBLB99]{LBLBa} 
Lynden-Bell, D. and R.M.,
	``Exact general solutions to extraordinary $N$-body problems,''
	\textit{Proc. R. Soc. Lond. A} \textbf{445}, 475--489 (1999).
\vskip-.3truecm \noindent
\bibitem[LBLB04]{LBLBb} 
Lynden-Bell, D. and R.M.,
	``Relaxation to a perpetually pulsating equilibrium,'' 
	\textit{J. Stat. Phys.} \textbf{117}, 199--209 (2004).
\vskip-.3truecm \noindent
\bibitem[ONR91]{oneilredner}
O'Neil, K., 
and 
Redner, R. A.,
        ``On the limiting distribution of pair-summable
        potential functions in many-particle systems,''
        \textit{J. Stat. Phys.} \textbf{62}, 399--410 (1991).   
\vskip-.3truecm \noindent
\bibitem[Ons49]{Onsager}
Onsager, L., 
	``Statistical hydrodynamics,''
	\textit{N. Cim. Sup.} \textbf{6},~279-287~(1949)
\vskip-.3truecm 
\noindent
\bibitem[Pen70]{PenroseBOOK} 
Penrose, O.,
	\textit{Foundations of Statistical Mechanics: A Deductive Treatment},
	Pergamon Press, Oxford (1970);	reprinted by Dover (2005).
\vskip-.3truecm
\noindent
\bibitem[Petal97]{PerezGetal}
P\'erez-Garrido, A.,
Dodgson, M.J.W.,
Moore, M.A.,
Ortu\~no, M.,
and 
D\'iaz-S\'anchez, A.,
	``Comment on \textit{Possible Global Minimum Lattice Configurations for Thomson's Problem
	of Charges on a Sphere}''
	\textit{Phys. Rev. Lett} \textbf{79}, 1417 (1997).
\vskip-.3truecm \noindent
\bibitem[ReSi80]{ReedSimon}
Reed, M.,
and
Simon, B.,
	\textit{Methods of Modern Mathematical Physics I},
	Academic Press, New York (1980).
\vskip-.3truecm \noindent
\bibitem[RoRu67]{RobinsonRuelle}
Robinson, D. W., 
and 
Ruelle, D., 
	``Mean entropy of states in classical statistical mechanics,'' 
	\textit{Commun. Math. Phys.} \textbf{5}, 288--300 (1967).
\vskip-.3truecm \noindent
\bibitem[Rue69]{RuelleBOOK} 
Ruelle, D.
	\textit{Statistical Mechanics: Rigorous Results},
	Benjamin, New York (1969);
	reprinted in the ``Advanced Book Classics'' series of
	Addison-Wesley, Reading (1989).
\vskip-.3truecm \noindent
\bibitem[Spo91]{SpohnBOOK} 
Spohn, H., 
	\textit{Large Scale Dynamics of Interacting Particles}, 
	Texts and Monographs in Physics, Springer (1991).
\vskip-.3truecm \noindent
\bibitem[SKS95]{StahlETal}
Stahl B., 
Kiessling M.K.H., 
and
Schindler K., 
	``Phase transitions in gravitating systems and the formation of condensed objects,''
	\textit{Planet. Space Sci.} \textbf{43}, 271--282 (1995). 
\vskip-.3truecm 
\noindent
\bibitem[Tho04]{Thomson}
Thomson, J.J.,
	``On the Structure of the Atom: an Investigation of the Stability and Periods of Oscillation 
	of a number of Corpuscles arranged at equal intervals around the Circumference of a Circle; 
	with Application of the results to the Theory of Atomic Structure,''
 	\textit{Philos. Mag.} \textbf{7}, 237--265 (1904).
\vskip-.3truecm 
\noindent
\bibitem[Tou08]{Touchette}
Touchette, H.,
	``Review on large deviation theory and statistical mechanics,''
	\textit{Phys. Rep.} \textbf{478}, 1--69 (2009).
\vskip-.3truecm 
\noindent
\bibitem[Var84]{Varadhan}
Varadhan, S.R.S.,
	\textit{Large deviations and applications},
	CBMS-NSF Regional Conf. Ser. in Appl. Math. \textbf{46} (1984).
\end{thebibliography}
\end{document}